\documentclass[preprint]{aastex62}
\usepackage{natbib}

\usepackage{color}
\shorttitle{Partial Covering in SDSS~J0850$+$4451}
\shortauthors{Leighly et al.}

\begin{document}


\title{The z=0.54 LoBAL Quasar SDSS~ J085053.12+445122.5: II. The
  Nature of Partial Covering in the Broad-Absorption-Line Outflow}


\author{Karen M.\ Leighly}
\affiliation{Homer L.\ Dodge Department of Physics and Astronomy, The
  University of Oklahoma, 440 W.\ Brooks St., Norman, OK 73019}

\author{Donald M.\ Terndrup}
\affiliation{Homer L.\ Dodge Department of Physics and Astronomy, The
  University of Oklahoma, 440 W.\ Brooks St., Norman, OK 73019}
\affiliation{Department of Astronomy, The Ohio State University, 140
  W.\ 18th Ave., Columbus, OH 43210} 

\author{Adrian B.\ Lucy}
\altaffiliation{LSSTC Data Science Fellow}
\affil{Columbia University, Department of Astronomy, 550 West 120th
  Street, New York, NY 10027}

\author{Hyunseop Choi}
\affiliation{Homer L.\ Dodge Department of Physics and Astronomy, The
  University of Oklahoma, 440 W.\ Brooks St., Norman, OK 73019}

\author{Sarah C.\ Gallagher}
\affiliation{Department of Physics \& Astronomy, The University of Western
  Ontario, London, ON, N6A 3K7, Canada}
\affiliation{Canadian Space Agency, 6767 Route de l'Aeroport,
  Saint-Hubert, Quebec, J3Y~BY9}
\affiliation{The Centre for Planetary and Space Exploration, The
  University of Western Ontario, London, ON, N6A 3K7, Canada}
\affiliation{The Rotman Institute of Philosophy, The University of
  Western Ontario, London, ON, N6A 3K7, Canada}

\author{Gordon T.\ Richards}
\affiliation{Department of Physics, Drexel University, 3141 Chestnut Street,
  Philadelphia, PA 19104}

\author{Matthias Dietrich}
\altaffiliation{Deceased 2018 July 19}
\affiliation{Earth, Environment, and Physics, Worcester State University,
  Ghosh Science and Technology Center, Worcester, MA 01602}

\author{Catie Raney}
\affiliation{Department of Physics and Astronomy, Rutgers, The State
  University of New Jersey, 136 Frelinghuysen Rd., Piscataway, NJ 08854}




\begin{abstract}
It has been known for 20 years that the absorbing gas in broad
absorption line quasars does not completely cover the continuum
emission region, and that partial covering must be accounted for to
accurately measure the column density of the outflowing gas.  However,
the nature of partial covering itself is not understood.  Extrapolation of
the {\it SimBAL} spectral synthesis model of the {\it HST} COS  UV
spectrum from SDSS~J0850+4451 reported by \citet{leighly18} to
non-simultaneous rest-frame optical and near-infrared spectra reveals
evidence that 
the covering fraction has wavelength dependence, and is a
factor of 2.5 times higher in the UV than in the optical and
near-infrared bands.   The
difference in covering 
fraction can be explained if the outflow consists of clumps that are
small and either structured or clustered relative to the projected
size of the UV continuum emission region, and have a more diffuse
distribution on size scales comparable to the near-infrared continuum
emission region size. { The lower covering fraction over the larger
  physical area results in a reduction of the measured total column
  density by a factor of 1.6 compared with the UV-only solution.}
This experiment demonstrates that we can compare rest-frame UV and
near-infrared  absorption lines, specifically \ion{He}{1}*$\lambda
10830$, to place constraints on the uniformity of absorption gas in
broad absorbing line quasars.    
\end{abstract}


\section{Introduction}\label{intro}

Broad absorption lines are found in the rest-frame UV spectra of a
significant fraction of quasars \citep[e.g.,][]{weymann91,gibson09}.
Most often, these lines are blueshifted with velocities as high as
tens of thousands $\rm km\, s^{-1}$, indicating the presence of
powerful outflows.  Optical spectra of $z\sim 2$ broad absorption line
quasars (BALQs) include absorption lines from  Ly$\alpha$, \ion{N}{5},
\ion{Si}{4}, and \ion{C}{4}, among others. Early on, scientists
recognized that it might be possible to use these lines to determine
the metallicity of the outflowing gas, thereby potentially
constraining the physical conditions in the quasar central regions 
and the potential for enrichment of the IGM \citep[see][and references
  therein]{hamann98a}.  However, they quickly
discovered that the implied metallicities were enormous \citep[e.g.,
  20--100 times solar;][]{hamann98a}.  Especially
problematic were quasars with \ion{P}{5}$\lambda 1118,1128$ absorption
lines, as phosphorus is a relatively rare 
element with an abundance only  $9.3\times 10^{-4}$ that of carbon
\citep{grevesse07}. \citet{hamann98a} proposed that, instead, the
absorber only partially covers the continuum source, so that an
absorption line from a high-abundance ion such as C$^{+3}$ can be
completely saturated without dropping to zero flux density, and a
low-abundance ion such as P$^{+4}$ can show significant optical depth.  

Additional support for partial covering comes from doublet analysis in
objects with relatively narrow absorption lines.  The ratios of
opacities of absorption lines from the same lower level are fixed by
atomic physics.  For example, due to the fine structure of the upper
level, the \ion{C}{4} absorption line at 1548\AA\/  (upper level
configuration $^2P_{3/2}$, with degeneracy 4) will have approximately
twice the opacity of the line at 1550\AA\/ (upper level configuration
$^2P_{1/2}$, with degeneracy 2).  If the apparent opacities are less
than 2:1, then the presence of partial covering is inferred
\citep[e.g.,][]{hamann01}.  In fact, partial covering is routinely
used to distinguish narrow absorption lines (NALs) intrinsic to the
quasar from absorption lines from  intervening gas which would
completely cover the continuum source
\citep[e.g.,][]{misawa07,rh11,ganguly13}.   

The spectropolarimetric properties of BALQs also provide evidence for
partial covering.  Often, the polarization is
stronger in the BAL troughs of polarized BALQs
\citep{ogle99,dipompeo13}, indicating that at least in some cases, 
scattered light fills in the troughs, or at least contributes to the
continuum in the trough.   

Once partial covering was recognized to be nearly ubiquitous in
quasars, investigators set about trying to account for it.
Fortunately, many of the most prominent absorption lines come from
lithium- or sodium-like ions (e.g., C$^{+3}$, N$^{+4}$, O$^{+6}$, Mg$^{+}$,
Al$^{+2}$, Si$^{+3}$, P$^{+4}$, Ca$^{+}$); all of these ions share the
atomic structure discussed above for C$^{+3}$, namely, a doublet
transition from the ground state, with the optical depth ratios of 2:1
fixed by atomic physics.  Optical depth measurements of both lines
yield two equations for two unknowns (the covering fraction and the
optical depth), implying that the true optical depth could be solved
for exactly \citep[e.g.,][]{hamann01}.  This method works well
\citep[e.g.,][]{arav05, arav08, borguet12, borguet13, chamberlain15, dunn10,
  finn14, gabel05, hamann97, hamann01, hamann11, hall03, moravec17,
  moe09, rh11}  as long as both lines are not saturated.  The fact
that the optical depths 
are 2:1 means that these lines have a rather limited dynamic range in
optical depth over which they are useful.  \citet{leighly11} discussed a
potentially very useful set of lines that arise from metastable helium,
especially \ion{He}{1}*$\lambda 10380$, the $2S \rightarrow 2P$
transition, and \ion{He}{1}*$\lambda 3889$, the $2S \rightarrow 3P$
transition, which have an opacity ratio of 23:1. { This high ratio
makes these lines ideal for very high column density outflows, which
are potentially the most interesting for identifying quasar activity
that is likely to affect the host galaxy}.  The true column
density of hydrogen can be estimated using the Lyman series lines
\citep[e.g.,][]{gabel05}, although this can be difficult in cases
where the Ly$\alpha$ forest is present and blended with the absorption
lines of interest { as is generally the case for quasars found at
  the epoch of peak quasar activity ($z=1$--3).}

The partial covering analysis discussed above implicitly assumes that
part of the continuum emission region is completely covered and the
other part is completely bare.  This ``step function'' \citep{arav05}
partial covering is not the only possibility, and indeed, early on it
was recognized that abundant ions tended to have higher covering
fractions than lower abundance ions  \citep{hamann01}.  A popular 
second model, called the inhomogeneous absorber \citep{dekool02c,
  arav05} or the power-law partial covering model \citep{arav05,
  sabra05}, posits that the optical depth has a power-law
distribution over the continuum emission region.  The power-law
partial covering model can naturally account for the difference in
apparent covering fractions among ions. The step function and
power-law partial covering models  can be distinguished if there are
more than two absorption lines from the same lower level, and detailed
analysis shows that the inhomogeneous absorber model is sometimes
preferred \citep{dekool02c,   arav05}.    

Most of these analyses ignore the interesting question of the physical
origin of partial covering.   Absorption lines are observed along the
line of sight to the continuum emission region in the central engine.
From the point of view of an observer on Earth, the continuum emission
region is spatially unresolved.  But from the point of view of the
absorber, the continuum emission region may be spatially
resolved.  Moreover, the continuum emission is assumed to come from an
accretion disk, hotter in the center and cooler at larger radii, which
means that the continuum emission region is resolved as a function of
wavelength too.  For example, a simple sum-of-blackbodies accretion
disk model has radial temperature dependence $T\propto R^{-3/4}$.
Therefore it is possible that the absorber, located in the vicinity of the
torus at $\sim 1\rm \, pc$, for example, may present a higher covering
fraction to the hot and compact central part of the accretion disk
than to cooler parts at larger radii.  Therefore, the covering
fraction measured in the UV bandpass refers to a much smaller
continuum emission region than the covering fraction in the optical
and near-infrared bands.   Analysis 
of partial covering as a function of wavelength would lead to an
enhanced understanding of the geometry of 
the absorber in BALQs as well as constrain the relative angular sizes
of the continuum emission regions as a function of wavelength.  So,
instead of only obtaining information along a single radial sight
line, we would be able to investigate the angular distribution as
well.   A caveat is that we assume that negligible flux is scattered
into our line of sight.   

To do this experiment, we clearly need to analyze lines widely
separated in wavelength to probe different size scales of the
accretion disk.  However, we cannot obtain this information
from just any pair of absorption lines, due to the fact that abundant
ions have higher covering fractions than less abundant ions:  the pair of
lines should have about the same optical depth in the gas.
\citet{leighly11} showed that this criterion is fulfilled over a wide
range of physical conditions by \ion{P}{5}$\lambda \lambda 1118, 1128$
and the metastable helium lines, in particular \ion{He}{1}*$\lambda
10830$. These two lines probe dramatically different size
scales of the accretion disk; for the sum-of-blackbodies model
\citep[e.g.,][]{fkr02}, the radius of the accretion disk emitting at 1
micron is a factor of 12 times larger than the radius of the accretion
disk emitting at 1100\AA\/ (see \S\ref{partial_covering}).  In terms
of area, this corresponds to a factor of 140, i.e., dramatically
different size scales. 

The low redshift ($z=0.5422$) LoBAL quasar 
SDSS~J085053.12$+$445122.5, hereafter referred to as SDSS~J0850+4451,
was discovered to have a \ion{He}{1}*$\lambda 3889$ absorption line in
its SDSS spectrum \citep[e.g.,][]{luo13}.  We obtained {\it Gemini} GNIRS and
LBT LUCI near-infrared spectroscopic observations and identified the
presence of a deep \ion{He}{1}*$\lambda 10830$ absorption line
(\S\ref{observations} below).  SDSS~J0850$+$4451 was
detected by {\it GALEX}, indicating that it was bright enough to be
observed by {\it HST} using COS.  The COS spectral analysis was
described in \citet{leighly18}, hereafter Paper I, using a novel spectral
synthesis program called {\it SimBAL}.   The results of that analysis
are summarized in \S\ref{recap}.  We extrapolated the {\it SimBAL}
best-fitting solutions to the optical and near-IR, finding that the
predicted 
absorption was significantly deeper than observed
(\S\ref{extrapolation}), and therefore apparently indicated differential
partial covering.  The {\it HST} and near-IR observations were not
simultaneous, and we investigate the potential impact of variability
on our result in Appendix~\ref{variability}.  In addition, the host
galaxy emits strongly at 1 micron, i.e., under the
\ion{He}{1}*$\lambda 10830$ absorption line, so we performed
spectral energy distribution (SED)
fitting and image analysis to show that the contribution of the host galaxy 
to the 1-micron continuum is negligible and is therefore not filling
in the absorption line (Appendix~\ref{host}).   A
quantitative analysis of the difference in covering fraction in the
UV, optical, and near-IR is
reported in \S\ref{quantifying}, and an analysis of the difference in
the covering fraction between the continuum and broad emission lines
is discussed in \S\ref{blr}.  { A discussion of the nature of the
power-law covering-fraction parameterization is given in
\S\ref{understanding}.}    The implications of our results on 
our understanding of the physical properties of partial covering are
discussed in \S\ref{discussion},  and the summary of our principal 
results is given in \S\ref{conclusions}. Vacuum wavelengths are used 
throughout. Cosmological parameters used depend on the context (e.g.,
when comparing with results from an older paper), and are reported in 
the text.

\section{Observations and Data Reduction}\label{observations}

We report data from taken at six different observatories.
We obtained near-IR spectra at Gemini (\S\ref{gemobs}) and LBT
(\S\ref{lbt}) to measure the properties of  the   \ion{He}{1}*$\lambda
10830$ line.  To mitigate against absorption-line variability
confounding the \ion{He}{1}* analysis,   we obtained new optical
spectra at MDM observatory (\S\ref{mdmobs})   contemporaneous with the
near-IR observations, as well as near-IR   photometry to estimate the
host galaxy contribution through SED   fitting.  Subsequent optical
spectra obtained   at APO (\S\ref{apoobs}) and KPNO (\S\ref{kpnoobs}),
combined with   the SDSS and BOSS spectra (\S\ref{sdss}), were used to
track   absorption-line variability. The log of all of the
observations of SDSS~J0850$+$4451 analyzed in 
this paper is given in Table \ref{obslog}.

\begin{deluxetable}{lcccc}
\tablewidth{0pt}
\tabletypesize{\scriptsize}
\tablecaption{Observations of SDSS~J0850$+$4451\label{obslog}}
\tablehead{
 \colhead{Observatory and Instrument} & 
 \colhead{Date} &
 \colhead{Exposure} & 
 \colhead{Rest Frame Band Pass or} & 
 \colhead{Resolution}\\ & & 
 \colhead{(s)} & 
 \colhead{Effective Wavelength (\AA\/)} & }
\startdata
SDSS & 2002 Nov 27 & 9000.0 & 2472--5975 & $100\rm \, km\, s^{-1}$ \\
HST (WFC3 IR) & 2010 April 9 & 905.9 & 12500 & 0.13 arc sec/pixel \\
LBT (LUCI) & 2010 Dec 12 & 1500.0 & 9512--15304 & $160 \rm \, km\, s^{-1}$\\
MDM (CCDS) & 2011 Feb 11 & 9600.0 &  3121--4108 & $210 \rm \, km\, s^{-1}$\\
{\it Gemini} (GNIRS) & 2011 Apr 23, 24; 2011 Jun 6 & 1520.0 &
5513--16466 & $240 \rm \, km\, s^{-1}$ \\
MDM (TIFKAM) & 2012 Dec 29 & 990, 720, 720 & 8105, 10700, 14265 & 1.0 arcsec \\
APO (DIS) & 2014 Apr 12 & 4500.0 & 2206--6353 & 380, $400 \rm \, km\, s^{-1}$\tablenotemark{a} \\
BOSS & 2015 Jan 20 & 3600.0 & 2345--6740 & $89\rm \, km\, s^{-1}$ \\
KPNO (KOSMOS) & 2015 Apr 24 & 3600.0 & 3804--6631 & $120 \rm \, km\, s^{-1}$ \\
\enddata
\tablenotetext{a}{{Resolutions at \ion{Mg}{2} and \ion{He}{1}*$\lambda
  3889$ absorption lines, respectively.}}
\end{deluxetable}

\subsection{{\it Gemini} GNIRS Observations}\label{gemobs}

SDSS~J0850$+$4451 was observed using
GNIRS\footnote{http://www.gemini.edu/sciops/instruments/gnirs} on the
Gillett Gemini 
Telescope using a standard cross-dispersed mode (the
SXD camera with the $31.7 \rm \, l/mm$ grating) and a $0{\farcs}45$
slit. 
Observations were made on 23 April 2011, 24 April 2011, 26 May 2011,
and 7 June 2011.  The 26 May observation was deemed unusable due to
detector noise, as the detector read mode had been mistakenly set at
``Very Bright/Acq./High Bckgrd'', rather than ``Very Faint Objects''
mode.  On 23 April 2011, $8\times 190$ second exposures were made, in
an ABBA pattern. On 24 April 2011, $8 \times 190$ second exposures
were made, in an ABBA pattern.   On 6 June 2011, $4 \times 190$ second
exposures were made, also in an ABBA pattern.  A0 stars were observed
at approximately the same airmass and adjacent to the object
observation for telluric correction.   The data were reduced
using the IRAF {\it Gemini} package, coupled with the GNIRS XD reduction
scripts, in the standard manner for near-infrared spectra, through the
spectral extraction step.  For telluric correction, the {\it Gemini} spectra
of the source and the telluric standard star were converted to a
format that resembled IRTF SpeX data sufficiently that the Spextool
{\tt xtellcor} package \citep{cushing04, vacca03} could be used.     

\subsection{LBT Observation}\label{lbt}

SDSS~J0850$+$4451 was observed using LBT
LUCI\footnote{http://abell.as.arizona.edu/$\sim$lbtsci/Instruments/LUCIFER/lucifer.html}
on 12 December
2010. Six exposures were made, with the object offset along the slit
between each observation.   An A0 star was observed adjacent to the
target observation at approximately the same airmass.  There is no
reduction pipeline for LUCI data, so the data were reduced by hand
using IRAF.  Because the target was much fainter than the sky lines,
special care was taken to straighten the object trace and sky lines.
Wavelength correction was obtained using sky lines.  The telluric
correction was performed using {\tt xtellcor\_general}, the
generalization of the {\tt xtellcor} procedure for 1-D (versus
cross-dispersed) spectra \citep{cushing04, vacca03}.

The LBT spectrum and the three {\it Gemini} spectra were combined.
First, the four
spectra were resampled onto a common wavelength range, and averaged
without weighting.  The GNIRS spectrum obtained on 23 April 2011
appeared to have the best signal-to-noise ratio and the best
calibration, and the other spectra were normalized and tilted to
conform with that one. 

In \S\ref{selection}, we discuss an {\it LBT} observation of the
quasar PG~1254$+$047.   It was observed using LBT {\it LUCI} on 2013 Jan 3
for 960 seconds in eight exposures using an ABBA configuration.  The
A0 telluric star HD 116960 was observed immediately after the
PG~1254$+$047 observation in four 12-second exposures for a total of 48
seconds.  Standard methods for extraction and wavelength calibration
were done using IRAF.  The telluric correction was done using IRTF
{\tt xtellcor\_general} \citep{vacca03}.

\subsection{KPNO Observation}\label{kpnoobs}

We obtained $3\times1200\rm \, s$ optical spectra of
SDSS~J0850$+$4451 on the night of UT 24 April 2015 
using the KOSMOS spectrograph \citep{martini14} on the Mayall
Telescope at the Kitt Peak National Observatory.  We employed the blue
VPH grism and center slit, which yielded spectra from 3804 --
6631\AA\/ at 0.69 \AA\/ ${\rm pixel}^{-1}$.  The slit width was 
$0{\farcs}9$, and typical seeing was about $1{\farcs}2$. The
resolution of the spectra, as measured by telluric emission lines, was
2.6 pixels at the center of the spectra and 2.8 pixels at either end.  

 The
output images of the spectrograph had dimension $320 \times 4096$
pixels, read out through two amplifier sections of size $160 \times
4096$ pixels.  All the data were contaminated by fixed pattern noise
which was symmetric on the two amplifiers. The spectra were positioned
along the slit so that they fully fell on one amplifier.  The first
step in the data processing was to apply an overscan correction on
each amplifier, and then to remove the pattern noise by flipping the
image section from the side not containing the spectrum and then
subtracting it from the side that did.  This also partially subtracted
the night sky lines, which extended across both amplifiers.  Other
calibration steps were the subtraction of zero-exposure frames and the
application of flat-field corrections; the latter were constructed
from a combination of quartz lamps in the spectrograph and lamps
illuminating a white spot on the inside of the telescope dome.  After
flat fielding, cosmic rays in the sky regions of the image were
removed by hand using a filter that replaced pixel values more than
5$\sigma$ from the median with the median value.  After cosmic-ray
cleaning, the three exposures were averaged and the spectrum extracted
in the usual fashion.  Noise as a function of counts in the spectrum
was estimated from the scatter after subtracting a highly smoothed
spectrum.  The spectrum was flux calibrated using
observations of Feige 34 that were taken the same night as
SDSS~J0850$+$4451. 

\subsection{MDM Observations }\label{mdmobs}

SDSS~J0850$+$4451 was observed using the Boller \& Chivens CCD
Spectrograph
(CCDS)\footnote{http://www.astronomy.ohio-state.edu/MDM/CCDS/} on the
Hiltner 2.4m telescope at MDM observatory on 11 Feb 2011 under
photometric conditions.  Eight 20-minute observations were made.
The data were reduced in a standard manner using IRAF.  

SDSS~J0850$+$4451 is a relatively faint target for 2MASS; the quality
flags associated with this in the catalog are ``BCB'' indicating that
the H band photometry is especially uncertain.  Therefore, we also
obtained deep JHK imaging observations in order to obtain the
photometry for SED fitting
with the goal of constraining the contribution of the host 
galaxy to the near-UV continuum (\S\ref{sed_fitting}).  We used
TIFKAM\footnote{http://www.astronomy.ohio-state.edu/MDM/TIFKAM/}
\citep{depoy93} 
at the 2.4m Hiltner Telescope of the MDM observatory. We employed the
f/5 reimaging camera, which delivered a field 5.1 arcmin over 1024
pixels at a frame scale of $0\farcs 30\ {\rm pixel}^{-1}$.  About 10\%
of each frame was slightly vignetted from an out-of-alignment internal
baffle, but this was corrected in the reduction process.

The observations were obtained on the night of UT 2012 December 29.
For each filter, we obtained a series of 90-second exposures with
position offsets varied irregularly between exposures.  The total exposure
time in $J$ was 990 seconds, while in $H$ and $K$ the total was 720
seconds.  After a small linearity correction, the pixel values in the
images from  each filter were scaled and then combined by a median to 
produce a sky 
frame; this was also corrected for dark current to generate a flat
field. The corrected images were then combined by averaging to
produce master images in each of $JHK$. Photometry of
SDSS~J0850$+$4451 was derived from these combined images with respect
to the 2MASS catalog values for the other objects on each frame, which
were almost always brighter than SDSS~J0850$+$4451. SDSS~J0850$+$4451
is one of the reddest objects in the nearby field, but its $J - H$ and
$J - K$ 
colors are within the range spanned by the other objects.  Evidence
for significant color terms in the transformation to the 2MASS system
was marginal, so in the end we derived a simple constant offset to
transform from the instrumental TIFKAM ``m'' to 2MASS ``M''
magnitudes: $M = m + {\rm const}$. Errors in the photometry were
derived from the standard deviation of the magnitudes on individual
frames and propagation of the 2MASS catalog error values.  The final
values from our photometry were $J = 16.202 \pm 0.024$, $H = 15.507
\pm 0.023$ and $K = 14.763 \pm 0.028$. 

\subsection{SDSS and BOSS Spectra}\label{sdss}

SDSS~J0850$+$4451 was observed using SDSS on 27 Nov 2002.  The
MDM and SDSS spectra had very similar emission and absorption lines,
so they were averaged over the segment including the
\ion{He}{1}*$\lambda 3889$ line, in order to increase signal-to-noise
ratio, after the MDM spectrum was tilted and scaled to match the SDSS 
spectrum.  

SDSS~J0850$+$4451 was observed again using BOSS on 20 Jan 2015.  As
will be discussed in \S\ref{obs_var}, the continuum shows an unusual
shape at the blue end of the spectrum (Fig.~\ref{fig13}).  Because
this observation was made relatively close in time to the KPNO
observation (within 3 months), and since the KPNO shows no such
unnatural shape, we suspect that a calibration problem is responsible
rather than a real change in continuum shape.  Note that this should
not be the atmospheric differential refraction problem known to plague
the BOSS spectrograph \citep{margala16} as correction for that issue
was included in the DR14 pipeline. 

\subsection{APO Observation}\label{apoobs}

We obtained spectra on the night of UT 2014 April 12 using the
Dual Imaging
Spectrograph
(DIS)\footnote{http://www.apo.nmsu.edu/arc35m/Instruments/DIS/} 
spectrograph on the ARC 3.5m telescope at the Apache Point  
Observatory.  This two-channel spectrograph uses a dichroic to
simultaneously obtain spectra in a blue and red channel.  On the blue
side we used the B400 grating, which delivered spectra from 3402--5564
\AA\/  at a dispersion of $1.82$ \AA\/ pixel$^{-1}$; the red side 
used the R300 grating, yielding spectra from 5281--9796\AA\/ at
2.66\AA\/ pixel$^{-1}$.  We observed with a $1\farcs5$ wide slit. The
dispersed images in both channels underfill the CCD detectors
spatially, so these wavelength ranges were set by determining the regions
where the locations of the SDSS~J0850$+$4451 spectrum could reliably
be traced. The  wavelength solution is not reliable for the first and
last $\sim 100$ \AA\/
of each spectrum because of a lack of arc lamp lines near the edges of
each image.  The spectral resolution on the blue and red sides was 3.0
and 2.9 pixels FWHM, respectively. Image processing and spectral extraction
were performed using the same techniques as for the KPNO spectra.  Flux
calibration was determined using spectra of Feige 34 obtained near in
time to the SDSS~J0850$+$4451 spectra. 

\section{Continuum Modeling}\label{contmod}

Our focus in this paper is on the absorption lines.  Therefore, we
model the continuum with several components including the emission
lines and divide by the result before performing the 
absorption-line modeling.  As in Paper I, we first correct the
spectrum for Milky Way reddening using $E(B-V)=0.024$ \citep{sf11},
and for the cosmological redshift  $0.5422$, estimated from the narrow
[\ion{O}{3}] line in the SDSS spectrum.

\begin{figure*}[!t]
\epsscale{1.0}
\begin{center}
\includegraphics[width=4.5truein]{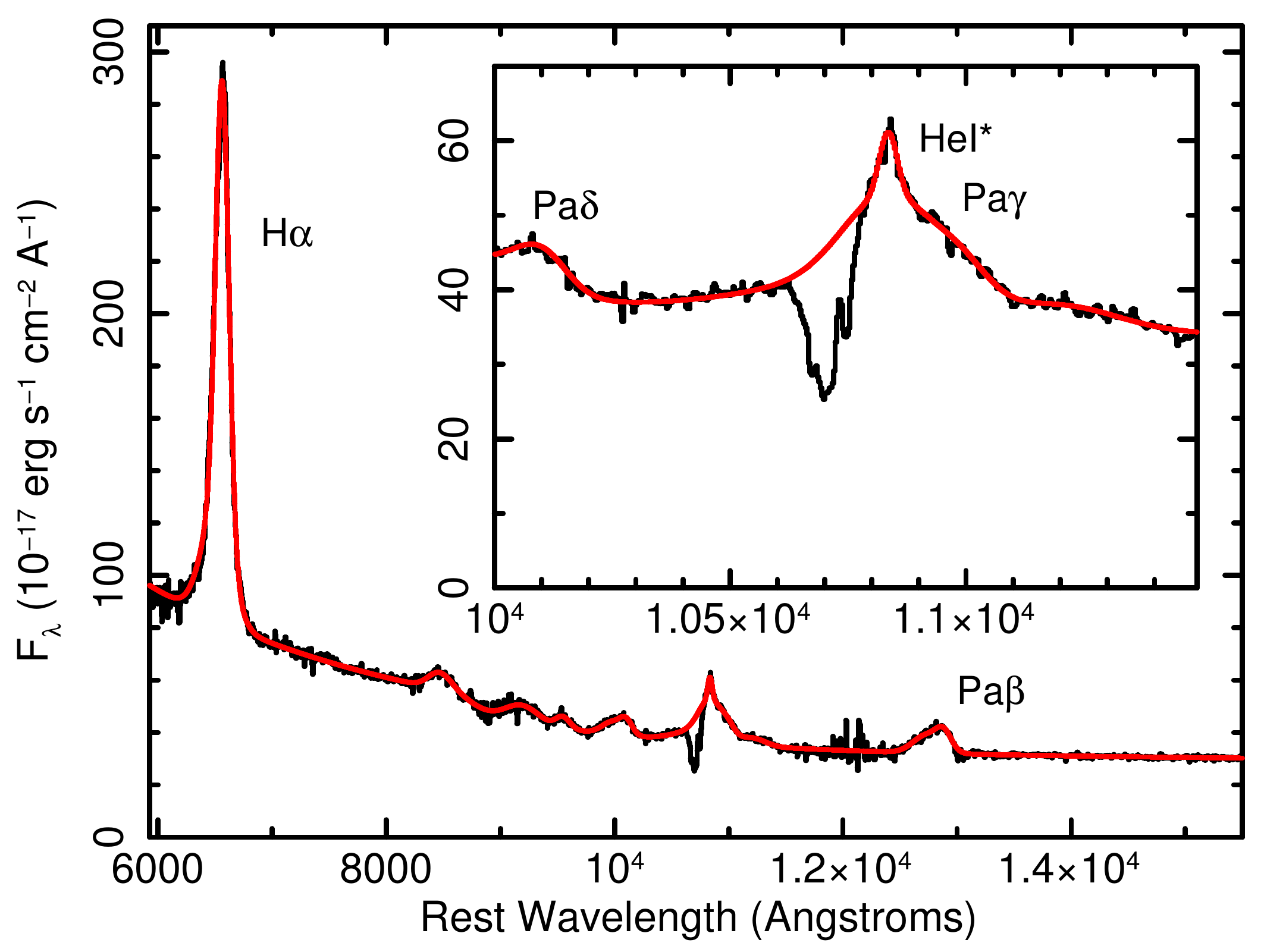}
\caption{The combined {\it Gemini} GNIRS and {\it LBT} LUCI spectrum
  of SDSS~J0850+4451, fitted with a continuum plus emission lines
  model (red curve).  The   inset shows   the bandpass that includes
  the \ion{He}{1}*$\lambda   10830$   absorption line.  Principal
  emission lines are marked. \label{fig1}}  
\end{center}
\end{figure*}

The SDSS~J10850$+$4451 near-infrared continuum shows the characteristic
break between the optical power law originating in the accretion disk 
and the near-infrared bump due to hot dust.  We used  {\it Sherpa}
\citep{freeman01} to  model the
entire continuum spectrum using a power law for the accretion disk
continuum and a black body for the thermal dust emission, plus a
modest contribution from Paschen recombination continuum  near
8000\AA\/. Both H$\alpha$ and Pa$\beta$ are present in the spectra.
Both of these lines could be modeled with two Gaussians.  H$\alpha$ is
somewhat broad and mostly symmetric, while Pa$\beta$ is very broad
with a prominent blue wing. Pa$\delta$ was also fit well with the same
profile as Pa$\beta$.  In the vicinity of the \ion{He}{1}*$\lambda
10830$ absorption line, the principal emission lines are
\ion{He}{1}*$\lambda 10830$, and 
Pa$\gamma\, \lambda  10941$, plus some low-level emission longward of
Pa$\gamma$, possibly attributed to \ion{O}{1}$\lambda 11290$.  We
found that we could only obtain a satisfactory fit if 
\ion{He}{1}*$\lambda 10830$ has a shape more similar to H$\alpha$,
i.e., as two Gaussians with FWHM and wavelength tied to the H$\alpha$
parameters, but with the flux free to vary, plus a  narrower 
\ion{He}{1}* component with FWHM $1210\rm \, km\, s^{-1}$.  It must be
noted that a prominent sky line falls at the wavelength of the
putative narrow component, so the properties and necessity of that
component are uncertain.   The resulting fit is shown in
Fig. \ref{fig1}.   

\begin{figure*}[!t]
\epsscale{1.0}
\begin{center}
\includegraphics[width=4.5truein]{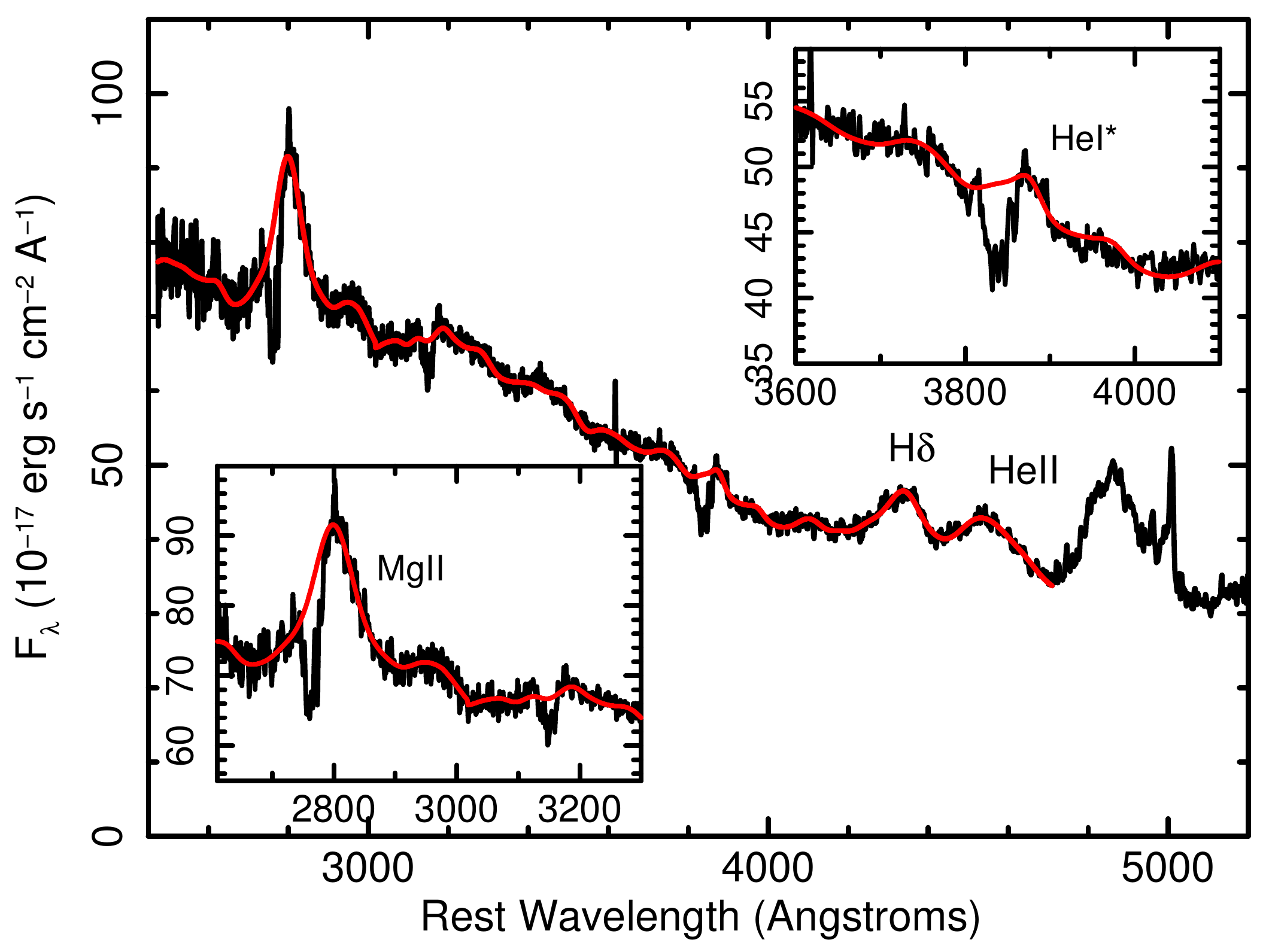}
\caption{The combined SDSS and MDM spectrum of SDSS~J0850+4451, fitted
  with a continuum model (red curve).  The insets show the bandpasses
  that include   the \ion{He}{1}*$\lambda 3889$ (upper right), and
  \ion{Mg}{2} and   \ion{He}{1}*$\lambda 3188$ absorption
  lines (lower left). Principal emission   lines are
  marked. \label{fig2}}  
\end{center}
\end{figure*}

The combined SDSS and MDM spectrum (Fig.~\ref{fig2}) shows that
  SDSS~J0850$+$4451 is a 
broad-line AGN with modest, broadened \ion{Fe}{2} emission.  
The absorption lines are easily identified against the continuum.
For the blue optical wavelengths, we used the composite \ion{Fe}{2}
spectra developed in \citet{leighly11}.  For the region around
\ion{Mg}{2}, we used an \ion{Fe}{2} emission spectrum extracted by us
from the {\it HST} observation of I~Zw~1 \citep{lm06}.  Both
were convolved with a Gaussian with a width of $2000\rm \, km\,
s^{-1}$.    We modeled the spectrum with the broad \ion{Fe}{2}, a
broken power law, a small amount of Balmer continuum, and broad
Gaussians for the \ion{Mg}{2} emission lines, as well as broad
Gaussians for H$\delta$ and \ion{He}{2}$\lambda 4865$.  This
continuum isolates the \ion{He}{1}*$\lambda 3889$,
\ion{He}{1}*$\lambda 3188$, and \ion{Mg}{2} absorption lines
(Fig.~\ref{fig2}).   

\section{Partial Covering Absorption in
  SDSS~J0850$+$4451}\label{absorption_modeling} 

\subsection{Summary of Paper I}\label{recap}

The goal of the multi-wavelength observations of SDSS~J0850$+$4451 was
to investigate the nature of partial covering in this object.  In
Paper I, we described the analysis of  the {\it HST} COS spectrum of
SDSS~J0850$+$4451 using our novel spectral synthesis code 
{\it  SimBAL}.  We briefly review the most relevant aspects of that 
analysis and the results to set the stage for the partial-covering
analysis described in this paper. 

The {\it SimBAL} analysis method uses large grids of ionic column
densities extracted from {\it Cloudy} \citep{ferland13} models to
create synthetic spectra as a function of velocity, covering fraction,
ionization parameter, density, and a combination parameter $\log N_H -
\log U$.  We use the Markov Chain Monte Carlo code {\tt
  emcee}\footnote{http://dan.iel.fm/emcee/current/} \citep{emcee} to
compare the continuum-normalized {\it HST} 
spectrum with the synthetic spectra, using $\chi^2$ as the likelihood
estimator.  The results of the modeling process are posterior
probability  distributions of the model parameters, which were used to
construct the best-fitting model spectrum and its uncertainties, and
to extract best-fitting model parameters and uncertainties.  From
these, the physical parameters of the outflow, including the total column
density, mass outflow rate, momentum flux, and kinetic luminosity were
derived.   

We developed an innovative method to model the velocity dependence of
the outflow parameters.  We divided the trough into a specified number
of velocity bins, where each bin is required to have the same width,
but the physical parameters of the gas were allowed to vary in each
bin.    The central velocity of the highest-velocity bin and the bin
width were fitted parameters.  For SDSS~J0850$+$4451, we ran models
with from 7 to 12 bins in order to investigate systematic uncertainty
associated  with the number of bins; we found that the dependence on
number of bins is small.  In addition, we considered two models for
the continuum that differ somewhat in the modeling of the Ly$\alpha$
and \ion{N}{5} emission line region; see Paper I for details.  We 
considered two {\it   Cloudy} input  spectral energy distributions, a
relatively soft one that may be characteristic of quasars
\citep{hamann11}, and a hard one that may be more suitable for
Seyferts \citep{korista97}.  Finally we considered two cases for the
metallicity, solar and $Z=3 Z_\odot$, both for the soft SED.  { For
  the enhanced metallicity models, we followed \citet{hamann02}: all
  metals were set to three times their solar value, while nitrogen was
  set to nine times the solar value, and helium was set to 1.14 times
  the solar value.}  As discussed in Paper I, the results were largely
independent of these differences in models.  

A number of results were robust to variations in our models.  The
trough spans $-6000$ to $-1000 \rm \, km\, s^{-1}$.   We found
significant structure in $\log N_H-\log U$ as a function of velocity,
namely an enhancement in the column density by a factor of three
around $-4000\rm \, km\, s^{-1}$.  We refer to the this
velocity-resolved feature as ``the concentration''.  Both the
ionization parameter and the column density were larger at higher
speeds.  The covering fraction showed a strong decrease with speed.

We estimated the bulk properties of the outflow from our results.
The total column density of the outflowing gas  $\log N_H$ lay between
$22.4$ and $22.9 \rm \, cm^{-2}$,   depending on the metallicity ($Z=3
Z_\odot$ and solar, respectively).  The  density-sensitive line
\ion{C}{3}*$\lambda 1175$  constrained the distance of the outflow
from the continuum-emission region to be between 1   and 3
parsecs. { \ion{C}{3}*$\lambda 1175$ arises from three
  fine-structure levels, each of which has its own critical
  density \citep[e.g.,][Fig.\ 5]{gabel05}.  While the $J=0$ level is
  populated at relatively low densities, the $J=1$ becomes
  significantly populated toward $\log n=6\rm \, [cm^{-3}]$, increasing
  the opacity of the transition significantly.}
Assuming that the whole outflow (i.e., including the velocity bins
that were not represented in the \ion{C}{3}* line) lies at
approximately the same distance from the central engine, we found that
the mass outflow rate is 17--28 solar masses per year, the 
 momentum flux is approximately equal to  $L_{Bol}/c$,  and the
ratio of the kinematic to bolometric luminosity is 0.8--0.9\%. This
range is greater than 0.5\% \citep{he10}, generally taken to be
the lower bound required for a quasar outflow to effectively
contribute to quasar feedback in galaxy evolution scenarios.  The
ability to model the velocity dependence of physical properties, as
well as extract the global outflow properties illustrates the power of
the forward-modeling methodology used by {\it SimBAL}.

\subsection{Extrapolation to Longer Wavelengths}\label{extrapolation}

\begin{figure*}[!t]
\epsscale{0.7}
\begin{center}
\includegraphics[width=5.5truein]{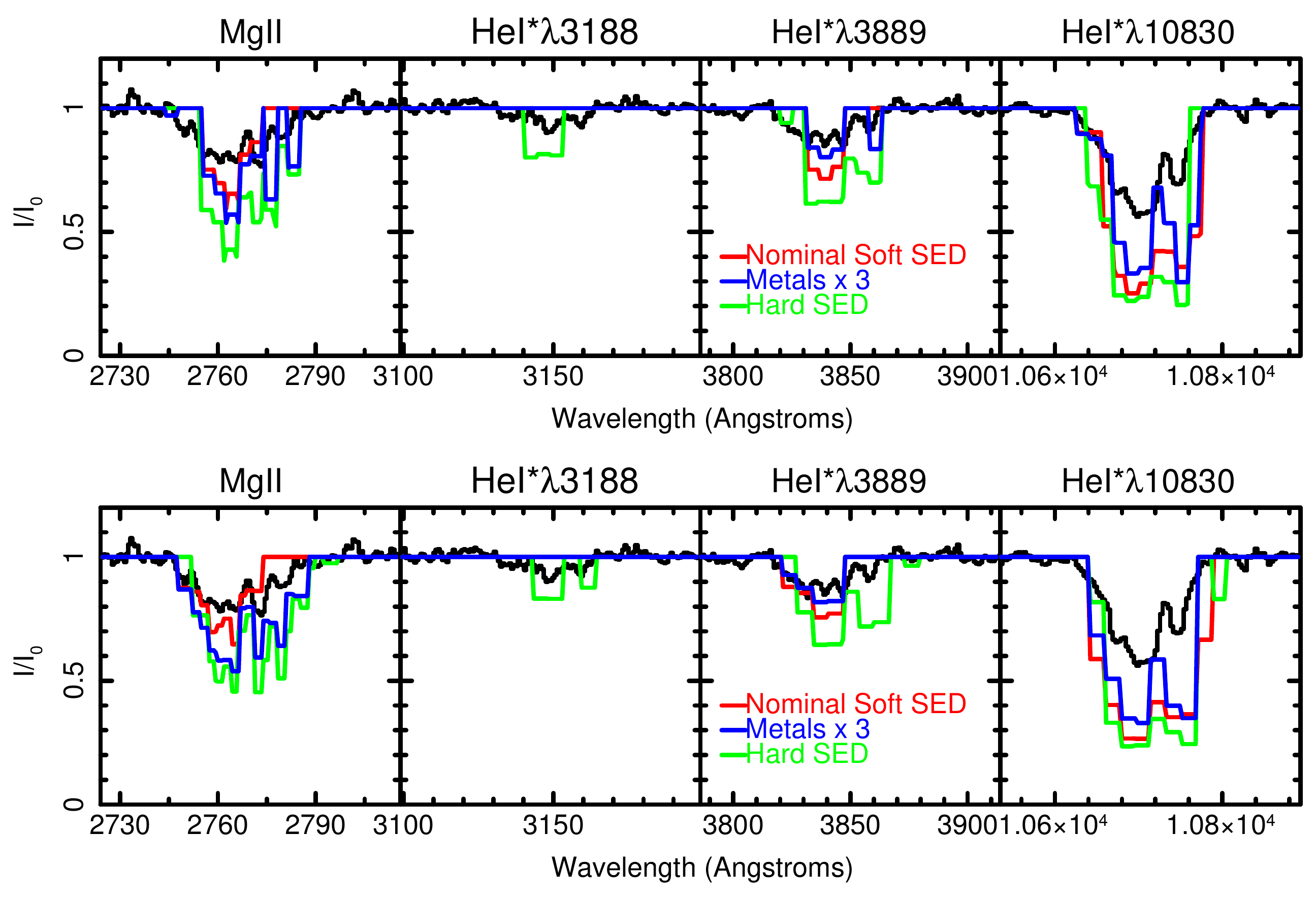}
\caption{The median spectra from the parameter posterior distributions
  for the 11-velocity-bin model fitted to {\it HST} spectrum
  and extrapolated to longer  wavelengths.
  The top panel shows the results using the first continuum model, and
  the   lower panel shows the results using the second  continuum
  model, characterized by stronger Ly$\alpha$ emission  (see Paper   I
  for details).   While   the overall correspondence between the shapes of the profiles is
  good (in particular, the ``mitten'' shape for \ion{He}{1}*$\lambda
  10830$), the model over-predicts the absorption line depth for the
  \ion{Mg}{2} and    \ion{He}{1}* lines, especially for the
  \ion{He}{1}*$\lambda   10830$ line.  \label{fig13}}    
\end{center}
\end{figure*}

In the near UV and optical spectrum, we observe absorption
lines from  \ion{Mg}{2}, \ion{He}{1}*3188, and \ion{He}{1}*3889
(Fig.~\ref{fig2}).  In the near-infrared spectrum, we observe
\ion{He}{1}*10830 (Fig.~\ref{fig1}).  We extrapolated 
best-fitting models from Paper I to longer wavelengths.  As the
solutions were largely independent of the number of bins, we chose
the 11-bin models from Paper I as representative, and plot the results
for the nominal soft SED, the hard SED, and the higher metallicity
(and nominal soft SED) for each continuum model. The flux-density
median model spectra are shown in Fig.~\ref{fig13}.  

This figure shows that the model that fits the UV  over-predicts the
\ion{Mg}{2} and  \ion{He}{1}* opacity.  Generally speaking, the hard
SED produces the worst fit, predicting far more opacity for all three
lines than observed.  { This is because a harder SED produces a
  thicker \citep[e.g.,][Fig.\ 13]{casebeer06} and hotter
  \citep[e.g.,][Fig.\ 14]{leighly07} \ion{H}{2} region; \ion{He}{1}*
  shows a mild dependence on temperature \citep{clegg87}.} 
 The enhanced metallicity model fits the
\ion{He}{1}*$\lambda 3889$ line rather well, but it over-predicts the
\ion{Mg}{2} absorption.  All models predict  much more absorption at
\ion{He}{1}*$\lambda 10830$ than is observed.  

At first glance, this result might imply that the
covering fraction of the longer wavelength continuum emission region
is lower than that of the shorter wavelength continuum emission
region.  This would allow more continuum emission to reach the
observer, producing a shallower line.  
However, there are two factors that we needed to consider before we
could draw this conclusion.  First, it turns out that SDSS~J0850+4451
has demonstrated absorption line variability, and our ground-based optical
and near-infrared observations were not simultaneous with the {\it
  HST} observation.  We explore the potential effects of variability
on our experiment in Appendix~\ref{variability}.  We conclude that 
variability is unlikely to have caused the difference between the 
observed line depths and the extrapolated model line depths, although
we cannot rule it out absolutely.  Second, the
\ion{He}{1}*$\lambda 10830$ line is located near 1 micron, the region
of the spectrum where the host galaxy is the brightest.  So it is
conceivable that the continuum is diluted by the 
presence of the host galaxy, making the line appear shallower than it
is.  We explore this possibility in Appendix~\ref{host}.  We conclude
that the host galaxy contribution to the continuum under the
\ion{He}{1}*$\lambda 10830$ line is negligible.

\subsection{Quantifying the Difference in Partial
  Covering}\label{quantifying}  

Having established that the difference in partial covering
implied from the extrapolated best-fitting UV spectrum is not an
artifact of variability or host galaxy contamination, we proceeded to
investigate it quantitatively.  As described 
in Paper I, we parameterized the partial covering using a power law, 
where $\tau = \tau_{max} x^{a}$.  Here, $\tau$ is the integrated
opacity of the line, and $\tau_{max}$ is proportional to  $\lambda f_{ik}
N({\rm ion})$, where $\lambda$ is the wavelength of the line, $f_{ik}$ is
the oscillator strength, $N({\rm ion})$ is the ionic column density
\citep[e.g.,][]{ss91}, $x \in (0,1)$ represents the fractional surface
area, and $a$, or more specifically $\log a$, is the parameter that is
modeled.  We chose this formalism because we compute the model
spectrum line by line, and we require a scheme that is mathematically
commutative.  The power-law partial-covering model has              
been explored by  \citet{dekool02c, sabra01, arav05}, and in several
cases is has been found to provide a better fit than the step-function
partial covering model \citep{dekool02c, arav05}.

As discussed in Paper I, the power-law covering fraction has the
property that the fraction of the continuum covered depends on
the opacity of the line, which means that the residual intensity can
vary dramatically among lines with different opacity for the same
value of $a$.  So a particular value of $\log a$ will produce lines
that are nearly black for a common ion, and lines that are quite
shallow for a rare ion.       In addition,
as discussed by \citet[][e.g., their   Fig.\ 1]{sabra01}, a value of
$a$ equal to 1 ($\log a=0$) corresponds to 50\% coverage (for a line
with total opacity equal to 1), while $a$ approaching zero corresponds
to full coverage, and high values of $a$ correspond to a small
fraction covered.  Thus, the fitting parameter $\log a$ has an inverse
behavior: it is smaller for a larger fraction covered, and larger for
a smaller fraction covered.  See \S\ref{understanding} for further
discussion of inhomogeneous partial covering and the power law
parameterization.  

\begin{figure*}[!t]
\epsscale{1.0}
\begin{center}
\includegraphics[width=6.5truein]{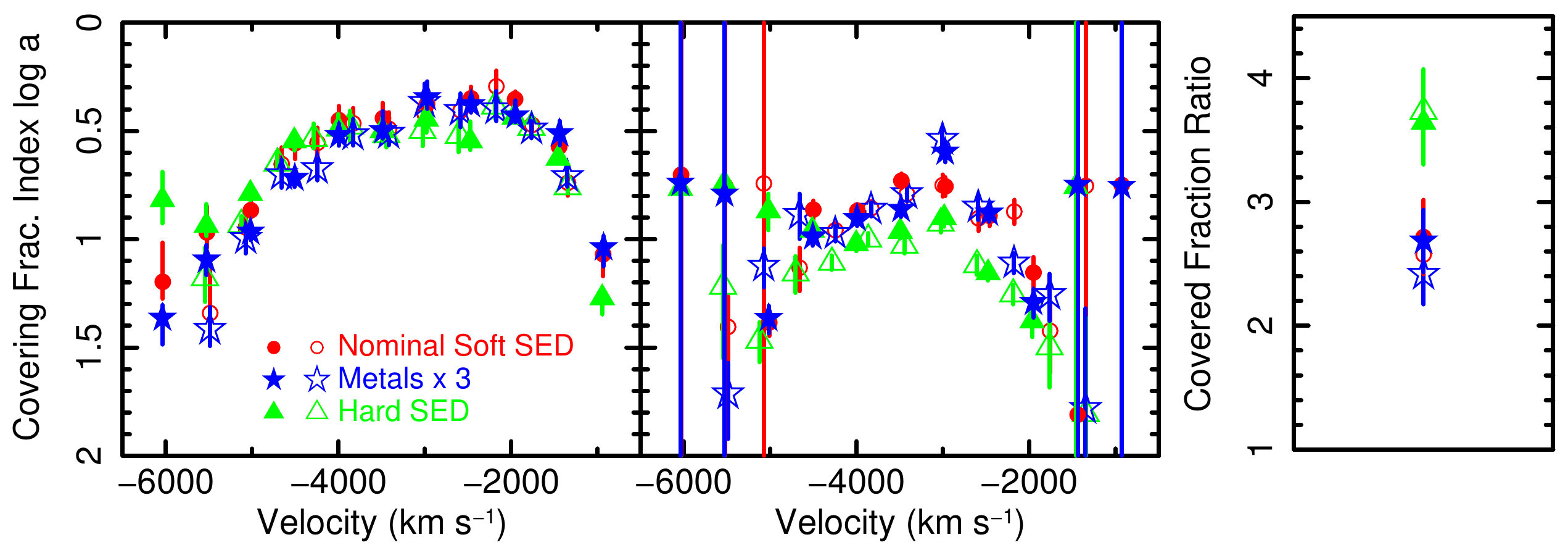}
\caption{The partial covering parameter $\log a$ constraints. The
  left panel shows the results for the full model fit to the {\it
    HST} spectra.  The open and solid points show the results from the
  first  and second continuum models, respectively (see the text and
  Paper I for details).  The middle panel shows the results from
  modeling the long wavelength spectra, fixing all parameters except
  the eleven covering fraction values at their best-fitting values from
  the UV-only model.  The value of the covering fraction parameter $\log
  a$ is systematically higher, indicating a lower covering fraction
  (note that the y-axis goes from larger to smaller values).
  The right panel shows the ratio of the mean $a$ values obtained by
  fitting a constant between $-4500$ and $-1500\rm \, km\, s^{-1}$.
  The ratio of   the fractions covered by a given opacity approaches
  these values as   $\tau$ approaches $\tau_{max}$.  We see that the
  UV-continuum-emitting region has a larger covering fraction by a
  factor  2.5 to 3.75 as $\tau$ approaches
  $\tau_{max}$.    \label{fig19}}  
\end{center}
\end{figure*}

\begin{figure*}[!t]
\epsscale{1.0}
\begin{center}
\includegraphics[width=5.5truein]{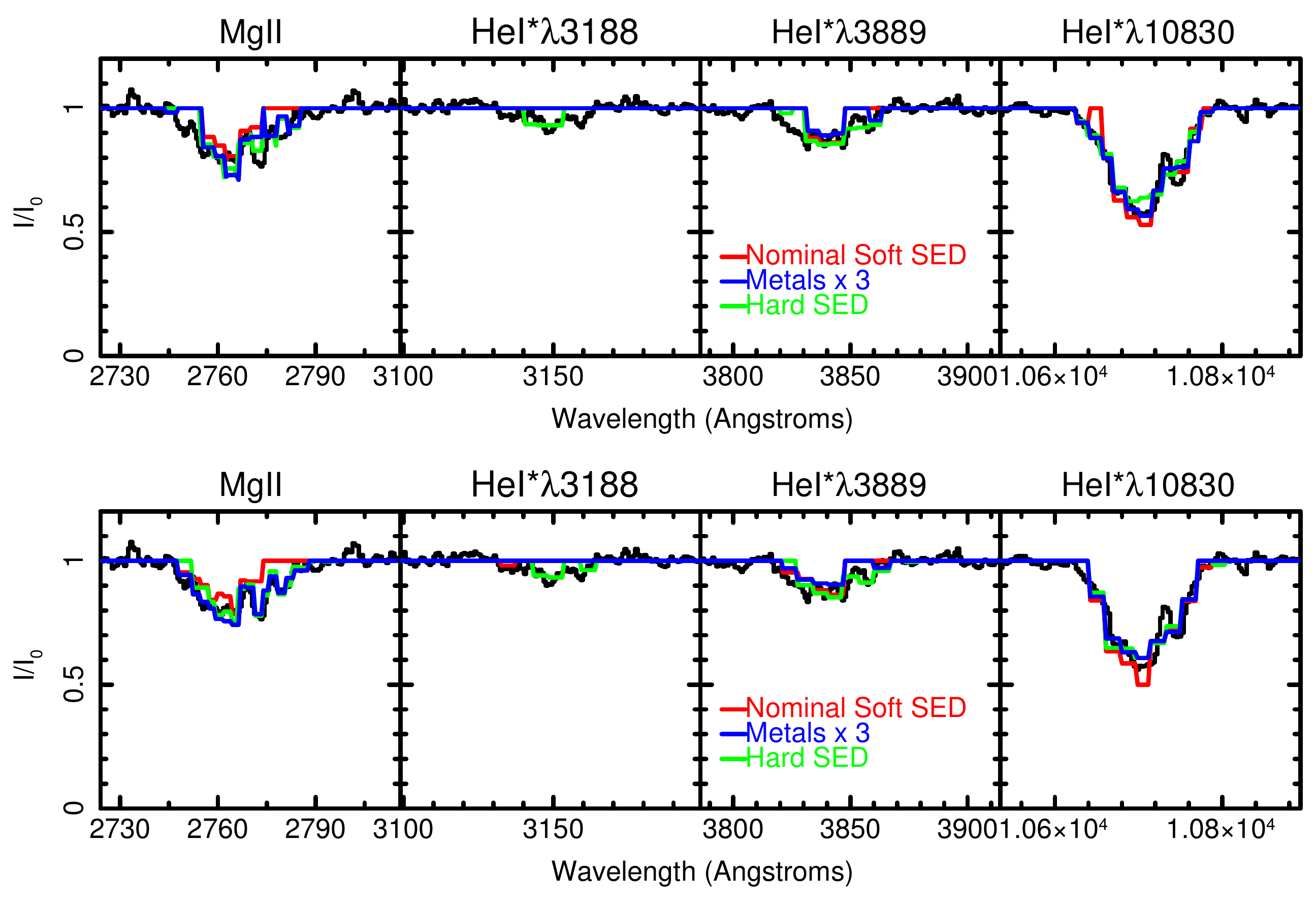}
\caption{The median spectra for the 11-bin  model fit to the optical
  through near-infrared wavelengths  in  which the eleven covering
  fraction values were allowed to vary, but the other parameters
  (ionization parameter $\log U$, column density   parameter $\log
  N_H-\log U$,  and density $\log n$) were fixed to   the best-fitting
  values obtained from fitting the  {\it HST} 
  spectrum (Paper I).  The reduced $\chi^2$ for these models ranges
  between 0.66 and 0.84, indicating an acceptable fit, and showing
  that these three models cannot be distinguished statistically.   \label{fits}}
\end{center}
\end{figure*}

To investigate the difference in covering fraction between the UV and
the long-wavelength spectrum, we performed a {\it SimBAL} analysis of
the continuum-normalized spectrum between 2500--4200\AA\/ and
9000--11500\AA\/.  We made the assumption that, of all the variables
required in the {\it SimBAL} analysis of the {\it HST} COS spectrum,
only the covering fraction varies.  As discussed in Paper I, the
resulting physical parameters of the outflow depend little on the
number of bins used to span the line profile, so we present the
results for the 11-bin case.  The variable parameters in this analysis
were the 11 values of $\log a$, i.e., the log of the covering fraction
index as a function of velocity.  The results
are shown in Fig.~\ref{fig19}.  The left panel shows, for reference,
the results from fitting the full model to the UV data from Paper
I. The partial covering parameter $\log a$ is plotted as a function of
velocity for the 11-bin model for  6 combinations of continuum
model, SED, and metallicity.  The error bars show the 95\% confidence
intervals from the posterior distributions obtained for each of the
model parameters. The middle panel shows the results for the fits of
covering fraction at optical and near-IR wavelengths.  The $\log a$ is
clearly shifted to larger values, indicating a lower covering
fraction.   The median models overlaid on the data are shown in
Fig.~\ref{fits}.  While the reduced $\chi^2$ for the extrapolated
models shown in Fig.~\ref{fig13} ranged from 1.6 to 3.3, indicating an
unacceptable fit, the reduced $\chi^2$ for these models are all less
than 1, indicating an acceptable fit.  Physically, this result implies
that the covering fraction along the line of sight to  the optical and
near-IR continuum emission region is lower than the covering fraction
along the line of sight to the UV continuum emission region.

{ We have measured the difference between the covering
  fractions in the UV and the optical through infrared bands.   In
  principle, there could be a continual decrease in covering fraction 
  as a function of wavelengths.   We tried to detect a difference in
  covering  fraction between the three bands: UV, optical (i.e.,
  \ion{Mg}{2} and   \ion{He}{1}*$\lambda 3889$) and the infrared
  (\ion{He}{1}*$\lambda   10830$).   We were unable to obtain any
  useful constraints because of limitations of the data; specifically,
  the lines are rather shallow and the signal-to-noise ratios are
  moderate. } 

To quantify the difference between the covering fractions in the UV
and at longer wavelengths, we fit a constant model to $\log a$  as a
function of velocity for each of the six models.  The $\log a$ varies
as a function of velocity, and is not well constrained at low and high
velocities where the absorption line is shallow, so we limited the
fitting to range between $-4500\rm \, km\, s^{-1}$ and $-1500\rm \, km\,
s^{-1}$.  Computing the power of 10 of  the resulting average values
of $\log a$ results yields six estimates of $a$ each for the UV models
and the long wavelength models respectively.  

How do we interpret the differences in $a$ between the UV and
the long wavelengths?  We want to know how much more of the continuum
emission source is covered in the UV compared with near-infrared and
optical  wavelengths.  To determine this, we return to the definition
of the power law covering fraction,  $\tau = \tau_{max}x^a$, where
$0 < x <1$ represents the fractional surface area, and ask, at a
particular value of $x$, what is the ratio of the fraction covered?
Solving this equation for  $x$ yields $x=(\tau/\tau_{max})^{1/a}$,
and the fraction covered for a particular value of $\tau/\tau_{max}$
is given by $1-(\tau/\tau_{max})^{1/a}$.   So in terms of $x$, we want
to determine
$$\frac{f_{UV}}{f_{long}}=\frac{1-(\tau/\tau_{max})^{1/a_{UV}}}{1-(\tau/\tau_{max})^{1/a_{long}}}$$
  where the ``long'' subscript refers to the optical through infrared
  wavelengths.  A limiting value is given by  $\tau \to \tau_{max}$,
  but the ratio becomes indeterminate.   It turns out that the ratio
  of the  fractions covered approaches the ratios of the $a$ values as
  $\tau$ approaches $\tau_{max}$\footnote{This is shown using
    L'H\^{o}pital's   Rule for 
$$\lim_{x \to c} \frac{f(x)}{g(x)}.$$
If the value $\frac{f(c)}{g(c)}$ is an indeterminate form, i.e.,
$\frac{0}{0}$ or $\frac{\infty}{\infty}$, then the following equality
holds:
$$\lim_{x \to c} \frac{f(x)}{g(x)} = \lim_{x \to c}
\frac{f^\prime(x)}{g^\prime(x)}.$$  Here, $f(x)$ and $g(x)$ are
  $1-(\tau/\tau_{max}^a)$ for the UV and long wavelength continua
respectively, and $x \to c$ corresponds   to $\tau \to \tau_{max}$.}.
That is,  for indices of $a_{UV}$ and $a_{long}$ in the UV and
near-infrared, respectively, the ratio of the fractions covered will
approach $a_{long}/a_{UV}$.  The results are shown in the right 
panel in Fig.~\ref{fig19}.

While the three different models (the nominal SED, the hard SED, and
the metals$\times 3$ case) yield covering fractions in both the UV and
at longer wavelengths that follow
essentially the same shape as a function of velocity (middle
panel in Fig.~\ref{fig19}), the 
normalizations for the different models are slightly different.
Specifically, the 
hard SED model yields a consistently larger covering-fraction index
parameter, indicating a lower covering fraction.  This is because  the
hard SED produces a thicker Str\"omgren sphere 
\citep[e.g.,][Fig.\ 13]{casebeer06} and a hotter \ion{H}{2} region
\citep[e.g.,][Fig.\ 14]{leighly07}, and given that the fraction of
neutral helium in the metastable state increases with temperature
\citep{clegg87}, more \ion{He}{1}* is predicted per metal ion from the
hard SED.  The near-infrared spectrum has better signal-to-noise ratio than
the optical spectrum, and the \ion{He}{1}*$\lambda 10830$ is deep
compared with \ion{He}{1}*$\lambda 3889$ or \ion{Mg}{2}, so
the \ion{He}{1}*$\lambda 10830$ drives the fit.  Therefore, it is no
surprise that the near-infrared covering fraction obtained from the hard
SED simulations is lower than the others.  
As discussed in Paper I, the hard SED produces the least satisfactory
fit to the {\it HST} COS spectrum.  Therefore, we reject the
relatively high covering-fraction ratio derived from the hard SED, and
take as the representative value of the ratio of fraction of the UV
continuum covered to the fraction of the optical through near-infrared
continuum covered to be 2.5.   

{

\subsubsection{Spatial Non-Uniformity of the Physical Conditions of
  the Gas}\label{nonuniform} 

We have assumed that the only difference between the
UV absorption lines and the optical/infrared absorption lines is the
covering fraction.  But because the infrared continuum emission region
is so much larger than the UV continuum emission region (we estimate
the area ratios to be $A_{10700}/A_{1100} = 140$ in
\S\ref{size_scales}), it is possible that the physical conditions of
the gas are also different.  The extrapolation of the UV solution to
the optical and infrared absorption lines shown in Fig.~\ref{fig13}
reveals the general shape is similar, and therefore the physical
conditions are probably not dramatically different. In particular, the
``mitten'' shape of the \ion{He}{1}*$\lambda 10830$ line is reproduced 
in the extrapolated solution.  However, the ``thumb'' of the mitten,
originating in absorption near $\sim -2000\rm \, km\, s^{-1}$ is 
longer in the extrapolated solution than in the data,
suggesting that on average, the outflowing gas with velocity near
$-2000\rm\, km\, s^{-1}$ covering the infrared continuum emitting
region has somewhat higher opacity than that covering the UV continuum 
emission region.

We attempted to quantify the possible difference in physical
conditions by fitting the optical and infrared $I/I_0$ spectrum with a
model in which the ionization parameter, $\log N_H-\log U$, and
covering fraction parameter $\log a$ were allowed to vary.  There are
no density diagnostic lines in that region of the spectrum, so we
froze those parameters at the best fitting values from the UV model.  We
also froze the velocity offset and velocity width of the bins.  Not
surprisingly, the results are not very conclusive because there is not
enough information among the \ion{Mg}{2} and \ion{He}{1}* lines to
constrain the physical conditions.  The ionization parameter is
particularly poorly constrained.   The $\log N_H-\log U$ is consistent
with the UV solution within the concentration (between $-4400$ and
$-3200\rm \, km\, s^{-1}$).  At lower velocities, the $\log N_H-\log
U$ is higher and the covering fraction parameter $\log a$ is larger
(lower covering fraction) in the long wavelength solution compared
with the UV solution, but with so few lines to constrain the solution,
it is clear that these parameters are highly covariant.  

Despite our failure to constrain the physical conditions at long
wavelengths, the similarities and differences between the extrapolated
UV solution and the observed long wavelength absorption lines suggest
intriguing constraints on the spatial uniformity of the absorbing gas.
}

\subsection{What About the Broad-line Region?}\label{blr}

\begin{figure*}[!t]
\epsscale{1.0}
\begin{center}
\includegraphics[width=6.5truein]{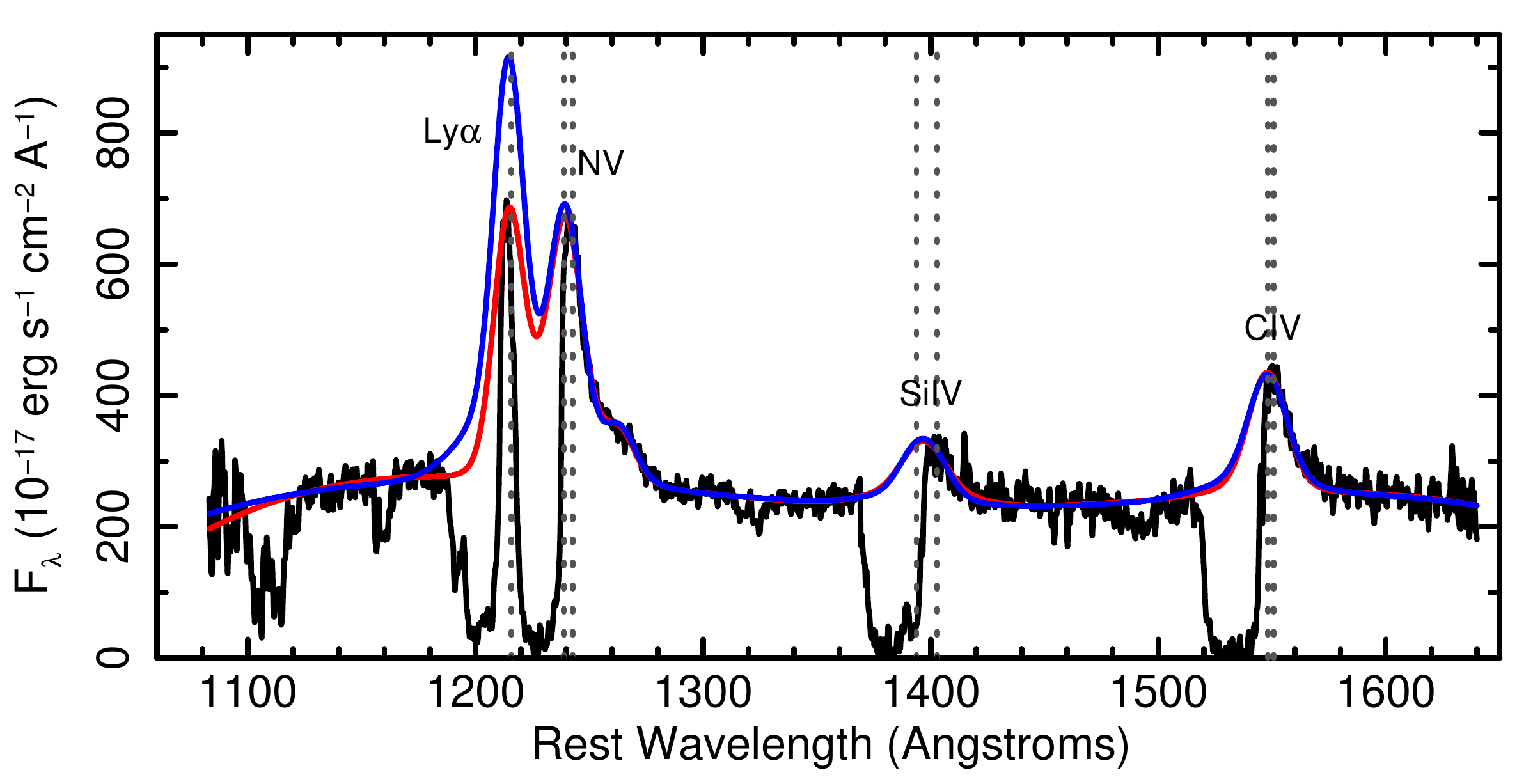}
\caption{The {\it HST} COS spectrum (black line), and continuum
  models, reproduced   from Paper I.  The red line shows the first
  continuum model and the blue line shows the second continuum model;
  see Paper I for details.   The
  rest wavelengths of prominent emission lines are marked.  The onset
  of the outflow is at low enough velocity, and the lines are deep
  enough that is clear that at the broad line region emission is
  absorbed.   \label{hst_spectrum}}   
\end{center}
\end{figure*}

We conclude that the  absorber in SDSS~J0850$+$4451
presents a larger covering fraction to the UV emission region compared 
with the near-infrared continuum emission region, indicating the
presence of structure in the absorbing outflow.  Size scales are
discussed in detail in \S\ref{size_scales}, but it is expected that
the broad line region should be located at a comparable or larger
radius than the near-infrared-emitting accretion disk.  The {\it  HST}
spectrum and continuum models, reproduced from Paper I, are shown for
reference in Fig.~\ref{hst_spectrum}.  The rest wavelengths of
prominent emission lines are marked.  The onset of the outflow is at
low enough velocity and the lines are deep enough that is clear that
the broad line region is substantially absorbed.  Comparison of
this figure with Fig.~\ref{fig13} or Fig.~\ref{fits} shows that the
near-UV, optical and near-infrared absorption lines are not as deep as the
UV absorption lines (e.g., \ion{C}{4}), giving the impression that the
broad line region is fully absorbed, i.e., has a higher covering
fraction than the near-infrared continuum emission region, a result that
does not make sense considering the relative expected size scales (see
\S\ref{size_scales}).   

This impression is mistaken, due to the nature of the power-law
covering fraction parameterization.  As discussed in Paper I, in the 
power-law covering fraction parameterization, the fraction of the
source covered, or alternatively, the residual intensity, depends on
the total opacity of the line { \citep[see also ][]{arav05}}.  The
prominent UV lines, including \ion{C}{4}, \ion{Si}{4}, and
\ion{N}{5}, have relatively high opacities, since the ions that
produce these lines are very abundant in the \ion{H}{2} region of the
ionized slab.  The ions producing \ion{He}{1}*$\lambda 10830$ and
\ion{He}{1}*$\lambda 3889$, which are also found in the \ion{H}{2}
region, are rarer, since they come from metastable helium
\citep[see][for a   discussion]{leighly11}. \ion{Mg}{2} is a
low-ionization line, and only starts to become commonplace as the
hydrogen ionization front is approached \citep[e.g., Fig.\ 10 in
][]{lucy14}, so it also has  relatively low opacity in
SDSS~0850$+$4451 since its LoBAL classification means that the
hydrogen ionization front is not present in the outflow (i.e., versus
FeLoBALs, where the hydrogen ionization front is expected to be 
present).  Therefore, \ion{Mg}{2} is also expected to not be a very
optically thick line. Therefore, it is possible that the broad-line
region has a lower covering fraction than the UV continuum, even
though casual examination of the spectrum suggests otherwise. We
  discuss   inhomogeneous partial covering and the power-law covering
  fraction   parameterization further   in \S\ref{understanding}.

\begin{figure*}[!t]
\epsscale{1.0}
\begin{center}
\includegraphics[width=6.5truein]{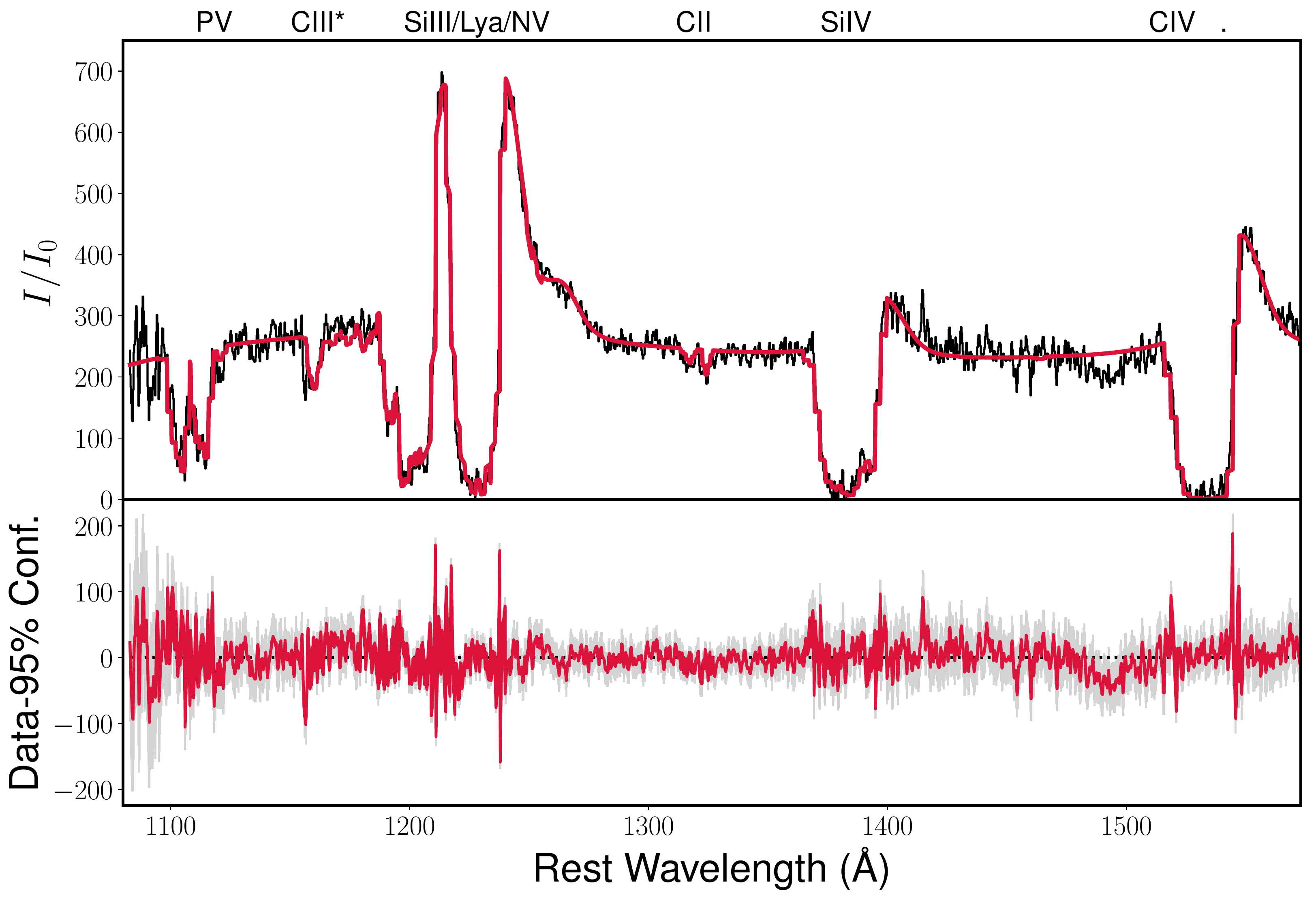}
\caption{Results from fitting the 11-velocity-bin enhanced-metallicity
  model to the spectrum between 1080\AA\/ and 11500\AA\/; the results
  for the UV   spectrum are shown.  In the top panel, the median
  synthetic spectrum   (crimson) is overlaid on the spectrum (black).
  The lower panel shows the spectrum  minus the median model and
  errors in gray, and the filled region between the spectrum and
  spectrum and plus and minus the 95\% confidence synthetic spectra in
  crimson, respectively.    Comparison with the
  enhanced metallicity model for the UV spectrum only (Fig.~14 in
  Paper I) shows that this model yields a comparable-quality fit.
    \label{uv_fit}}

\end{center}
\end{figure*}

We test this scenario by fitting all of the spectra: the {\it HST} COS
spectrum analyzed in Paper I that samples the UV band, the combined
SDSS and MDM spectra described in \S\ref{mdmobs} and \S\ref{sdss}
(sampling the near-UV and optical, between 2500\AA\/ and 4000\AA\/),
and the combined LBT and Gemini spectra described in \S\ref{gemobs}
and \S\ref{lbt} (sampling the near-IR, between 9000\AA\/ and
11500\AA\/).  Although we now have developed a method to fit the 
continuum and line emission simultaneously with the absorption
(Leighly et al., in preparation), for direct comparison with Paper I,
we separate the line and continuum contributions to our continuum
models and fit with the normalizations of these components fixed.  As
shown in Paper I, there is little  dependence on the number of bins
used to span the troughs, so the 11-bin model was chosen as
representative.  Three sets of 11 parameters modeled the covering
fractions of the UV, the long wavelengths, and the broad line region,
respectively.  The UV continuum covering fraction was modeled using
$\log a$ as in Paper I.  The long wavelength continuum was 
modeled using $\Delta \log a_{long}$, and a prior was used to
constrain these parameters to be  greater than zero, i.e., making the  
physically reasonable assumption that the covering fraction of the
longer wavelength continuum is lower than the covering fraction of the
UV continuum (as shown in \S\ref{quantifying}), and keeping in mind
that a larger value of $a$ corresponds to a smaller covering fraction.
To be specific, the covering fraction parameter in a particular
velocity bin applied to the long wavelength continuum was $\log a +
\Delta \log a_{long}$, where $\log a$ is the value applied to the same
velocity bin in the UV, and $\Delta \log a_{long}$ is the model
parameter. Finally, the broad lines were modeled with an additional
$\Delta \log a_{lines}$, thereby making the physically reasonable
assumption that the fraction covered is at least as small as that of
the long wavelength continuum.  Thus, the covering fraction applied to
the line emission was $\log a +\Delta\log a_{long}+\Delta\log
a_{lines}$.  

Overall, the fits are good despite the increase in bandpass. The
reduced $\chi^2$ computed over the points where the median model
experienced opacity (see Paper I; this modified $\chi^2$ is used
because the continuum is not allowed to vary) are found to be, for the
first and second continuum models, respectively: 1.41 and 1.59 for
solar metallicity and soft SED, 1.55 and 1.54 for the hard SED, and
1.15 and 1.18 for the soft SED and $\times 3$ metallicity.  The values
for the solar metallicity and hard SED are larger than the ones
obtained for the UV-only models of Paper I
\citep[see][Fig.~5]{leighly18}, but are comparable for the enhanced
metallicity model, indicating that the $\times 3$ metallicity and soft
SED model is preferred.  Despite the additional constraints imposed by
the inclusion of the long-wavelength spectra, the fit in the UV band
is still good (Fig.~\ref{uv_fit}).

\begin{figure*}[!t]
\epsscale{1.0}
\begin{center}
\includegraphics[width=5.5truein]{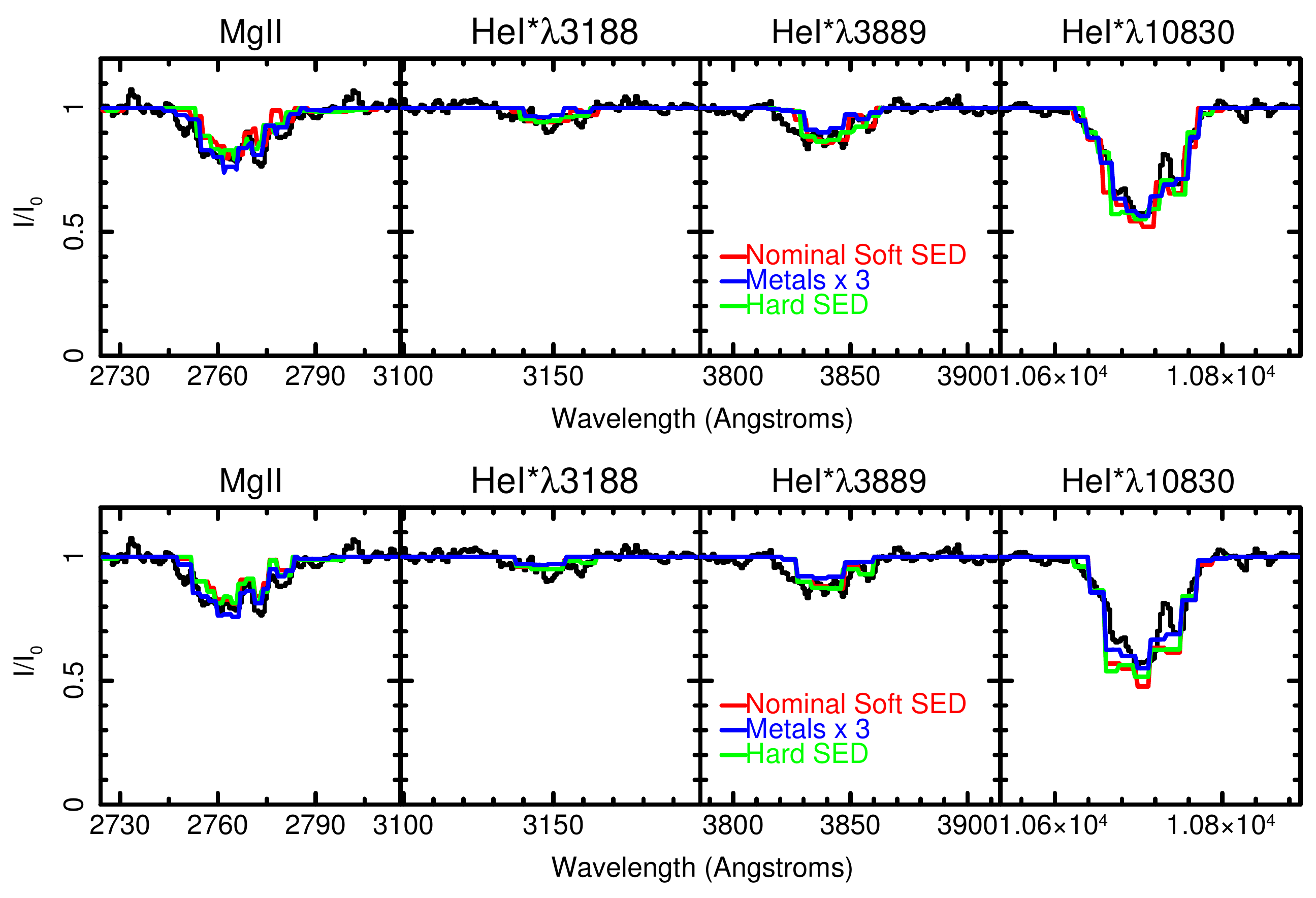}
\caption{The long wavelength results from fitting the 11-velocity-bin
  model to the   spectra between 1080\AA\/ and 11500\AA\/.  The models
  have been divided by the continuum to facilitate comparison with
  Fig.~\ref{fits}.  Overall, the fits are comparable, with minor
  differences from line to line.  We conclude that the model for the
  full wavelength range provides an excellent description of the
  spectra.  
    \label{long_wavelengths}}     
\end{center}
\end{figure*}

Fig.~\ref{long_wavelengths} shows the results in near-UV to near-IR
spectra.  Here, the spectra have been normalized by the continuum
model to facilitate comparison with extrapolation analysis presented
in \S\ref{quantifying}.   Comparison with Fig.~\ref{fits} show that
the fits are good and overall very similar to  one another, although
small differences are found from line to line.  We conclude that the
model presented in this section describes the full bandpass well.

\begin{figure*}[!t]
\epsscale{1.0}
\begin{center}
\includegraphics[width=6.5truein]{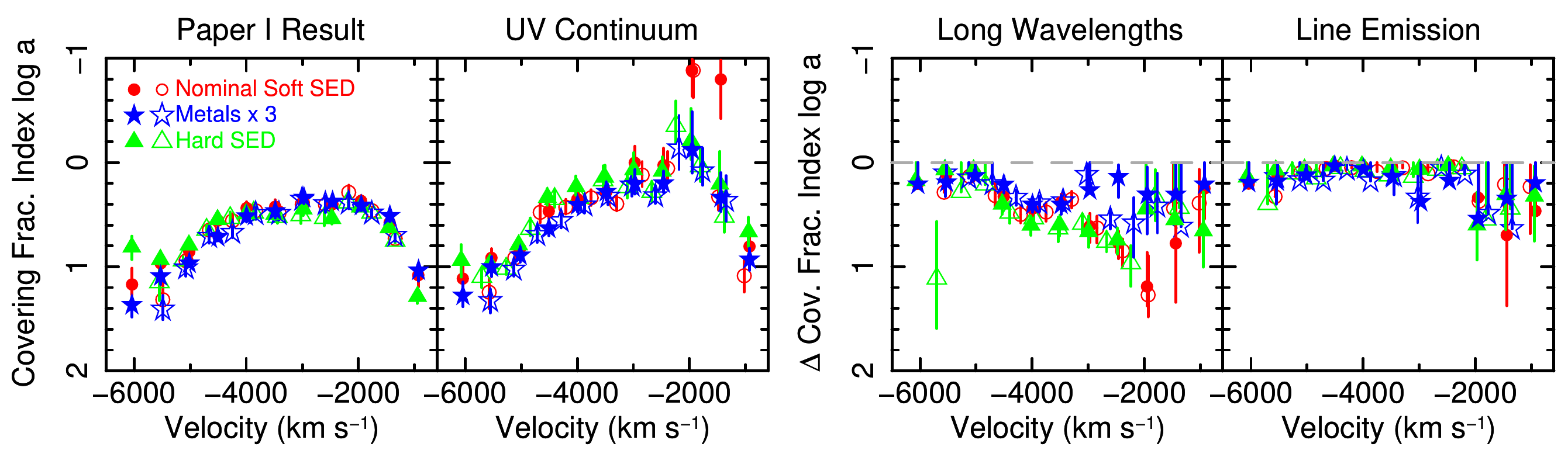}
\caption{Results from fitting the 11-velocity-bin models to the
  UV through near-IR spectra. The open and solid symbols correspond the first
  and second  UV continuum models shown in Fig.~\ref{hst_spectrum}.
  Points at the end of an error bar denote upper limits.   The
  left-most   panel shows the results   from Paper I,
  where the   {\it HST} spectrum was modeled with a single   set of
  covering-fraction parameters.  The right three panels show   the
  results from   the model presented in this paper, where the UV
  continuum was   assumed to have the largest covering fraction
  (corresponding to a lower value of $\log a$), the near-UV through
  near-IR continuum was assumed to have an equal or lower covering
  fraction,  parameterized by $\Delta \log a_{long}$ required to be
  greater   than or   equal to zero, and the emission line region was
  parameterized by $\Delta \log a_{lines}$, also required to be greater
  than zero.  The results show an offset for the long-wavelength
  continuum of $\Delta \log a_{long}$ close to 0.4, consistent with
  the   value derived in \S\ref{quantifying}.  The results for the
  broad line region show that $\Delta \log a_{lines}$ is essentially
  consistent with zero, indicating that there is no evidence that the
  broad line region has a lower covering fraction than the
  long-wavelength continuum.
    \label{fig9}}     
\end{center}
\end{figure*}

The covering fraction results are shown in Fig.~\ref{fig9}.   The
left-most panel shows the results for fitting the UV continuum and
lines together from Paper I.  The results from the new
model presented in this paper are shown in the right three panels.
The second-from-the-left panel shows the covering fraction for the UV
continuum alone.  The covering fraction index is somewhat smaller than 
the Paper I result, indicating a somewhat larger covering fraction for
the UV continuum than found in Paper I.  This is especially true
around $-2000\rm \, km\, s^{-1}$, where the line emission is
prominent.    

The second-from-the-right panel shows the $\Delta \log a$ for the
near-UV, optical, and near-IR wavelengths.  The difference is particularly
strong and robust near $-4000\rm \, km\, s^{-1}$, the location of the
enhanced region of $\log N_H -\log U$ referred to in Paper I as ``the
concentration''.   This result makes sense, since the ions that produce
the long-wavelength lines are found deeper in the photoionized slab
and are therefore  most prominent in the velocities defined by the
concentration.   They are also coincident with the
  \ion{C}{3}*$\lambda 1175$ feature discussed in Paper I, e.g., Fig. 6.
The $\Delta \log a$ value is close to 0.4, the value obtained in 
\S\ref{quantifying}.  

The right-hand panel shows the $\Delta \log a_{lines}$ for the
emission-line spectrum.  These are, for the most part, consistent with
$\Delta \log a_{lines}$ equal to zero.  This can be interpreted as
evidence that the broad line emission has the same covering fraction
as the long wavelength continuum emission region.     However, we note
that in this model each velocity bin is fit by 3 covering fraction 
parameters.   It seems reasonable to suspect that the data are over-fit,
i.e., there are potentially too many covering-fraction degrees of
freedom in each velocity bin, resulting in covariance among model
parameters.   

\begin{figure*}[!t]
\epsscale{1.0}
\begin{center}
\includegraphics[width=5.0truein]{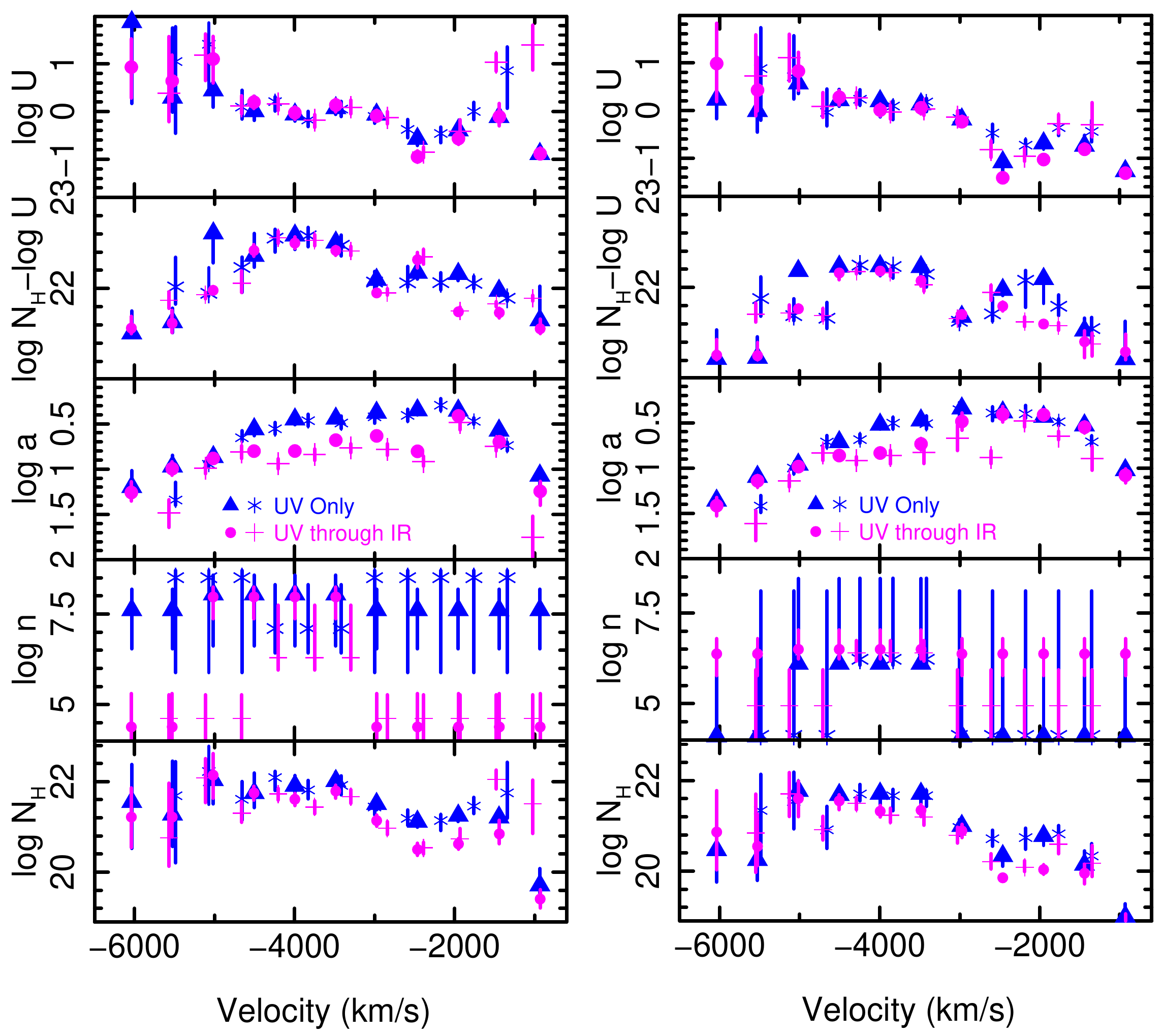}
\caption{Results from fitting the 11-velocity-bin models to the
  spectra. The median values and 95\% confidence regions from the
  posterior probability distributions are shown.  The left (right)
  panel shows the results for solar ($Z=3Z_\odot$) metallicities,
  respectively. The   star and plus markers (triangle and
  filled-circle markers) show the   results for the first (second)
  UV continuum models, respectively.  The blue symbols show the results
  from Paper I  obtained from modeling the UV spectrum 
  with a single set of covering fractions.  The magenta symbols show
  the results presented in this paper, from fitting the UV through near-IR
  spectra with three  covering fractions.  { Note that the density
    $\log n$ is constrained by the presence of \ion{C}{3}*$\lambda
    1175$ only in the concentration, i.e, between $-4500$ and
    $-3000\rm \, km\, s^{-1}$ (see Paper I for details.  Overall, no
  significant shifts in the solutions are observed, with the exception
  of the covering fraction, and the covering-fraction-weighted column
  density. } 
    \label{fig10}}     
\end{center}
\end{figure*}

Allowing the covering fractions for the UV continuum, the long
wavelength continuum, and the broad-line region continuum to vary
independently causes the solution to shift compared with the UV-only
models presented in Paper I.  We find that these shifts are minor and
the physical parameters describing the outflow are nearly the same.
Fig.~\ref{fig10} shows the outflowing-gas physical parameters as a
function of velocity; the results for the UV-only model fits from
Paper I are reproduced for comparison.  The results for the fitted  
ionization parameter $\log U$, the column density parameter $\log
N_H-\log U$, and the derived parameter $\log N_H$ are roughly
consistent between the two models, with small changes at low
velocities where the broad emission lines dominate.  The density $\log
n$ appears to be much different for velocities higher and lower than
that of the concentration (centered near $-4000 \rm \, km\, s^{-1}$),
but as discussed in Paper I, there are no density-dependent lines at
those velocities and the density is unconstrained.  

\begin{figure*}[!t]
\epsscale{1.0}
\begin{center}
\includegraphics[width=2.5truein]{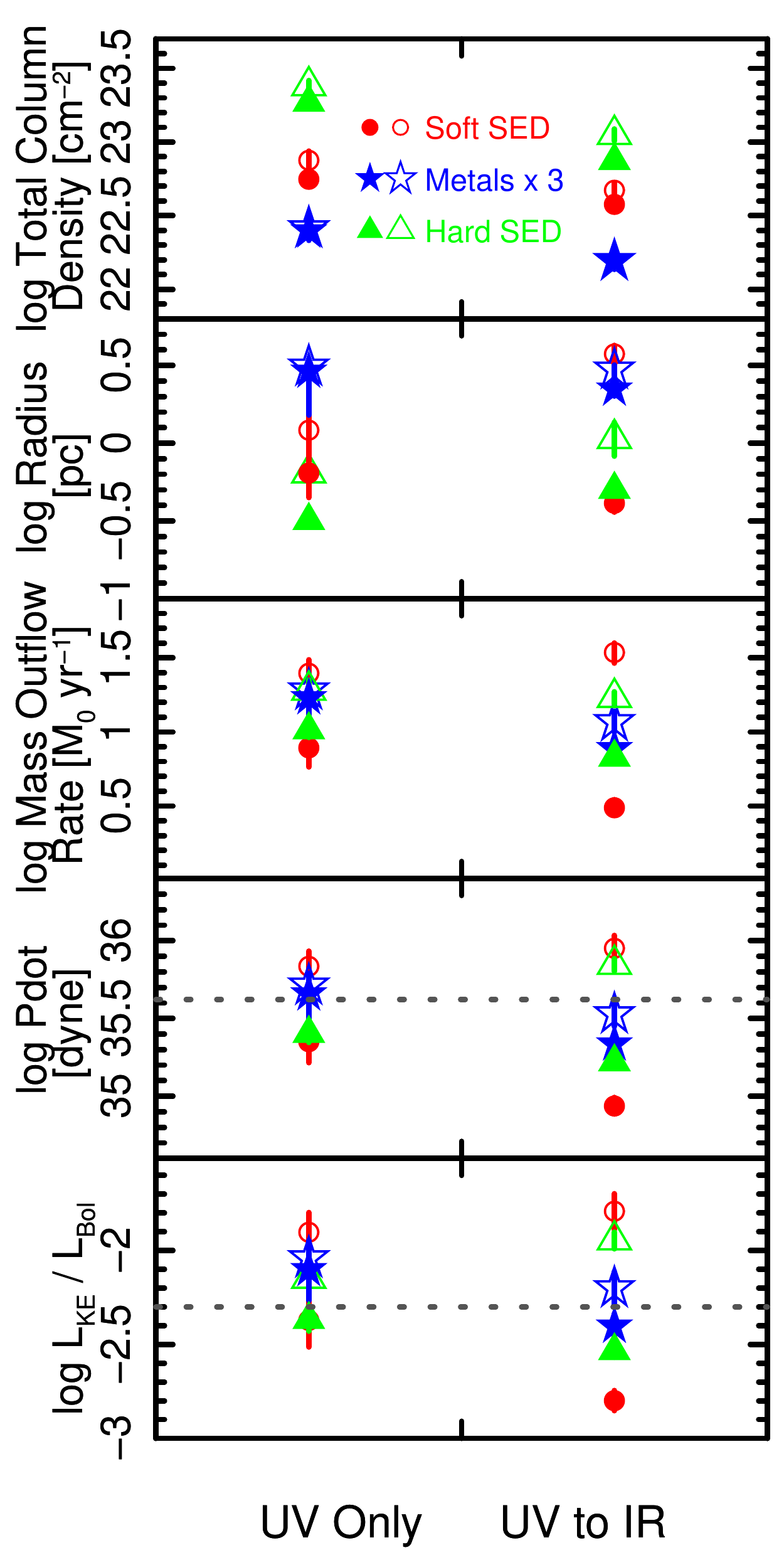}
\caption{The derived quantities for the 11-velocity-bin model for the
  UV spectrum only (left side), and from the new analysis of the UV through
  near-infrared   spectra presented in this paper (right side).  In
  each   case, the open   (filled) symbols show the results using the
  first (second) continuum   model.  The top panel shows the total 
  hydrogen column density in the outflow, weighted by the covering
  fraction in each bin. { The column densitiess are lower for the
    UV through 
  near-infrared models because the covering fraction is lower for
  larger size scales.}  The   second-from-the-top panel shows the radius
  of the outflow obtained   using the density constraints in the
  concentration. The next panel   shows the mass outflow rate,
  assuming that the radius of all outflow   components is the same as
  the radius of the concentration.  The   fourth panel shows the
  momentum flux $\dot P$, applying the same   radius assumption, while
  the dashed line shows $L_{bol}/c$. Finally,   the bottom   panel
  shows the ratio of the kinetic luminosity to the   bolometric
  luminosity, again with the same radius assumption. See   text for
  further details. In each case, the median and 1 sigma   error bars
  obtained from the MCMC model are plotted. { The UV--near-IR modeling
  reported in this paper finds a weaker outflow than the UV-only
  modeling, principally due to the lower covering fraction observed on
  larger size scales. }
    \label{derived}}     
\end{center}
\end{figure*}

Finally, we show the results of the derived parameters including the
total column density, the radius of the outflow, the mass outflow
rate, the momentum flux, and the ratio of kinetic to bolometric
luminosity in Fig.~\ref{derived}.  For comparison, we show the results
from Paper I as well.  { In \S\ref{size_scales} we show that the
infrared continuum emission region is 140 times larger than the UV
continuum emission region.  Therefore, the appropriate covering
fraction to use for the UV--to--near-IR model is the one for the
largest size scale, i.e., $\log a_{long}$ or $\log a_{lines}$, since the
properties relevant for the larger area should dominate the outflow,
at least as far as we can tell from the information
we have.  Therefore, we use $\log a_{lines}$ to weight the column
densities, resulting in a reduction in the estimated total column
density for the enhanced-metallicity models by a factor of $\sim 1.6$
compared with the results of Paper I to $\log
N_H=22.19^{+0.058}_{-0.056}$ and $22.18^{+0.045}_{-0.043}\rm \,
[cm^{-2}]$ for the first and second  continuum models respectively
(1-sigma errors).  The factor of 1.6 is lower than than the ratio of
the two covering fractions which was estimated in \S\ref{quantifying}
to be 2.5.  The difference is that the value of 2.5 was extracted from
the well-sampled data in the center of the velocity profile, from
$-4500$ to $-1500\rm \, km\, s^{-1}$, while the column density was
obtained from the whole profile.  If we extract the column density
from those that range of velocities only, the difference is a factor
of 2.5, as expected.  

Other parameters shift due to the reduction in column density and
small shifts in the best fit. Specifically, the radius of the outflow
is found to be $\log R=0.47^{+0.03}_{-0.04}$ and $0.34\pm 0.04 \rm \,
[pc]$, the mass outflow rate is $\log \dot M= 1.07 \pm 0.07$ and 
$0.88^{+0.05}_{-0.06} \rm \, [M_{\odot}\, yr^{-1}]$, and the log of
the ratio of the kinetic to bolometric luminosity is $-2.20\pm 0.09$
and $-2.41\pm 0.07$, for the first and second continuum
enhanced-metallicity models,  respectively.  Notably, the kinetic
luminosity for the enhanced metallicity models decreases to
0.39--0.63\% for the second and first continuum models.  This range
straddles the 0.5\% value taken to be a conservative cutoff for
effective galaxy feedback \citep{he10}.  Therefore, SDSS~J0850$+$4451
does not appear to be undergoing strong feedback from the BAL outflow.}

To summarize, we have shown that there exists in SDSS~J0850$+$4451 a
hierarchy of partial covering.  The spectra are  consistent with a
model in which the covering fraction parameter $\log a$ to the optical
and near-IR continuum is about 0.4 higher than to the UV continuum (i.e.,
consistent with the analysis presented in \S\ref{fits}, and implying
a covering fraction that is lower by a factor of about 2.5).  The
covering fraction to the broad line region is mostly consistent with
that of the long-wavelength continuum, and therefore the broad line
region has a lower covering fraction than the UV continuum.  In
addition, while in Paper I we found only mild support for the
preference for high metallicity, the support is much stronger here,
given that the reduced $\chi^2$ values evaluated over the non-zero
opacity portions of the spectra are larger than 1.2 for the solar
metallicity and hard SED models, and only the models with $Z=3Z_\odot$
are acceptable. { Finally, taking into account the lower covering
fraction over the larger area results in a reduction in the total
column density and other outflow parameters including the kinetic
luminosity. }

\section{Understanding the Power-law Partial Covering
  Parameterization of Inhomogeneous Partial
  Covering}\label{understanding}    

The traditional form of partial covering, wherein a fraction of the
emission region is covered uniformly by the absorber and the
remainder is not covered, is easy to understand intuitively: one
needs to only imagine an eclipse.  Inhomogeneous partial covering is
much less intuitive.  Because partial covering seems to be extremely
important in shaping the spectrum of SDSS~0850$+$4451, as well as
other objects modeled using {\it SimBAL}, we explore the nature of
partial covering in this section.  

Four factors must be considered in order to understand how 
absorption lines are shaped: the concept of inhomogeneous
partial covering itself, the mapping of the output of the
photoionization models (ionic column densities) to the power-law
parameterization, the opacity of the particular line, and the relative
brightness of the background source. We will explore each of these in
turn.   Note that substantial previous discussions of inhomogeneous
partial covering are given in \citet{dekool02c, arav05, sabra05}.

\subsection{Inhomogeneous Partial Covering and the Power Law
  Parameterization} 

\begin{figure*}[!t]
\epsscale{1.0}
\begin{center}
\includegraphics[width=6.5truein]{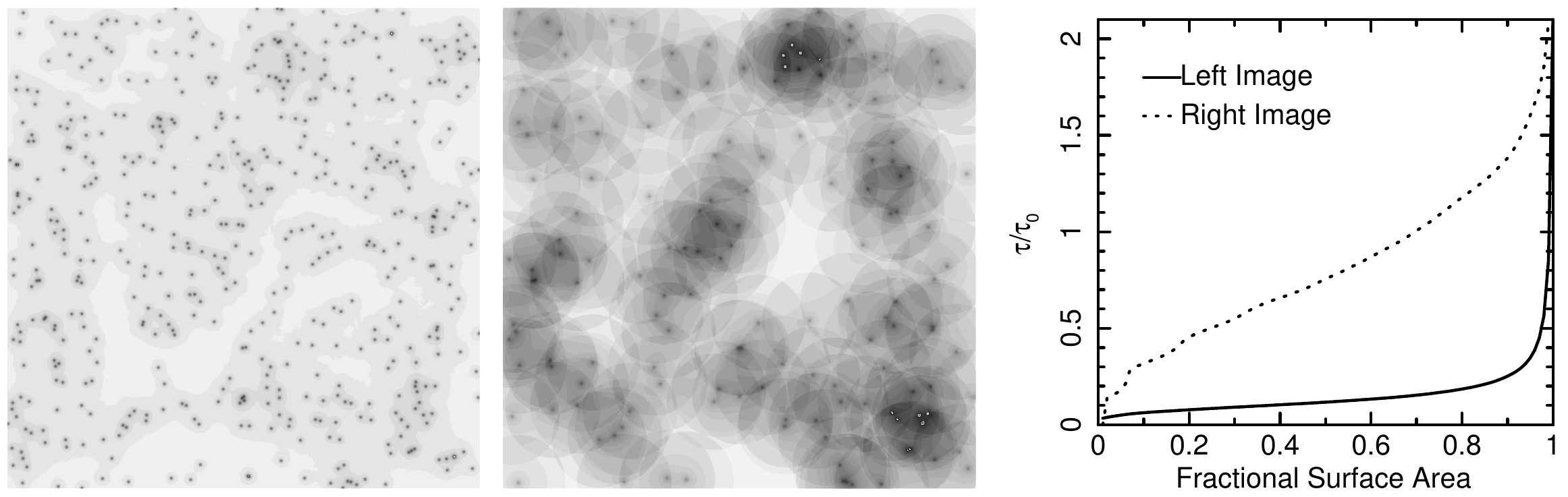}
\caption{A toy model illustrates inhomogeneous partial covering
  \citep[see also][]{dekool02c}.  The left images illustrate two
  scenarios of opacity profile and density of ``clouds'', with the far
  left (middle) image illustrating the case of many (few) clouds with steep
  (shallow) opacity profiles.  The  gray-scale is linear, and in each
  case, the opacity in the center  of each cloud, $\tau_0$, is equal to
  1.  The right plot shows the   resulting optical depth distribution
  over the whole image,   normalized by $\tau_0$.  The power law
  parameterization for   inhomogeneous partial covering is intended
  to model a wide range of   distributions of optical depth across a
  continuum   source.  \label{toy}}    \end{center} 
\end{figure*}

The concept of inhomogeneous partial covering can be illustrated using
a toy model \citep[e.g.,][]{dekool02c}.  Fig.~\ref{toy} shows linear
gray-scale images for two examples of distributions of ``clouds''.
Each cloud was constructed with opacity in the center of the
two-dimensional cloud projection set to $\tau_0=1$.  The left image
illustrates the case where there are many clouds (500) and each cloud
has a steep radial opacity profile ($r^{-1.5}$).  The right image
illustrates the case where there are fewer clouds (150) and each cloud
has a flat opacity profile ($r^{-0.5}$).  The distribution of optical
depths is given in the right panel.  As might be expected, many clouds
with a steep opacity profile yield low opacity across a large fraction
of the continuum source, and a small fraction of the continuum is
covered by a high opacity.  In contrast, few clouds with flat opacity
profiles yield significant opacity across a large fraction of the
continuum source. 

The toy model is useful for illustrating the concept of partial
covering, but given that we do not know anything about the ``clouds''
except their approximate size (\S\ref{partial_covering}), we use a power-law
parameterization for fitting.  The power law parameterization is given
by $\tau=\tau_{max} x^a$ where $x$ is the fractional surface area, as
above, and $a$ is the fit index. Fig.~\ref{indices} illustrates the
opacity for different values of $a$.  A small value of $a$ corresponds  
to relatively high opacity over a large fraction of the continuum
emission region.  A large value of $a$ corresponds to a low opacity
over most of the continuum emission region, and a high opacity over a
small fraction.  

\subsubsection{{\it Cloudy} and the Power Law Parameterization}

As discussed by \citet{sabra05}, the power-law opacity profile
$\tau(x)=\tau_{max} x^a$ yields the following residual intensity
  equation:  
$$I(\lambda)=\frac{1}{a}\frac{1}{\tau_{max}^{1/a}}\Gamma(1/a)P(1/a,\tau_{max}),$$
where $\Gamma$ and $P$ are the complete and incomplete Gamma
functions, respectively.  This is the equation that is used in {\it
  SimBAL}.  

{\it Cloudy} computes photoionization equilibrium in a slab of gas;
there is no provision in the software for partial covering. How the
ionic column densities produced by the {\it Cloudy} simulations map to
the power-law opacity profile is a matter of interpretation.   There
are at least two possibilities: the opacity of an ion 
calculated using {\it Cloudy}  corresponds to the  {\it average}
opacity across the continuum emission region (i.e., $N_{ion}
\Rightarrow \bar\tau$, where $\bar\tau=\int_0^1 \tau_{max} x^a dx =
\tau_{max}/(1+a)$), or 
the opacity of the ion maps to the maximum opacity (i.e., $N_{ion}
\Rightarrow \tau_{max}$).  These two methods produce indistinguishable
results when the covering fraction is high ($a$ is low), but lead to
somewhat different interpretations of partial covering, somewhat
different implementations in {\it   SimBAL}, and different line profile
behaviors, as we discuss below.     

\begin{figure*}[!t]
\epsscale{1.0}
\begin{center}
\includegraphics[width=4.0truein]{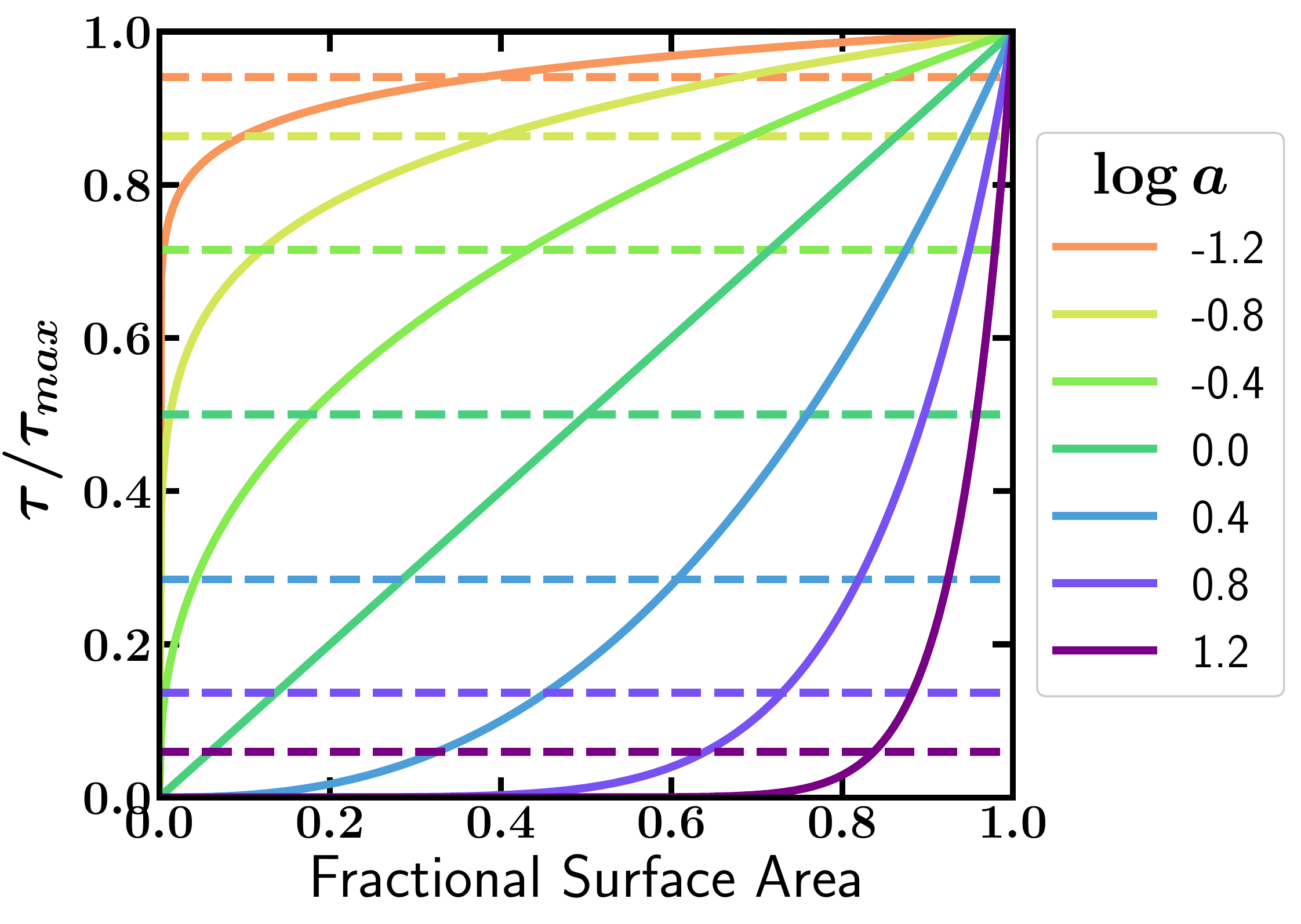}
\caption{The effects of power-law optical depth parameterization as a
  function of fractional surface area. The solid lines show
  the normalized power-law opacity $\tau/\tau_{max}=x^a$ for a range of
  values of $\log   a$ shown in the legend. The dashed lines show the
  average opacity for each value of $\log a$.  For large values of
  $a$, $\tau_{max}$ and the average opacity diverge.   $\log a   =0.4$
  is a value typical of the UV   continuum at $-4000\rm \, km\,
  s^{-1}$ (i.e., in the   concentration), and $\log a=0.8$ is a
  value typical of the wing of the   emission lines at $-2000 \rm \,
  km\,     s^{-1}$.    \label{indices}}    \end{center}   
\end{figure*}

For the $N_{ion} \Rightarrow \bar\tau$ case, we must first obtain
$\tau_{max}$ using 
$\tau_{max}=\bar{\tau}(1+a)$.  Thus, the opacity of an ion computed by   
{\it Cloudy} is multiplied  by $1+a$ {\it before} the spectrum is
computed in {\it SimBAL}.  For the $N_{ion} \Rightarrow \tau_{max}$
case, the opacity computed by {\it Cloudy} is used directly as
$\tau_{max}$ by {\it SimBAL} to compute the spectrum, and the fitted
column density is then corrected for the portion that is not covered
by dividing by $1+a$ {\it after} the {\it SimBAL} computation
(referred to as the covering-fraction-weighted column density here and
in Paper I).  There is no difference when $a$ is small, simply because
$\bar{\tau}$  approaches $\tau_{max}$.  But when $a$ is large,
$\bar{\tau}$ is  much less than $\tau_{max}$.  This fact is
illustrated in Fig.~\ref{indices}, where the run of opacity
as a function of fractional surface area is shown by the solid lines
for a range of $a$ values, and the average opacity is shown by the
dashed lines.  For large values of $a$, the average value is much less
than the maximum value.

If the proportions of ions were uniform as a function of column
density of the {\it Cloudy} slab, it might seem 
that there would be no difference between the two interpretations: 
either the average opacity is scaled up by $1+a$ before the spectrum
is constructed, or the inferred column density is corrected by dividing
by $1+a$ after the spectrum is constructed.  The proportionally of the
ionic populations is the assumption that is implicitly made by the
$N_{ion} \Rightarrow \bar\tau$ method, since it assumes that the
optically thickest part of the inhomogeneous partial covering is
adequately modeled by $\tau_{max}=(1+a)\bar\tau$.  However, it is
readily apparent that the ionic column densities do not increase in
proportion with the hydrogen column density \citep[e.g.,][their
  Fig.\ 1]{hamann02}. 
As ionizing photons are removed from the photoionizing continuum by 
transmission through the gas, the proportions of different types of
ions change.  This is especially true when approaching the hydrogen
ionization front where  low-ionization ions such as Mg$^+$ start to
become common.  These low-ionization lines can be very important in
constraining the column density.  Indeed, in SDSS~J0850$+$4451, it is
the \ion{C}{3}* that constrains the $\log N_H-\log U$ of the
simulation \citep[see Fig.\ 10   in] [in particular, see the
  accompanying animation]{leighly18}.  For large $a$, it is more
important to model the ionic proportions in the high-column density
centers of the ``clouds,'' which is done by the $N_{ion}
\Rightarrow \tau_{max}$ method, but not the $N_{ion} \Rightarrow
\bar\tau$ method.

\begin{figure*}[!t]
\epsscale{1.0}
\begin{center}
\includegraphics[width=5.0truein]{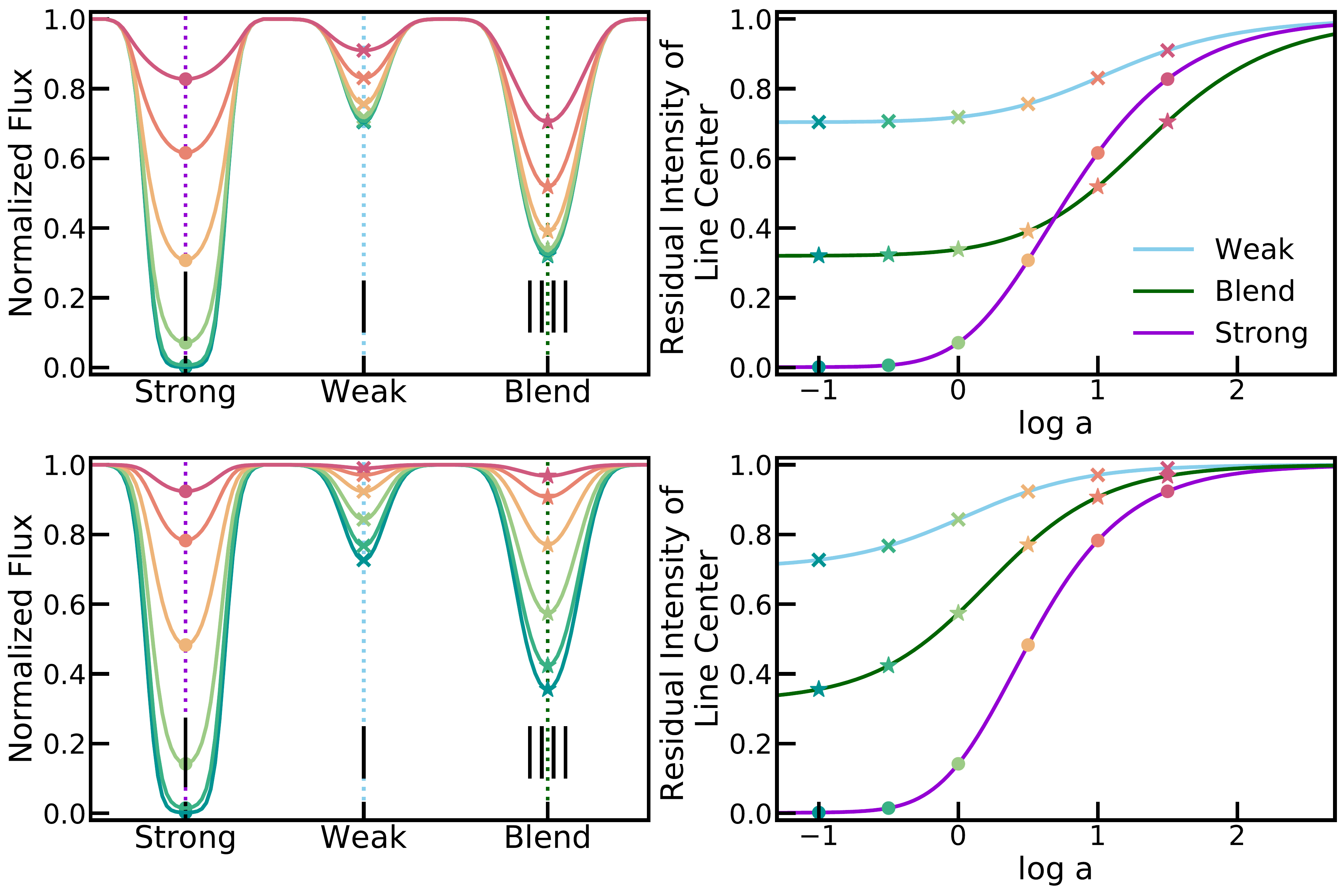}
\caption{The influence of the choice of mapping of {\it  Cloudy}
  output results to the power law partial covering parameters.  The top panel
  shows the   results for $N_{ion} \Rightarrow \bar\tau$ method, and
  the lower   panel shows the $N_{ion} \Rightarrow \tau_{max}$ method,
  the method   currently used in {\it SimBAL}.  {\it
    Left:} A strong, a weak, and a blend of four weak lines were
  simulated for a range of covering fraction parameters $\log a$,
  where the total opacity of the strong line is 20 times that of the
  weak line. {\it Right:} the depth at the lowest point for the
    simulated lines shown in the left panel. For the $N_{ion}
    \Rightarrow \tau_{max}$ case (lower panel), the line
    depth tracks the total opacity.  That is, as $\log a$ is increased
    (covering fraction decreased), all lines become shallower, but the
    blended line is always shallower than the strong line.  For the
    $N_{ion} \Rightarrow \tau_{max}$ case (upper panel), the blended
    line is deeper than the strong line for $\log a > 0.7$, even
    though the total opacity for the blend is five times smaller than
    for the strong line.  This is unphysical, and therefore we reject
    the $N_{ion} \Rightarrow \bar\tau$ mapping.
    \label{investigate}}    \end{center}   
\end{figure*}

Further tests show subtle but significant differences in behavior that
lead us to reject the $N_{ion} \Rightarrow \bar\tau$ interpretation.
We  created a mock line list to test the differences between the two
methods. The mock line list includes a strong line, a weak line, and
a blend of four weak lines (Fig.~\ref{investigate}).  The weak lines all 
have the same line strength (i.e., same $\lambda f_{ik} N_{ion}$), and
the strong line is a factor 20 times larger.  Thus, the total opacity
of the blend is 5 times smaller than that of the strong line.  The
left panel shows the synthetic line profiles for a range of $\log a$
values for the $N_{ion} \Rightarrow \bar\tau$ method (top panel) and
the $N_{ion} \Rightarrow \tau_{max}$ method (bottom panel).   The
right panel shows the depth of each feature as a function of $\log
a$.   As expected, the depths of all features decrease with the
increase of $\log a$.   The difference is seen in the relative
decrease of the features for the two methods.  For the $N_{ion}
\Rightarrow \tau_{max}$ method, the depths of the lines decrease
together, maintaining the order of the total opacity. That is, the
strong line is always deeper than the blend, which is always deeper
than the weak line.  This makes sense, because the total opacity of
the strong line is 5 times that of the blend, which is in turn 4 times
 that of the weak line.  However, for the $N_{ion} \Rightarrow
 \bar\tau$ case and $\log a > 0.7$, the depth of the blend is larger
 than the depth of the strong line.  This is unphysical, since the
total opacity of the blend is smaller than the opacity of the strong
line.   This result occurs because, as
 mentioned above, in this method, opacities from {\it Cloudy} are
 multiplied by $1+a$ to obtain $\tau_{max}$ before 
the spectrum is made, and the $1+a$ factor dominates over the actual
opacity of the lines for sufficiently high $a$.   The same result is
obtained if line equivalent width is measured instead of line depth.
This problem is most noticeable when modeling overlapping-trough
FeLoBALs, where a large $\log a$ means that blends of iron multiplets
that are predicted to have low opacity still produce significant
optical depth due to the dominance of the $1+a$ factor.  

{\it SimBAL} uses the second method, i.e., $N_{ion} \Rightarrow
\tau_{max}$.   We have run a few tests using $N_{ion} \Rightarrow
\bar\tau$ on SDSS~J0850$+$4451, and we obtain commensurate total
column densities (so the derived  parameters do not change
significantly), but slightly lower log likelihoods (worse fits).  This 
preference for the $N_{ion} \Rightarrow \tau_{max}$ method 
makes sense for SDSS~J0850$+$4451, as the high opacity cores of the
clouds that yield sufficient opacity in weak lines such as \ion{C}{3}*
strongly constrain the $\log N_H -\log U$ best fit.  But given the
unphysical results produced by the  $N_{ion} \Rightarrow \bar\tau$
method for blended lines as discussed above, we see no reason to
investigate this method further.   

\subsection{The Effect of the Total Optical Depth of a Line}

In a slab of ionized gas, the column densities of different ions can
be dramatically different.   The opacity to \ion{C}{4} can be very
large, since this ion is abundant and the transition is easily
excited.  The opacity to other ions can be very low.  In the case of
\ion{P}{5}, the opacity is low because phosphorus has low elemental 
abundance compared with carbon.  Other ions may have low opacity
because they are found at the very back of the matter-bounded slab; for
example, for SDSS~J0850$+$4451, \ion{Mg}{2} and \ion{C}{3}* fall into
this category. 
Finally, other ions may have low opacity because they have low
oscillator strengths. An example of this category is
\ion{He}{1}*$\lambda 3889$, which has $f_{ik}=0.064$.  Many of the 
lithium-like ions have oscillator strengths that are much higher; e.g.,
\ion{Mg}{2} has $f_{ik}=0.608$ and $0.303$ for its doublet lines.

These different total optical depths combine with inhomogeneous
partial covering to yield different {\it effective} covering fractions
for different ions.  Physically, we can interpret the dependence of
effective covering fraction on line opacity in the power-law partial
covering parameterization if we imagine an inhomogeneous absorber
distributed over the emission region, which is resolved from the point
of view of the absorber.   For this thought experiment, we do not need
to specify the physical form of the inhomogeneity. C$^{+3}$ is a very
common ion in photoionized gas, and so it is probable that any line of  
sight through the inhomogeneous absorber would encounter an optically
thick column of \ion{C}{4}. Thus the covering fraction to \ion{C}{4}
would be close to 100\%.  In contrast, P$^{+4}$ is rare in
photoionized gas, due to its low abundance, and only a few lines of 
sight through thicker clumps would encounter sufficient P$^{+4}$ to
produce significant absorption.  So the effective covering fraction of
\ion{P}{5} would be smaller.   The same would hold true for other ions
that are rare. 

\begin{figure*}[!t]
\epsscale{1.0}
\begin{center}
\includegraphics[width=5.5truein]{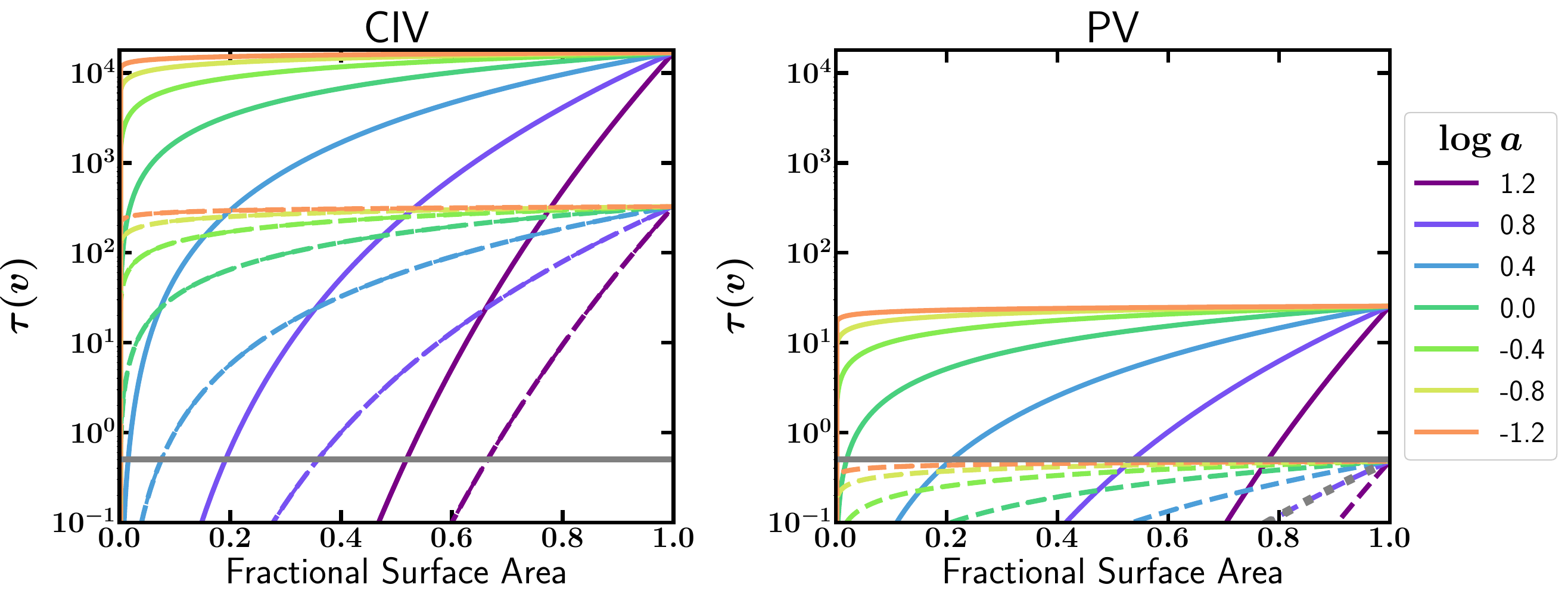}
\caption{The power-law optical depth parameterization convolved with 
  the line optical depth yields the effective covering fraction of the
  line.  {\it Left panel:}  The opacity of \ion{C}{4} as a
  function of   velocity at $-4000\rm \, km\, s^{-1}$ and $-2000\rm \,
  km\, s^{-1}$ (solid and dashed lines respectively).  The solid gray
  line shows $\tau(v)=0.5$, a value proposed by \citet{arav05} to be
  characteristic of the effective covering fraction (see text for
  details).    {\it Right panel:}  The same for \ion{P}{5}.    $\log a   =0.4$
  is a value typical of the UV   continuum at $-4000\rm \, km\,
  s^{-1}$ (i.e., in the   concentration), and $\log a=0.8$ is a
  value typical of wing of the   emission lines at $-2000 \rm \, km\,  
  s^{-1}$.   These plots show that for the same covering fraction
  parameter $\log   a$, the effective covering fraction is much
  greater for \ion{C}{4}   than for \ion{P}{5} because the total
  optical depth is much
  higher.      \label{opacities}}    \end{center}    
\end{figure*}

We illustrate this behavior by plotting the opacity $\tau(v)$ as a
function of fractional surface area for \ion{C}{4} and \ion{P}{5} in 
Fig.~\ref{opacities}.  We show the results for the {\it
  SimBAL} fit solutions shown in Fig.~\ref{uv_fit} for two bins
corresponding to offset velocities  $-4000 \rm  \, km\,
s^{-1}$ (i.e., in the concentration) and $-2000\rm \, km\, s^{-1}$ (on
the flank of the broad emission lines).  We plot $\tau(v)$ because the
opacity from the  {\it Cloudy} simulation is distributed evenly across
a velocity bin in the tophat opacity model; here we use $\Delta
v=511\rm \, km\, s^{-1}$, the value obtained as the best
fit. \citet{ss91} relate $\tau(v)$ to the ionic column density
$N_{ion}(v)$ through  $\tau(v)=2.654\times 10^{-15} f\lambda
N_{ion}(v)$ where $f$ is the oscillator strength of the transition,  
$\lambda$ is the wavelength of the line in Angstroms, and $N_{ion}(v)$
is in $\rm atoms\, cm^{-2} (km \, s^{-1})^{-1}$.  
The results are seen in Fig.~\ref{opacities}.  \citet{arav05} suggest
that the 
fraction of the surface area with opacity greater than 0.5 provides a
good fiducial number for the {\it effective} covered fraction.  At
$-4000 \rm \, km\, s^{-1}$ (solid colored lines), the UV continuum
covering fraction is $\log a \sim 0.4$, and the effective covered
fraction is 100\%.  At $-2000 \rm \, km\, s^{-1}$ (dashed colored
lines), the UV continuum covering fraction parameter is still $\log a
\sim 0.4$, but the long wavelength and broad-line-region covering
fractions are $\log \sim 0.8$.  The figure indicates that at $-2000
\rm \, km\, s^{-1}$, effective covering fraction of the emission line
region is only about 60\%. However, the absorption line appears much
deeper in the spectrum because the continuum  is effectively
completely covered, and the wing  of the line makes up only 25\% of
the total flux at $-2000\rm \, km\, s^{-1}$.

The right panel of Fig.~\ref{opacities} shows the results for
\ion{P}{5}.  Because $P^{+4}$ is a rare ion, the opacity for
\ion{P}{5} smaller than that of \ion{C}{4}.  At $-4000\rm \, km\,
s^{-1}$, the effective covering fraction is about 80\%; \ion{P}{5} is
a shallower line than \ion{C}{4}.  The same would be true for other
rare ions.  At $-2000\rm \, km\, s^{-1}$, the opacity is lower than
the fiducial minimum, and no line is observed.  This is expected; the
solution found by {\it SimBAL} yields a lower $\log N_H - \log U$ at
$-2000 \rm \, km\, s^{-1}$ compared with $-4000 \rm \, km\, s^{-1}$;
the gas is not optically thick enough to produce significant
\ion{P}{5}.  So the \ion{P}{5} line is observed to be narrower than
the \ion{C}{4} line.  

\begin{figure*}[!t]
\epsscale{1.0}
\begin{center}
\includegraphics[width=2.5truein]{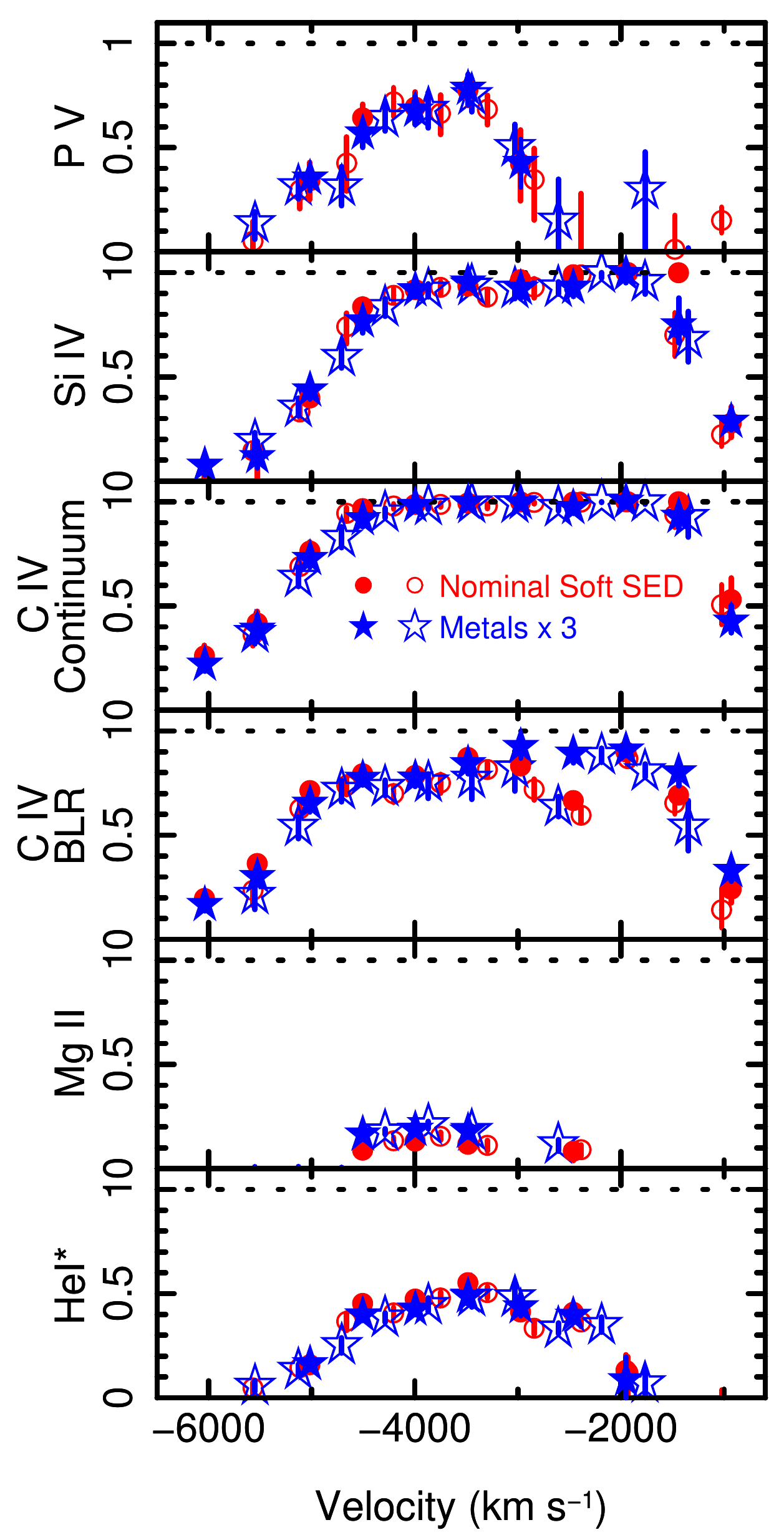}
\caption{The effective covering fraction as a function of velocity for
  five lines obtained using the $\tau=0.5$ criterion discussed in the
  text.  For \ion{P}{5}, \ion{Si}{4}, \ion{C}{4}, and
  \ion{Mg}{2}, only the results from the stronger doublet line is
  shown.   The open (filled) 
  symbols   show the results using the   first (second) continuum model.  The
  median values and 95\%   confidence regions from the posterior
  probability distributions are   shown.  The UV continuum $\log a$
  was used for \ion{P}{5},   \ion{Si}{4}, and the \ion{C}{4}
  continuum, the   long-wavelength $\log a$ was   used for
  \ion{Mg}{2}   and   \ion{He}{1}*, and the broad-line region  
  $\log a$ was used for \ion{C}{4} BLR.   This plot
  demonstrates that the covering   fraction parameter $\log a$ and
  the line opacity interact to produce   the effective covering
  fraction.  For example, although the value   of $\log a$ for the BLR
  is larger than for the UV continuum, the   greater opacity of
  \ion{C}{4} compared with \ion{P}{5} means that   the   effective
  covering fraction of \ion{C}{4} BLR is larger than for 
  \ion{P}{5}.    \label{effective}}    \end{center}    
\end{figure*}

Finally, we display the effective covering fractions for several lines
in Fig.~\ref{effective}.  In this figure we used the MCMC results and
computed the effective covering fraction for each of five absorption
lines using the $\tau=0.5$ criterion proposed by \citet{arav05}, and
the appropriate $\log a$ value: the UV continuum $\log a$ for
\ion{P}{5}, \ion{Si}{4}, and \ion{C}{4} continuum, the long-wavelength
$\log a$ for \ion{Mg}{2} and \ion{He}{1}*, and the broad-line region
$\log a$ for \ion{C}{4} BLR.  This plot shows that although the BLR
has a higher $\log a$ (lower covering fraction) than the UV continuum
or the long wavelengths, the effective covering fraction for the
\ion{C}{4} BLR is larger than that of \ion{P}{5}, \ion{Mg}{2} or
\ion{He}{1}* due to the much greater opacity of \ion{C}{4}.

\subsection{The Effect of the Brightness of the Background Source}

The rest-UV quasar spectrum is composed of the continuum emission,
presumably from an accretion disk, and emission lines.   Depending on
 the object, most of the lines have moderate equivalent widths, with
 the exception of Ly$\alpha$, which can be very strong.
 \ion{N}{5}$\lambda 1238$  and Ly$\alpha$ are separated by $\sim 5500
\rm \, km\, s^{-1}$, so for outflows with velocities much larger than 
this value, the \ion{N}{5} line will have the Ly$\alpha$ emission
lines as well as the accretion disk continuum as a background source.
Thus, the \ion{N}{5} 
line can be filled in by Ly$\alpha$, or Ly$\alpha$ can 
appear as a spike in the \ion{N}{5} trough, simply as a consequence of
the large intensity of the Ly$\alpha$ line. Different covering
fractions for the continuum and emission lines, as might be expected
for a relatively compact outflow, also contributes to Ly$\alpha$
leakage.   An example of these phenomena is seen in a composite
spectrum of strong \ion{P}{5} quasars \citep[][ their
  Fig.\ 9]{capellupo17}.  This composite spectrum shows deep 
absorption in \ion{C}{4}, \ion{Si}{4}, Ly$\alpha$, \ion{P}{5}, and
\ion{O}{6}, but the \ion{N}{5} absorption is quite shallow.  Taken at
face value, this result might suggest that the absorber is
characterized by low ionization, given that \ion{N}{5} is a
high-ionization line; however, the presence of strong \ion{O}{6} and
especially \ion{P}{5}, which is known to indicate a high ionization
parameter \citep{leighly09} refutes that idea.  {\it SimBAL} modeling
of individual objects explicitly demonstrates that 
the \ion{N}{5} absorption line can be diluted by a strong Ly$\alpha$
emission line \citep[][also 
  Hazlett et al.\ in prep]{leighly_aas19}.  

In summary, inhomogeneous partial covering, modeled here
using the power-law parameterization, produces a range of
covering-fraction phenomenology that depends both on the covering
fraction parameter, but also on the brightness of the background
source, as well as the abundance of the ions which in turn depends on
the physical conditions in the gas which are solved for using {\it
  SimBAL}.

\section{Discussion}\label{discussion}

\subsection{Size Scales in SDSS~J0850$+$4451}\label{size_scales}

We have demonstrated that the fraction of the continuum emission
region covered is about 2.5 times smaller in the near-infrared compared
with the UV in SDSS~J0850$+$4451.   We have also found that
the fraction of the broad line region covered is consistent with the
fraction of the near-infrared continuum emission region covered.  
To understand the implications of these results on the structure of
the broad absorption line outflow, we first examine the size scales of 
the continuum emission region, the broad line region, and the torus,
and compare those with the location of the absorber, established in
Paper I to be 1--3 parsecs from the central engine.

We used a simple  sum-of-black-bodies accretion disk model
\citep{fkr02} to estimate 
the sizes of the continuum emission regions.   The black hole mass was
shown to be $1.6\times 10^9\rm \, M_\odot$ in Paper I.  The log   
bolometric luminosity was estimated by \citet{luo13} to be $46.1 \rm
\, [erg\, s^{-1}]$.  We assumed a standard accretion efficiency of
$\eta =0.1$.  Using these values, we estimated an accretion rate of
$\dot M_{acc}=2.2 \rm \, M_\odot \, yr^{-1}$.  { This was reported
  to be smaller than the outflow rate from the wind by a factor of
  $\sim 8$ by \citet{leighly18}, but that value is revised to a factor
  of $\sim 4$ from the results presented in this paper as a
  consequence of the lower covering fraction over the larger region
  (\S\ref{blr}). }

We first estimated the continuum emission region sizes (radii) for four
wavelengths: 1100\AA\/ and 1600\AA\/ (values that span the {\it HST}
spectrum), 2770\AA\/ (corresponding to the \ion{Mg}{2} absorption line),
and 10700\AA\/ (corresponding to the \ion{He}{1}* line).  We used the 
Wien displacement law and the $T(R)=3GM$\.M$/8\pi R^3 \sigma$ run of
temperature for a sum-of-blackbodies accretion disk  to estimate the 
temperatures at which the Planck function should be a maximum at these
wavelengths.  These values are 0.0016, 0.0027, 0.0056, and 0.034 pc
respectively.  We find that the radius increases with wavelength as
a powerlaw with an index of $4/3$ as expected for a sum-of-blackbodies
accretion disk.  Thus, the radius of the continuum emission region
absorbed by \ion{P}{5} is 21 (441) times smaller than the radius
(area) of the continuum emission region absorbed by
\ion{He}{1}*$\lambda 10830$.    

\begin{figure*}[!t]
\epsscale{1.0}
\begin{center}
\includegraphics[width=6.5truein]{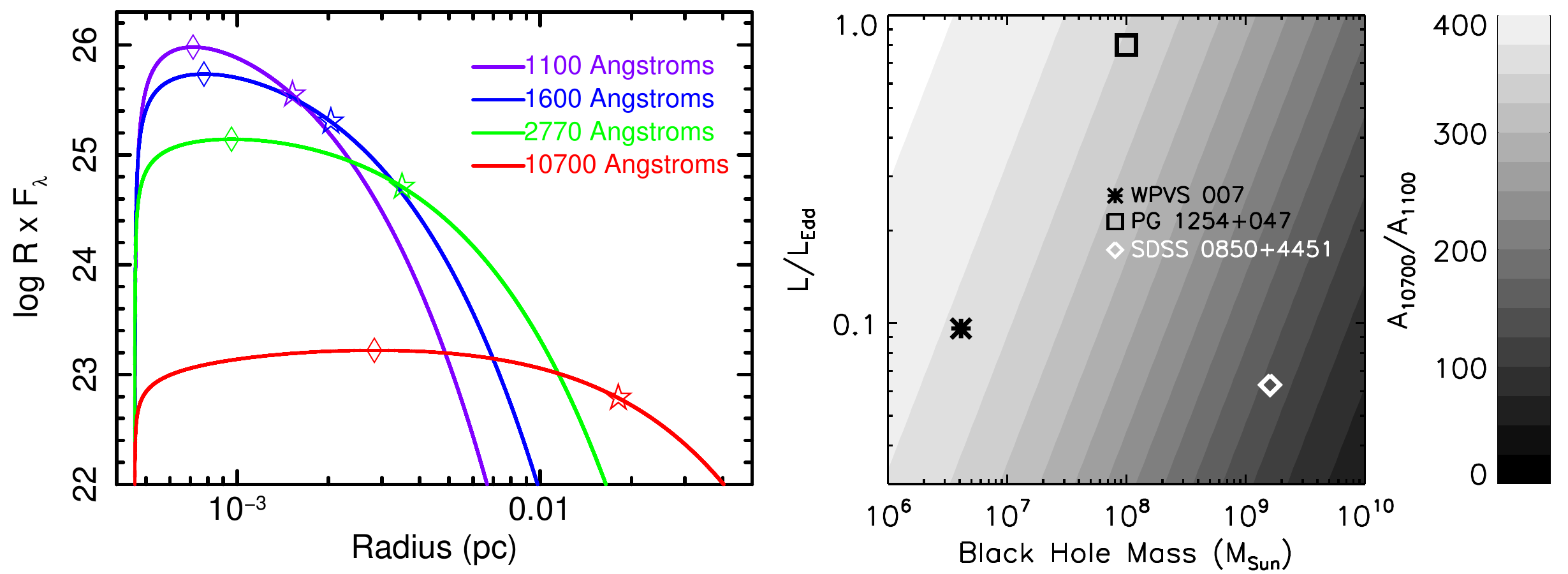}
\caption{{\it Left:} The log of the radius-weighted flux density as a
  function of   radius for the $M_{BH}=1.6\times 10^9 \rm M_\odot$ and $\dot M=2.2
  \rm \, M_\odot \, yr^{-1}$ black hole at four wavelengths for the
  sum-of-blackbodies accretion disk. 
  The location of maximum emission is shown by a diamond, and the star
  marks the outer radius at which the radius-weighted flux density is
  $1/e$   times the maximum. { {\it Right:} The ratio of the area of
  the accretion disk emitting $10700$\AA\/ to the area emitting
  $1100$\AA\/, where the areas are defined by the radius at which the
  radius-weighted flux density drops to $1/e$ times the maximum value,
  and as a function of black hole mass and accretion rate.
  SDSS~0850$+$4451 has relatively low accretion rate and large black
  hole mass, yielding a cool disk and relatively small area
  ratio.}  \label{accretion_disk}} 
\end{center}
\end{figure*}

This analysis ignores the fact that blackbodies at other
temperatures will contribute to the flux at any given wavelength, and
that the emission at a given wavelength in a disk needs to be weighted
by the radius \citep[e.g., Fig.\ 5.7,][]{fkr02}.
Fig.~\ref{accretion_disk} shows the radius-weighted 
flux density at the four wavelengths.   The radius at which the
emission is maximum is marked, but since there is considerable
emission outside of that radius, we identified the size of the
accretion disk at each wavelength to be the radius at which the flux
density falls to $1/e$ of the maximum value (i.e., roughly the
half-light radius).   Those radii are 0.0015, 0.0020, 0.0035, and
0.018 parsecs at 1100, 1600, 2770, and 10700\AA\/ respectively.  These
values are comparable to although smaller than the Wein
displacement-estimated values above, with radius increasing with
wavelength as a powerlaw with an index of $1.16$. 
{ Using the radii defined above, we find that the ratio of the area
  of the accretion disk emitting substantially at 10700\AA\/ to the
  area emitting 1100\AA\/ is 140.}

These values do not fully account for the difference in continuum
emission as a function of radius, because the radius-weighted flux
density falls off faster with radius for shorter wavelengths.  For
example, the slope of the power law tangent to the
$1/e$ point increases from $-2.44$ at 1100\AA\/ to $-1.45$ at
10700\AA\/.  This means that is there is quite  a large region of
accretion disk where the near-infrared continuum is emitting strongly
but the far-UV continuum emission is negligible, but  the same cannot
be said of the near-UV (near 2770\AA\/) versus the far-UV
($<1600$\AA\/).  So, while we can expect to be able to measure a
difference in the covering fractions between  1100\AA\/ and
10700\AA\/, there is too much emission overlap between the 1100\AA\/
and the 2770\AA\/ continuum-emitting regions to be able to detect a
difference in covering fraction. Thus, we need the long wavelength
absorption from  \ion{He}{1}*$\lambda 10830$ to do these
covering-fraction experiments.    

{ These sizes depend on the black hole mass and accretion rate
  relative to Eddington.  SDSS~0850$+$4415 has perhaps a relatively
  large  black hole mass and relatively low accretion rate compared with the
  expectation for broad absorption line quasars
  \citep[e.g.,][]{boroson02}; either a lower black hole mass or a
  large accretion rate relative to Eddington would predict a hotter
  accretion disk.  The dependence of the ratio of the 10700\AA\/
  emission region area to the 1100\AA\/ emission region area is
  explored in Fig.~\ref{accretion_disk}.  We find that hotter disks
  predict a much larger ratio of areas, perhaps implying that a more
  significant difference in the UV versus \ion{He}{1}* covering
  fractions and/or physical conditions might be expected for smaller
  black hole masses and higher accretion rates.  We return to this
  point in \S\ref{selection}.}

It is interesting to visualize how the accretion disk would appear from the
perspective of an observer at the location of the absorber.  
In Paper I we found that the absorber is constrained to lie in the
vicinity of the torus, about 1--3 parsecs from the central engine.
At this distance, the 1100\AA\/ continuum emission region (diameter)
would subtend 3.5--10.5 arcminutes, while the 10700 \AA\/ emission
would subtend 0.7--2.1 degrees, a bit larger than the full moon.

{ The size scales and other results computed based on the
  sum-of-blackbodies accretion disk should be used with some caution,
  as it can only approximately model the broad-band spectral energy
  distribution of quasars.  It would be interesting to estimate size
  scales using more sophisticated accretion disk models, such as the
  one by \citet{done12}.  }

In Paper I, we estimated the radius of the
H$\beta$ emission to be $0.13^{+0.024}_{-0.021}\rm \, pc$, using the
reverberation-mapping regression measured by \citet{bentz13}.
As discussed in Paper I, \citet{luo13} fit the H$\beta$ emission-line
profile with a relativistic Keplerian disk model, obtaining inner and
outer radii of 450 and 4700 $r_g$ respectively.  For our derived black
hole mass, these values correspond to $r_{in}=0.035\rm\, pc$ and 
$r_{out}=0.37 \rm \, pc$ respectively, consistent with the
reverberation-mapping estimate.   For reference, $r_{in}$ is $10\times$
larger than the $1/e$ radius of the 2770\AA\/ emitting region.

We can also estimate the location of the \ion{C}{4} emission region
using the regression presented by \citet[][Eq.\ 1]{lira18}.  The flux
density at 1345\,\AA\/ is $2.4\times 10^{-15}\rm \, ergs\, s^{-1}\,
cm^{-2}$\AA\/$^{-1}$,  corresponding to $\lambda L_\lambda = 3.7\times
10^{45} \rm \, erg\, 
s^{-1}$.   The \ion{C}{4} radius is therefore estimated to be
$0.028^{+0.039}_{-0.019}\rm \, pc$, where the errors come from
the uncertainty on the regression parameters.  Using these values, we
find that the H$\beta$ emission region is 4.6 times larger than the
\ion{C}{4} emission region.  This is somewhat larger than but
comparable to the values found from reverberation mapping.   The
Seyferts NGC~5548, NGC~3783, and NGC~7496 show an H$\beta$ lag about
1.8 to 2.8 times the \ion{C}{4} lag
\citep{pw00,op02,wanders97,collier98}.  The exception is the
double-peaked object 3C~390.3, which showed an inverted relationship,
with the \ion{C}{4} lag about twice the H$\beta$ lag.   

\begin{figure*}[!t]
\epsscale{1.0}
\begin{center}
\includegraphics[width=4.0truein]{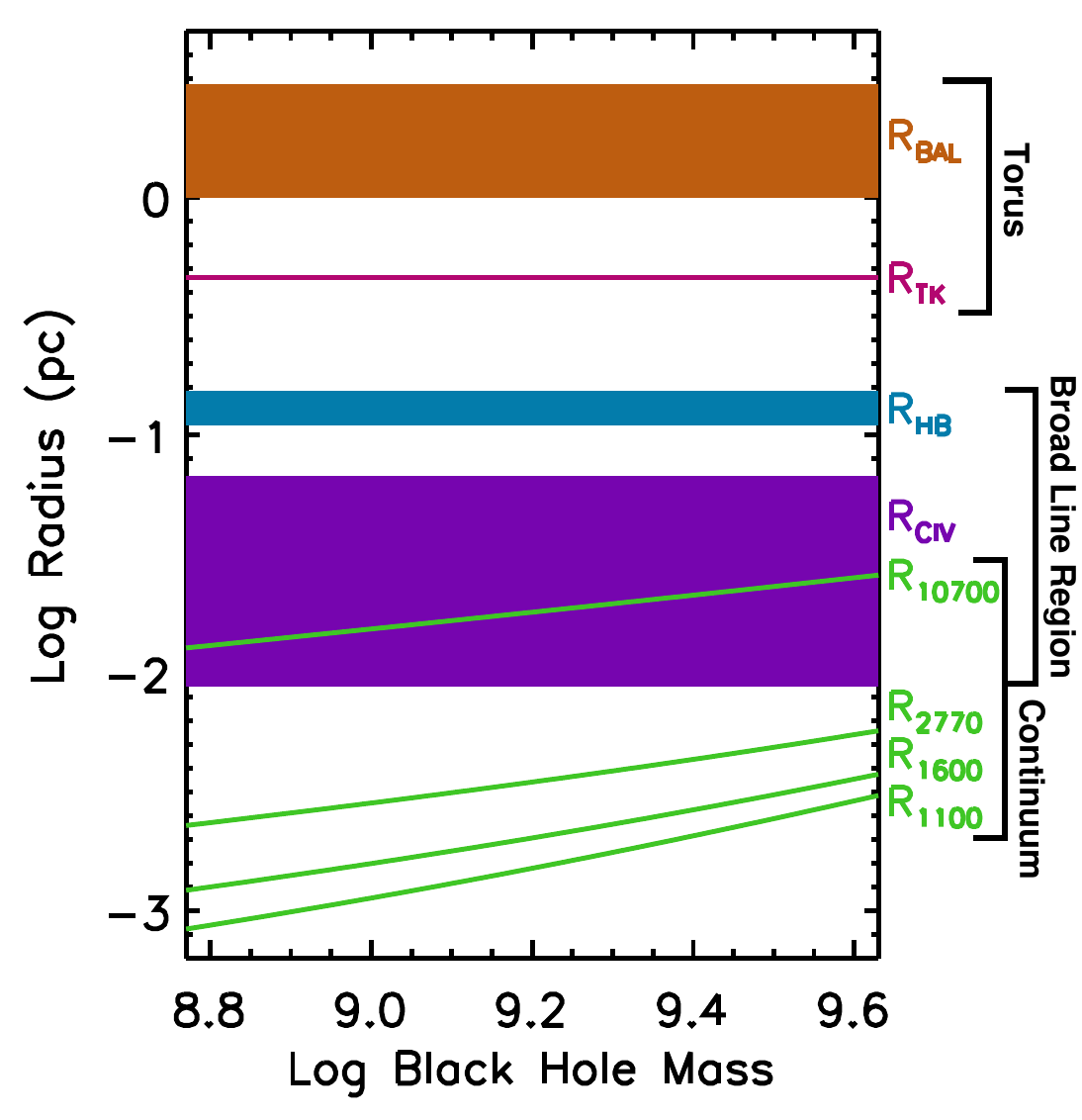}
\caption{Size scales for SDSS~J0850$+$4451 as a function of black hole
  mass, where the uncertainty in the single-epoch black hole mass
  estimate is assumed to be 0.43 dex \citep{vp06}.  
  From bottom to top, the green lines show the radius at which the
  radius-weighted 
  flux density drops to $1/e$ of its maximum value (see text for
  details).   The broad line region radii, $R_{H\beta}$ (teal) and
  $R_{CIV}$ (purple),   were computed using the reverberation
  regressions from   \citet{bentz13} and \citet{lira18}, respectively,
  where the color   bands show the uncertainty in the regression
  parameters.  The inner   edge of the torus, $R_{\tau_K}$ (magenta),
  was estimated based on the   regressions from \citet{kishimoto07},
  while $R_{BAL}$ (orange)   is the radius 
  of the broad absorption line outflow measured in Paper I.\label{fig20a}} 
\end{center}
\end{figure*}

The size scales for SDSS~J0850$+$4451 are shown in Fig.~\ref{fig20a}
as a function of black hole mass, where we have assumed that the systematic
uncertainty in single-epoch black hole mass is 0.43 dex \citep{vp06}.
Besides the continuum and emission-line radii discussed above, we have
also graphed the estimated location of $R_{\tau_K} = 0.46 \rm \, pc$,
 the hot inner edge of the torus \citep{kishimoto07}
computed in Paper I, as well as the estimated radius of the outflow
measured in Paper I.  The plot shows the expected hierarchy of size
scales. Interestingly, the near-infrared continuum emission overlaps the
\ion{C}{4} emission region.

\subsection{Partial Covering in SDSS~J0850+4451}\label{partial_covering}

Armed with the quasar size scales, we can discuss the implications of
our results on simple scenarios for partial covering in
SDSS~J0850$+$4451 and BALQs in general.  

In one  possible scenario for partial covering  (left panel of
Fig.~\ref{fig20}), the size scale of the outflow is large compared 
with the UV continuum emission region. It would then essentially
completely cover the far UV-continuum emission region, but only
partially cover the near-IR emission region. This scenario is ruled out
because it predicts that covering fraction to the UV continuum would
be 100\%, and that is not the case. 

Alternatively, the absorbing clumps are small, but have internal
structure on the size scale of the 1100\AA\/ continuum emission
region, and the clumps are diffusely distributed on large size scales 
(middle panel of Fig.~\ref{fig20}). In this scenario, similar to the
one posited by \citet[][their Fig.\ 6]{hamann01}, each
clump might present a distribution of column densities to the
continuum source, as would be expected for, e.g., a spherical clump.
Each clump would behave as a photoionized slab, with the effective
column density and covering fractions of various ions depending on
both the abundance of the ion, and where the ion is located within the
clump (e.g., on the surface, as might be expected for a
high-ionization ion, or buried deep, as expected for a low-ionization
ion).  Thus, partial covering to the UV continuum is achieved by the
structure of the clumps (presenting a range of thicknesses to the
illuminating continuum); such a model seems
roughly consistent with the power law covering fraction
\citep{dekool02c}.  A lower covering fraction to the
infrared continuum is achieved by assuming that these clumps are
sparsely distributed on larger size scales.  

{ If this scenario is correct, we can estimate the sizes of the
  individual clumps by dividing the covering-fraction-weighted column
  density by the density, assuming that the clumps are approximately 
  spherical.  There are several additional assumptions that need to be
  made, however. First, it is probably not   reasonable to assume that
  one clump produces the absorption spanning   the whole outflow,
  i.e., $5000\rm \, km\,  s^{-1}$, especially since   the covering
  fraction for a single ion is observed to vary across   the trough
  profile.  We assume, somewhat arbitrarily, that a clump   spans one
  velocity bin.  We also assume, again somewhat arbitrarily,   that
  velocity bin with   the thickest outflow (at $-4000\rm \, km\,
  s^{-1}$) is most representative.  The other bins that have similar
  covering fraction may be the same size but physically thinner (lower
  column density and more pancake-like).  The bins that have lower
  covering fractions   may have  have the same size but a sparser
  distribution, i.e., a smaller number of clouds across the continuum
  emission region.  This scenario is by no means unique; other
  configurations could be constructed that are consistent with the
  analysis results.  

The spectra are not very sensitive to the density; this is illustrated
by \citet{leighly18} in their Figure 11.  All we can say with
any certainty is that the presence of \ion{C}{3}* means that the
density is greater than the critical density for that ion, $\log n
\sim 10^6\rm \, cm^{-3}$.  Thus, the cloud size is highly
degenerate with density.   We illustrate this covariance for the
second continuum model and enhanced metallicity run via the corner
plot shown in Fig.~\ref{corner}.  For that simulation, a typical cloud
size (diameter) is $5\times 10^{-4}\rm \, pc = 1.5 \times 10^{15}\rm \,
cm$.

\begin{figure*}[!t]
\epsscale{1.0}
\begin{center}
\includegraphics[width=3.5truein]{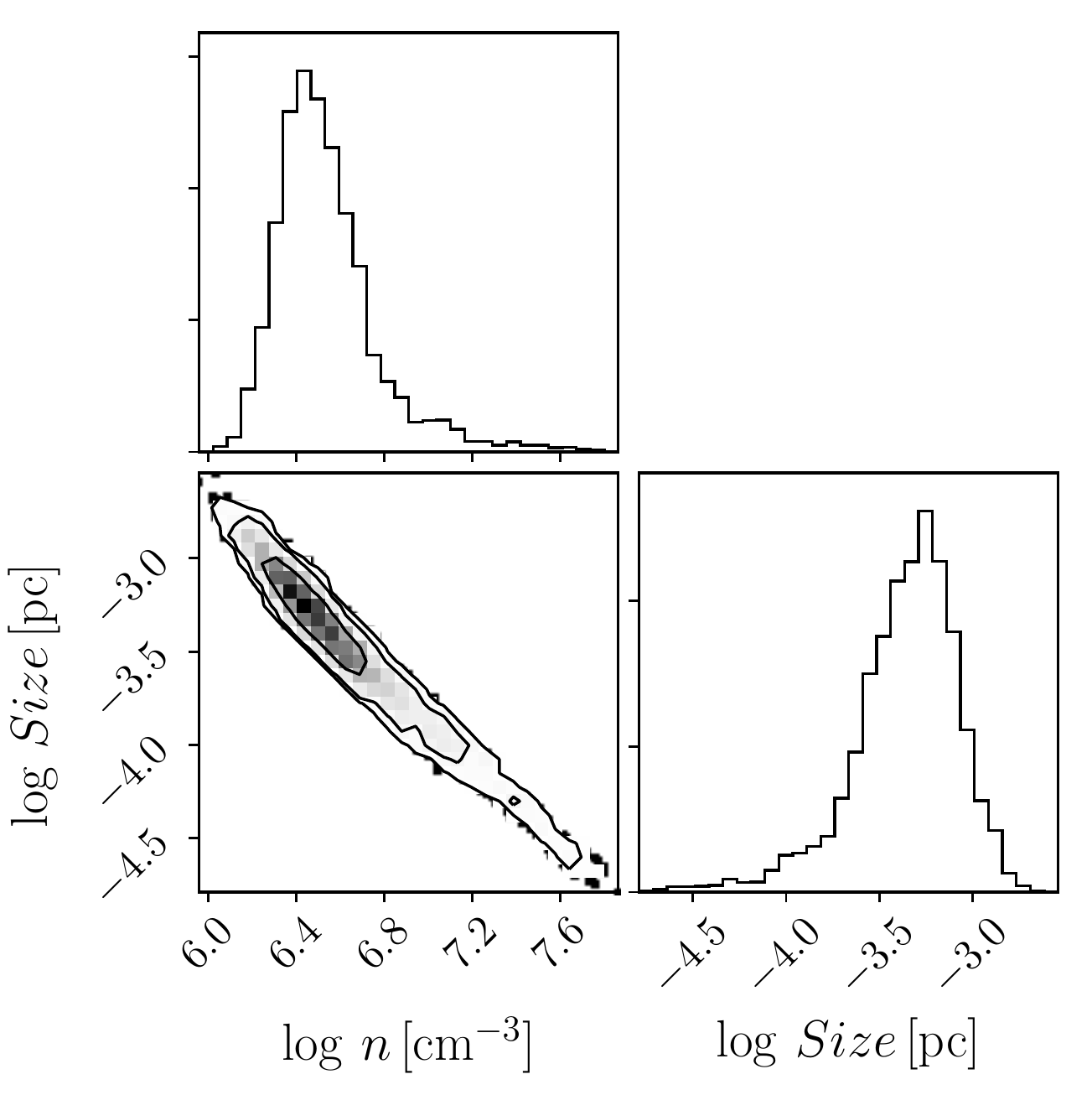}
\caption{Representative size scales for spherical clouds for the
  $-4000\rm \, km\, s^{-1}$ velocity bin for the enhanced metallicity
  and second continuum as a function of density.  The contours show
  the 68, 95, and 99\% uncertainty levels.  The spectra are
  generically not as sensitive to density as other parameters.  Since
  the size scale is the covering-fraction-weighted column density
  divided by the density, the cloud size scale is strongly covariant
  with the density. \label{corner}}  
\end{center}
\end{figure*}

The size scales for the 1100\AA\/ and 10700\AA\/  continuum emission
regions were found in \S\ref{size_scales} to be 0.0015 and $0.018\rm
\, pc$, respectively.  Assuming that we see the accretion disk face
on, and that the inhomogeneous partial covering for the 1100\AA\/
continuum emission region is accounted for by weighting the column
density with the covering fraction, we find that for the $-4000\rm \,
km\, s^{-1}$ bin, approximately 35 clouds are sufficient to cover the
emission region.  The 10700\AA\/ continuum emission region was shown
to be 140 times larger, but the covering fraction was found to be 2.5
times lower (\S\ref{quantifying}), indicating that $\sim 2000$ clouds
would be sufficient to cover the 10700\AA\/ emitting accretion disk,
if observed face on.} 

Another possibility is that the clumps are very small, much smaller
than the UV continuum emission region (right panel of
Fig.~\ref{fig20}).  In this scenario, the individual clumps should be
clustered on size scales smaller than the optical--infrared continuum
emission region because if uniformly distributed, identical covering
fractions in the UV and long wavelengths would be observed.   We note
that the difference in distribution is rather modest for this object;
in \S\ref{quantifying}, we inferred that the UV continuum is a factor
of only 2.5 times more densely covered than the optical through
near-infrared continuum.  

\begin{figure*}[!t]
\epsscale{1.0}
\begin{center}
\includegraphics[width=6.5truein]{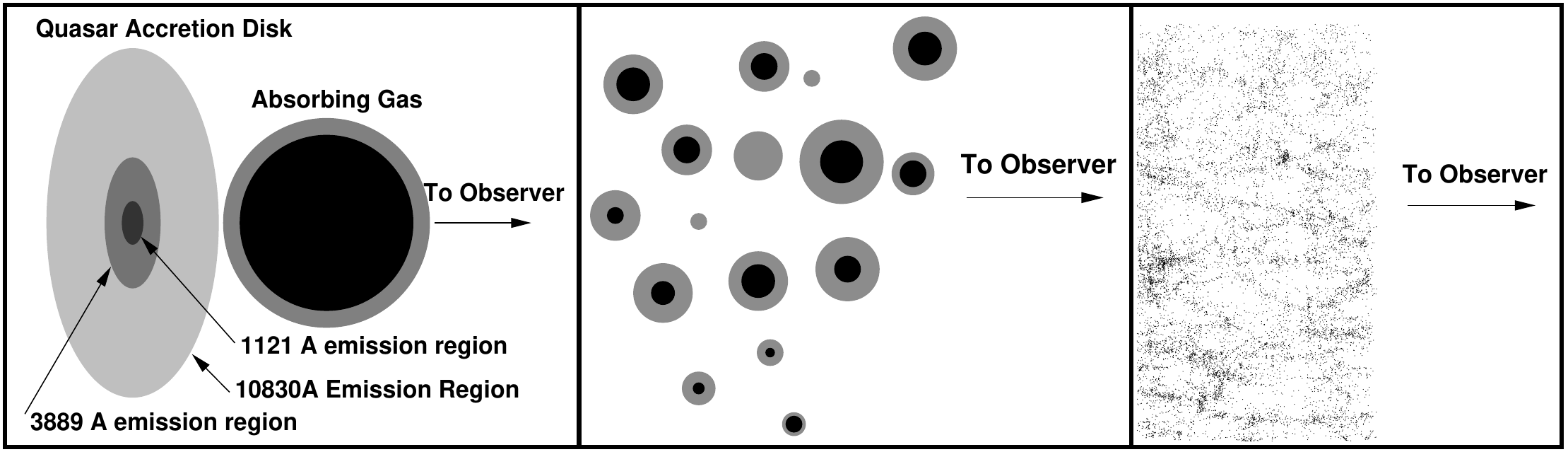}
\caption{Partial covering scenarios.    {\it Left:} Geometrical
  partial covering, in which the absorber only partially covers the
  emission region (not to scale). Here, the \ion{P}{5} absorption line
  would be expected to fully cover the continuum emission region,
  while the \ion{He}{1}* would show partial covering; this scenario is
  excluded because the UV continuum is observed to be partially
  covered. {\it Middle:} 
  Inhomogeneous covering, in which the   absorber   consists of many
  small clouds that, together, do not   completely occult the
  continuum   emission region (adapted from   \citet{hamann01}). Black
  represents the higher column  density cores   of the absorbing
  clumps, so that \ion{C}{4} absorption (from an   abundant ion) will
  come from the whole cloud (gray + black), while    \ion{P}{5}   and
  \ion{He}{1}* absorption only occurs in the   highest column density
  cores (black).    {\it Right:}
  Partial covering   characterized by  clumping of the inhomogeneous
  absorbers;     the data to create this plot were borrowed from the
  Two Degree Field   Galaxy Redshift Survey \citep{colless01}.  The
  clustering of small absorbers predicts both partial covering of  UV
  continuum emission region, and a lower covering fraction for the
  near-infrared continuum emission region, as we infer for 
SDSS~J0850$+$4451.\label{fig20}} 
\end{center}
\end{figure*}

Additional information can be obtained from expected time scales of
variability.  At a distance of 1--3 parsecs from a $1.6\times
10^{9}\rm \, M_\odot$ black hole, the Keplerian velocity is between
$1500$ and $2600\, \rm km\, s^{-1}$.  Considering the 2770\AA\/ $1/e$
continuum diameter as a crossing size scale, we find that an
orbiting cloud could cross the continuum emitting region in
between 2.6 and 4.6 years.  This seems to be similar to the time scale 
of the variability inferred from the \ion{Mg}{2} absorption line
changes (Appendix~\ref{obs_var}), although we note that the pattern of
variability perhaps suggests that ionization changes may be
responsible (Appendix~\ref{predicted}), rather than covering fraction
changes.  In contrast,  it would take 13--24 years to cross the $1/e$
near-infrared continuum emission region, possibly predicting less
variability in the \ion{He}{1}*$\lambda 10830$ absorption line.

Finally, while our principal focus has been on the difference in
covering fraction between the UV and near-infrared continuum emission
regions, we also made measurements of the covering fraction of the
broad-line region.  We found that there is no strong evidence for a
difference in covering fraction between the broad-line region and the
long-wavelength-continuum emission region.  
While it is difficult to know how robust this result is given the
large number of degrees of freedom in the models, we note that this
result is consistent with the size scales shown in Fig.~\ref{fig20a},
where the radius of the near-infrared continuum emission region is
completely included in the range of \ion{C}{4} emission regions
estimated based on the the \citet{lira18} regressions.

\subsection{Selection Effects and Other Objects}\label{selection}

We have demonstrated, using {\it HST} COS and optical and
near-infrared spectra of SDSS~J0850$+$4451, that a difference in
covering fraction to the UV and near-infrared continuum emission
regions is observed and is quantifiable.  We note, however, that this experiment had
significant selection effects built in.  { We chose SDSS~J0850$+$4451
for {\it HST} COS observations because it showed strong \ion{He}{1}*
absorption, and because of the similarity in opacity between
\ion{He}{1}*3889 / \ion{He}{1}*10830 and \ion{P}{5} predicted over a broad range
of physical conditions \citep[][Fig. 15,
  Section 4.4.1]{leighly11},
we could practically guarantee that \ion{P}{5} would be present.   }

The ratio of the covered portion of the UV continuum-emitting
  region to the covered portion of the near-infrared continuum-emitting
  region, was found to be 2.5.  In principle, the covered portion of
  the infrared continuum-emitting region could be much lower. That
is, since the $1/e$ radius of the near-infrared continuum 
emission region (under \ion{He}{1}*) is 140 times larger than the 
radius of the UV continuum emitting region (under \ion{P}{5}), objects
could exist that have significant \ion{P}{5} absorption, but no
\ion{He}{1}* absorption.   In fact, there are two examples where this 
seems to be the case.

\begin{figure*}[!t]
\epsscale{1.0}
\begin{center}
\includegraphics[width=3.5truein]{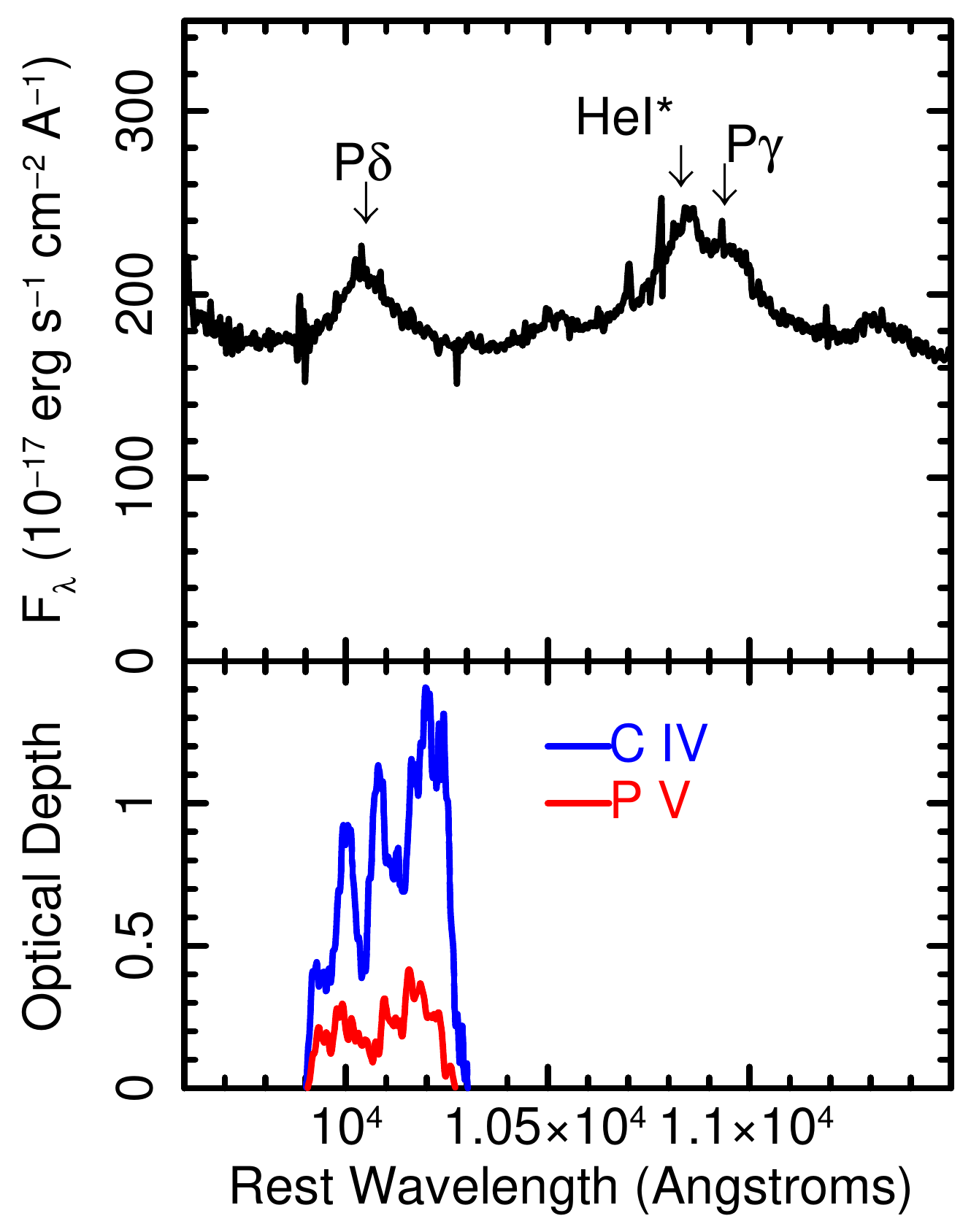}
\caption{The top panel shows a segment of the LBT {\it LUCI} spectrum
  of the $z=1.010$   quasar PG~1254$+$047.   The bottom panel shows
  the apparent optical depths of the \ion{C}{4} and \ion{P}{4}
  absorption lines from the {\it HST} FOS spectrum \citep{hamann98a}
  shifted to the wavelengths appropriate for \ion{He}{1}*$\lambda
  10830$ absorption. 
  No evidence for \ion{He}{1}* absorption is seen.  This could mean
  that the absorber does not occult a significant portion of the
  near-infrared continuum emission region, or that variability in the
  absorption has occurred between the 1993 {\it HST} observation and
  the 2013 LBT {\it LUCI} observation.
\label{fig21}}
\end{center}
\end{figure*}

The $z=1.010$ quasar PG~1254$+$047 is known to host  \ion{P}{5}
absorption; it is best known as the original \ion{P}{5} quasar
\citep{hamann98a}.  Our {\it LBT} observation of this object is
described in \S\ref{lbt}.  
A segment of the PG~1254$+$047 spectrum is shown in Fig.~\ref{fig21},
along with the apparent optical depths of the \ion{C}{4} and
\ion{P}{5} lines, digitized from \citet{hamann98a} Fig.\ 3 and shifted
to the wavelengths appropriate for \ion{He}{1}*$\lambda 10830$
absorption.  No evidence for absorption is observed, perhaps implying
that the UV absorber does not occult an appreciable amount of
the near-infrared-emitting continuum, and indicating a relatively
compact absorber.   However, the {\it HST} observation was
made in 1993, and the LBT {\it LUCI} observation was made 20 years
later, and it is quite possible that the broad-line absorption has
changed its properties, becoming optically thin enough that appreciable
\ion{He}{1}* is not expected, or that the absorber has completely
disappeared.  

Another example is provided by the low-luminosity broad absorption
line object WPVS~007.  \citet{leighly09} reported the emergence of a
broad absorption line outflow in the 2003 November 6 {\it FUSE}
spectrum of this object; a broad \ion{P}{5} absorption line was among
the lines that were observed. \citet{grupe13} presented an
near-infrared spectrum 
obtained in 2004 September 12.  Quantitative analysis of the lack of a
\ion{He}{1}*$\lambda 10830$ line yielded a conservative estimate of
the apparent metastable helium log column density of 12.9 [$\rm
  cm^{-2}$].  In comparison, the apparent log column 
density of \ion{P}{5} in the {\it FUSE} spectrum was 15.4 [$\rm
  cm^{-2}$] \citep{leighly09}.  The opacity represented as a 
function of velocity is given by $\lambda f_{ik} N_{ion}(v)$
\citep{ss91}, implying that the opacity due to \ion{P}{5} at least 41
times the opacity due to \ion{He}{1}*$\lambda 10830$, rather than
comparable as predicted by photoionization models \citep{leighly11}.
This result seems to indicate a dramatic difference in the coverage of
the UV compared with the near-infrared regions in WPVS~007. However,
WPVS~007 is known to have variable absorption lines \citep{leighly09,
  leighly15}, leading to some doubt about this robustness of this
conclusion.  The near-infrared observation was made after the 
emergence of the broad absorption lines, and broad absorption lines
were observed in subsequent {\it HST} observations \citep{leighly15}
and so it seems unlikely that variability explain the lack  of
\ion{He}{1}* absorption.   

{ A potential interesting difference between WPVS~007 and
  PG~1254$+$047 versus SDSS~0850$+$4451 is illustrated in
  Fig.~\ref{accretion_disk}.  We found that because the black hole
  masses are smaller in these two objects, their disks are hotter, and
  the ratio of the area emitting 10700\AA\/ to the area emitting
  1100\AA\/ is larger.    The black hole mass and accretion rate
  for WPVS~007 were estimated by \citet{leighly09} to be
  $M_{BH}=4.1\times   10^6\rm \, M_\odot$ and $L/L_{Edd}=0.096$.  The
  black hole mass and   accretion rate for PG~1254$+$047 were
  estimated by \citet{sabra01}   to be $M_{BH}=1\times 10^8\rm \,
  M_\odot$ and $L/L_{Edd}=0.8$.   Using the sum-of-blackbodies disk
  model   described in   \S\ref{size_scales} we found that the ratio
  of the   accretion disk   area   emitting 10700\AA\/ to the area
  emitting   1100\AA\/ is 370 and   350 for WPVS~007 and PG~1254$+$047   
  respectively, much larger than   the value of 140 found for
  SDSS~0850$+$4451. Despite the caveats regarding the
  sum-of-blackbodies accretion disk model discussed in
  \S\ref{size_scales}, these numbers suggest that it might be
  reasonable to expect that if outflows are not uniform, then we might
  be more likely to observe this lack of uniformity in
  objects  with larger area ratios.  It is possible that this fact
  explains  the lack of \ion{He}{1}*$\lambda 10830$ in WPVS~007 and 
  PG~1254$+$047.  On the other hand, both of these objects have
  smaller black hole masses than SDSS~0850$+$4451,   and therefore
  smaller size scales.  If {\it SimBAL} analysis of their spectra were
  to yield a similar solution as for SDSS~0850$-$4451, then the
  clouds might have similar sizes, and fewer would cover the emission
  regions.  Detailed analysis of more objects would be  necessary to
  deconvolve these effects.  }   
                
\section{Summary and Future Prospects}\label{conclusions}

\subsection{Summary of Results}

This is the second of two papers investigating the outflow in
the low-redshift LoBAL quasar SDSS~J0850$+$4451.  The first paper
described application of the novel spectral synthesis code {\it
  SimBAL} to the {\it HST} COS spectrum. We found that the absorber is
located about 1--3 parsecs from the central engine, among other
results. This paper describes extrapolation of the {\it SimBAL}
solution to long wavelengths, and the implications for the nature of
partial covering in this object. Our principal results follow. 

\begin{enumerate}
\item  In \S\ref{extrapolation}, we showed that the extrapolation of
  the   best-fitting spectral synthesis model of the UV spectrum of
  SDSS~J0850$+$4451 obtained in Paper I to long wavelengths indicates
  that the   \ion{Mg}{2},   \ion{He}{1}*$\lambda   3889$, and
  \ion{He}{1}*$\lambda   10830$   lines are all predicted to be
  significantly   deeper than   observed, implying that the smaller UV
  continuum emission region experiences a  higher covering fraction
  than the larger  optical / near-infrared continuum   emission
  region. 
\item  In Appendix \ref{variability}, we discussed the observed variability 
  in the absorption lines, and concluded that the variability is
  unlikely to have produced the difference in UV and  optical/near-infrared
  covering fractions,  but it cannot be ruled out   absolutely.   In
  Appendix \ref{host} we   presented analysis of   broad-band 
  photometry and an archival {\it     HST} near-infrared-band   image of
  SDSS~J0850$+$4451, and showed that   the  contribution of the   host
  to the near-infrared continuum is   negligible.  Therefore, dilution of
  the   \ion{He}{1}*$\lambda 10830$ absorption line by the host galaxy
  continuum cannot be responsible for the difference in UV and optical
  / near-infrared covering fractions.  
\item In \S\ref{quantifying}, we found that the absorber
  covers about 2.5   times more of the far UV   continuum emission
  region than the optical through near-infrared continuum emission
  region.   
\item In \S\ref{blr}, we performed {\it SimBAL} modeling of the
  UV-through-infrared spectra, using three sets of covering fractions: 
  one for the UV continuum, one for the near-UV through near-infrared
  continuum, and one for the broad emission lines.  We found that the
  near-UV through near-infrared continuum covering fraction results
  were consistent with the constrained modeling presented in
  \S\ref{quantifying}, and that the covering fraction of the broad-line
  region is mostly consistent with the covering fraction of the
  optical    through near-infrared continuum.  { Considering that
    the projected size of the infrared continuum emission region is
    much larger than the UV continuum emission region, we revise the
    estimated bulk outflow properties from Paper I downward to
    account for the lower covering fraction.  For the
    statistically-preferred enhanced metallicity model, the estimated
    column density of the outflow is $\log N_H=22.19\rm \, [cm^{-2}]$,
    the radius of the outflow is $2.2$--$3.0\rm \, pc$, and the mass
    outflow rate is $\dot M=8-12\rm \, M_\odot\, yr^{-1}$.  Finally,
    the ratio of the kinetic to 
    bolometric luminosity is 0.4--0.6\%.  This range
straddles the 0.5\% value taken to be a conservative cutoff for
effective galaxy feedback \citep{he10}.  Therefore, SDSS~J0850$+$4451
does not appear to be undergoing strong feedback from the BAL outflow.} 
\item  In \S\ref{understanding}, we discussed inhomogeneous partial
  covering and the power-law parameterization used by {\it SimBAL}.
  Four factors that must be considered in order to understand how
  absorption lines are shaped: the concept of inhomogeneous partial
  covering itself, the mapping of the output of the photoionization
  models (ionic column densities) to the power-law parameterization,
  the opacity of the particular line, and the relative brightness of
  the background source.  In particular, we show how the   observed
  absorption lines depend on the value of the covering fraction
  parameter $\log a$ but also on the abundance of the ions.  So, a
  rare ion such as P$^{+4}$ can produce a shallower absorption line
  against the UV continuum than a common ion such as C$^{+3}$ against
  the broad-line region, even though the covering fraction is lower
  for the latter.   
\item  In \S\ref{size_scales} and \S\ref{partial_covering}, we examined
  the size scales of the continuum emission region (accretion disk),
  the broad-line region, the torus, and the outflow (estimated to be
  1--3 parsecs in Paper   I).  To explain the partial covering in the
  UV (established in Paper I), and the difference in covering fractions
  between the UV and long wavelengths, we suggest a model in which the
  outflow consists of clumps that are individually structured or very
  small  relative to the UV continuum emission region size scale, and
  are themselves   clustered on size scales comparable to the
  near-infrared continuum   emission region size scale.  
\item In \S\ref{selection}, we note SDSS~J0850$+$4451 was chosen for
  this experiment based on the previous observation of
  \ion{He}{1}*$\lambda  10830$ absorption, and that in principle,
  there may be objects which show strong UV absorption but no infrared
  absorption.  We discuss two examples where this seems to be the
  case, but note that the UV and infrared observations were not
  simultaneous (and in the case of PG~1254$+$47 were separated by 20
  years), so variability cannot be ruled out.
\end{enumerate}

\subsection{Conclusions and Future Prospects}\label{future}

In this paper, we show that broad absorption lines widely separated
in wavelength can be used to investigate the nature of partial
covering.  This experiment shows that we need not be limited to
knowledge about outflows along the radial light of sight to the
continuum emission region, but we can also learn about the angular 
distribution of the outflowing clumps.     

It would be interesting to investigate other objects using this
method.  The ideal experiment would involve an object known to have
\ion{P}{5} absorption, so that absorption from \ion{He}{1}*$\lambda
10830$  would also be predicted \citep{leighly11}.  This requires the
object to have a  redshift of less than $\sim 1.2$
(depending on the velocity of the outflow) so that
\ion{He}{1}*$\lambda 10830$ can be observed from the ground. The
redshift requirement means that UV spectrum would need to be observed
using {\it HST}.    The UV and near-infrared
observations should be contemporaneous, in order avoid uncertainty due
to variability.  Depending on the quasar luminosity, an near-infrared image
may be advisable to quantify the host galaxy contribution to the
1-micron continuum. 

\acknowledgements

KML acknowledges very useful conversations with Carolin Villforth and 
Hermine Landt about the host galaxy.   KML acknowledges useful
discussion with the current {\it SimBAL} group:  Hyunseop Joseph Choi, 
Collin Dabbieri, Amy Griffin, Francis MacInnis, Adam Marrs, and
Cassidy Wagner.  KML thanks OU undergraduate Collin McLeod for working
out the limit in \S\ref{quantifying}.  KML gratefully acknowledges
John Wisniewski's donation of APO time to the OU astronomy group, and
thanks him for taking the 2014 observation as part of the Advanced
Observatory Methods class.  Support for {\it SimBAL} development was   
provided by NSF Astronomy and Astrophysics Grant No.\ 1518382.
Support for program 13016  was provided by NASA through a grant from
the Space Telescope Science Institute, which is operated by the
Association of Universities for Research in Astronomy, Inc., under
NASA contract NAS 5-26555. {
  The {\it SimBAL} team acknowledges partial funding for the server
  ``Balthazar'' from the OU Research Council and the Homer L.\ Dodge
  Department of Physics and Astronomy.}  DT
acknowledges the Homer L.\ Dodge Department of Physics and Astronomy
of the University of Oklahoma for graciously hosting his sabbatical
visit in 2017.  ABL is
supported by NSF~DGE-1644869 and Chandra DD6-17080X.  ABL thanks the
LSSTC Data Science Fellowship Program; their time as a Fellow has
benefited this work. SCG thanks the Natural Science and Engineering
Research Council of Canada.

Based on     observations obtained at the {\it Gemini}
Observatory, which is operated     by the Association of Universities
for Research in Astronomy,     Inc., under a cooperative agreement
with the NSF on behalf of the     {\it Gemini} partnership: the National
Science Foundation  (United States), the National Research Council
(Canada), CONICYT (Chile),     the  Australian Research Council
(Australia),  Ministério da Ciência, Tecnologia e Inovação (Brazil)
and  Ministerio de Ciencia, Tecnología e Innovación Productiva
(Argentina).  The LBT is an international collaboration among
institutions in the  United States, Italy and Germany. LBT Corporation
partners are: The University of Arizona on behalf of the Arizona
university system; Istituto Nazionale di Astrofisica, Italy; LBT
Beteiligungsgesellschaft, Germany, representing the Max-Planck
Society, the Astrophysical Institute Potsdam, and Heidelberg
University; The Ohio State University, and The Research Corporation,
on behalf of The University of Notre Dame, University of Minnesota and
University of Virginia. This     work is based on     observations
obtained at the MDM Observatory,     operated by     Dartmouth
College, Columbia University, Ohio State     University,     Ohio
University, and the University of      Michigan.  TIFKAM was funded by
The Ohio State University, the MDM consortium, MIT, and NSF grant
AST-9605012. The HAWAII-1R array upgrade for TIFKAM was funded by NSF
Grant AST-0079523 to Dartmouth College.  Based  on
observations obtained with      the Apache  Point Observatory
3.5-meter telescope, which is owned   and operated by the
Astrophysical Research Consortium.  Based in  part on observations at
Kitt Peak National Observatory, National Optical Astronomy Observatory
(through time exchange with Ohio State University), which is operated
by the Association of Universities for Research in Astronomy (AURA)
under cooperative agreement with the National Science Foundation. The
authors are honored to be permitted to conduct astronomical research
on Iolkam Du'ag (Kitt Peak), a mountain with particular significance
to the Tohono O'odham.

\facility{Mayall (R-C CCD Spectrograph),
  Gemini:Gillett (GNIRS), Hiltner,ARC: 3.5m (DS), LBT (LUCI)}

\software{emcee \citep{emcee}, Cloudy \citep{ferland13}, Sherpa
  \citep{freeman01}, SimBAL \citep{leighly18}}

\appendix

\section{Absorption Line Variability}\label{variability}

\subsection{Observed Variability}\label{obs_var}

\citet{vivek14} presented the results of time variability studies of 
\ion{Mg}{2} and \ion{Al}{2} absorption lines in a sample of 22 LoBAL
quasars.  Their sample included SDSS~J0850$+$4451.   They obtained
several spectra at the IUCAA Girawali Observatory (IGO) in 2010 and
2011.  While there was no variability among the 2010 and 2011
observations, there was significant variability when compared with the
2002 SDSS spectrum: the equivalent width of the absorption line
increased from $12.6 \pm 0.4$\AA\/ to (weighted mean) value of $18.9
\pm 0.6$\AA\/.   

This result spurred us to obtain two additional spectra in April
2014 (APO), and April 2015 (KPNO) to measure the
\ion{He}{1}*$\lambda 3889$ and \ion{Mg}{2} absorption lines.  The
descriptions of these data sets are given in \S\ref{apoobs} and
\S\ref{kpnoobs}.  Combined with the MDM spectrum from 2011, a
digitized \citet{vivek14} spectrum from 2010, and the BOSS spectrum
from  January 2015, we can examine the absorption line variability
observed over 12.5 years (observed frame, 8.1 years rest frame).  We
focus on the \ion{Mg}{2} and \ion{He}{1}*$\lambda 3889$ lines, as they
are deepest, and model the two regions separately. With the goal of
quantifying the absorption variability, we fit all spectra containing
each line simultaneously using {\it Sherpa} \citep{freeman01}.  The
continuum model is similar to the one described in \S\ref{contmod} in
most cases.  The exception was the 2015 BOSS spectrum which displays
an unusual continuum shape at shortest wavelengths.  We modeled the
continuum of that spectrum with a 3rd order polynomial.  The KPNO
spectrum, taken three months after the BOSS spectrum, shows a normal
AGN continuum, and therefore, we suspect that the BOSS spectrum
suffered bad flux calibration.   As noted above, this should not be
the well-known  BOSS spectrograph differential refraction problem
\citep{margala16}, as that is now corrected for in the pipeline.  

For the \ion{Mg}{2} region, we tie the widths of the \ion{Mg}{2}
emission lines together between spectra, and model the absorption with
a single Gaussian opacity profile. Our first model left the central
wavelength and width of the opacity model independent among the epochs.
The reduced $\chi^2$ was 0.85 for 7138 degrees of freedom.  In order
to make the simplest comparison of opacity (the factor we are most
interested in), we try a second model with the position and widths
tied together. The $\chi^2$ was slightly worse (0.87 for 7146 degrees
of freedom), but an application of the F-test indicated only 17\%
chance that the difference was significant.  

For the \ion{He}{1}*$\lambda 3889$ region, we tie the widths of the
[\ion{Ne}{3}]$\lambda 3870$ emission lines together among the spectra,
and model the \ion{He}{1}* absorption line with a single Gaussian
opacity profile.  In this case, the fits are indistinguishable when
the absorption line widths and positions are free or tied together (in
both cases, $\chi^2_\nu=0.96$ for 7156 and 7164 degrees of freedom,
respectively).   

\begin{figure*}[!t]
\epsscale{0.7}
\begin{center}
\includegraphics[width=6.5truein]{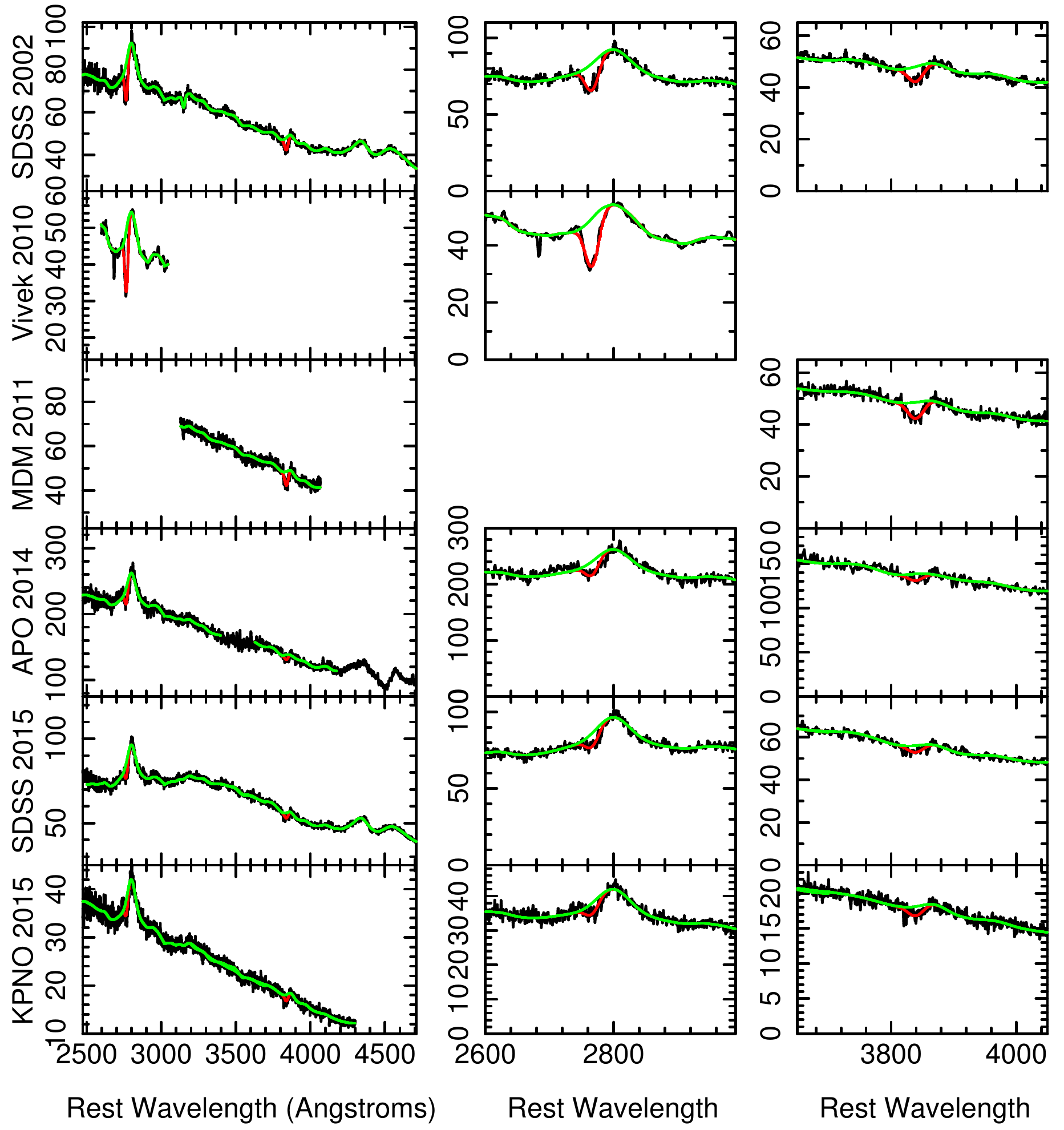}
\caption{Results of spectral fitting used to extract the opacities of
  the \ion{Mg}{2} and \ion{He}{1}*$\lambda 3889$ lines.  The spectra
  are plotted as a function of rest wavelength and flux density
  $F_\lambda\, (10^{-17}\rm \, erg\, s^{-1}\, cm^{-2}$\AA\/$^{-1})$.
 The spectrum
  labeled ``Vivek 2010'' was digitized from figures shown in
  \citet{vivek14}.  Note   that the ``MDM 2011'', ``APO 2014'', and
  ``KPNO 2015'' spectra are not accurately fluxed.  The 
  spectra were fit simultaneously in the \ion{Mg}{2} and \ion{He}{1}*
  regions separately.  The best fitting model is overlaid in red, 
  while the model without the line absorption is overlaid in green.
  The variability of the \ion{Mg}{2} and \ion{He}{1}*$\lambda 3889$
  lines is clearly seen.      \label{fig14}}    
\end{center}
\end{figure*}

\begin{figure*}[!t]
\epsscale{0.7}
\begin{center}
\includegraphics[width=6.5truein]{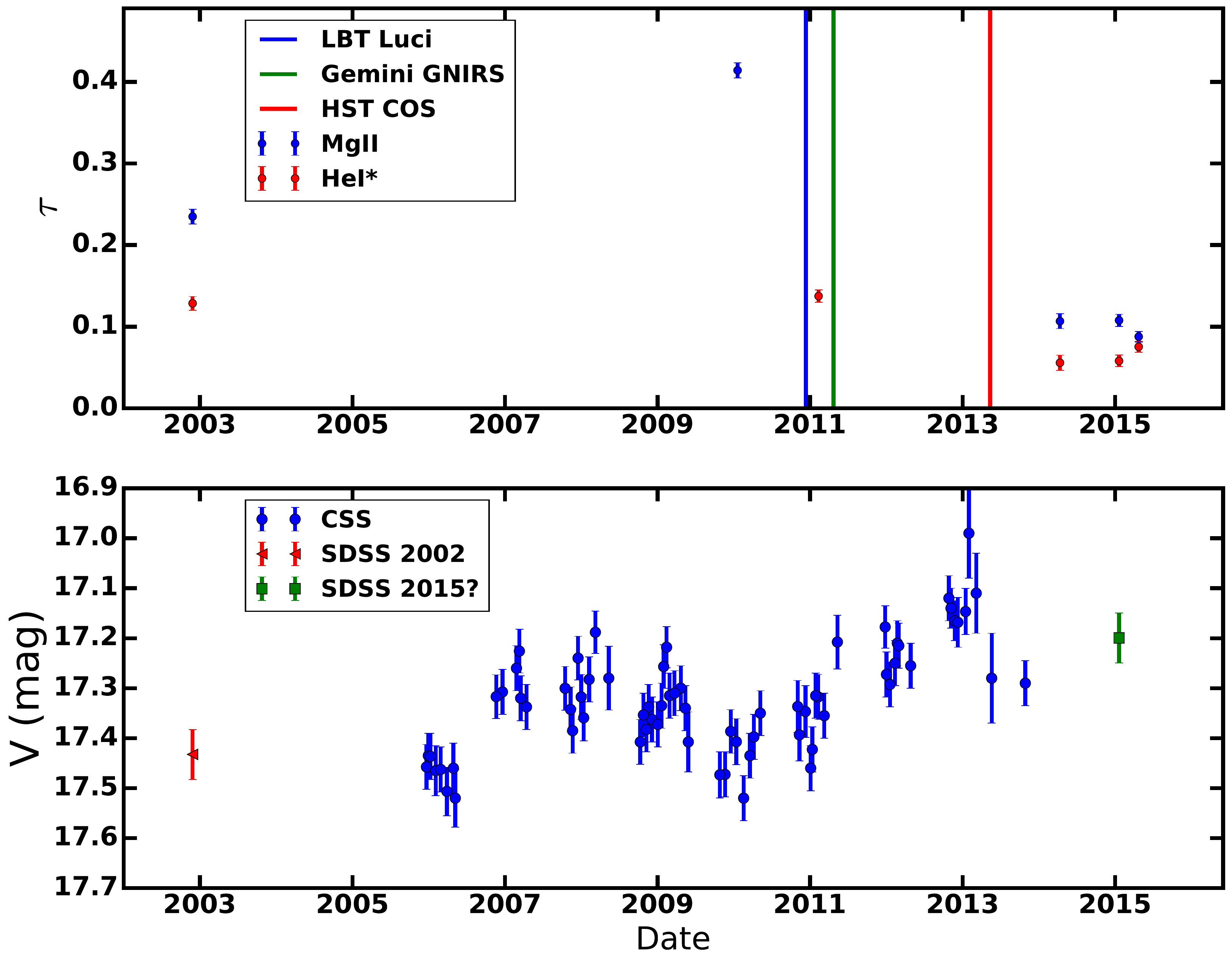}
\caption{{\it Upper panel:} The evolution of the \ion{Mg}{2} and
  \ion{He}{1}*$\lambda   3889$ optical depths.  The absorption lines
  were modeled using a   Gaussian opacity profile, and the values
  shown are the   normalization. Since the widths and  positions were
  tied together   for     each species in the spectral fitting, the
  normalization is    proportional to the apparent column density /
  optical depth.  The   \ion{He}{1}*   optical depth varied by a
  factor of two. The   \ion{Mg}{2} line varied   by a larger factor
  among the five   observations than the   \ion{He}{1}* line (by a
  factor of 4), but   this result is not   conclusive given that we    do not  have a
  measurement of   \ion{He}{1}*$\lambda 3889$   absorption when   the
  \ion{Mg}{2} line was 
  strongest.    {\it Lower Panel:}  The v-magnitude variability
  observed using SDSS and the Catalina Sky Survey.  The trends
  suggest that the absorption lines are shallower when the object is
  brighter, and were therefore shallow when the {\it HST} observation
  was performed. \label{fig15}}     
\end{center}
\end{figure*}

The resulting fits are shown in Fig.~\ref{fig14}.   The resulting apparent
optical depths for \ion{Mg}{2} and \ion{He}{1}*$\lambda 3889$  are
presented in Fig.~\ref{fig15}.  Also marked are the dates of the near-IR
observations (using LBT LUCI and {\it Gemini}  GNIRS) and UV
observation ({\it HST} COS).  This plot clearly shows that while the
MDM observation of \ion{He}{1}*$\lambda 3889$ was contemporaneous with
the near-infrared observations using LBT and {\it Gemini}, there are
unfortunately no ground-based observations within one year of the {\it
  HST} COS observation.

After an initial
increase in opacity by a factor of 1.8, the \ion{Mg}{2} absorption
line decreased to a value slightly less than half the value observed
in 2002.  Thus, the apparent opacity of the \ion{Mg}{2} line was
observed to vary by a factor of more than four. The
\ion{He}{1}*$\lambda 3889$ displayed less variability (factor of 2).
However, we do not have a measurement of the \ion{He}{1}*$\lambda
3889$ line in 2010 when the \ion{Mg}{2} was so strong, and when we
compare the only the results from observations  where both lines were 
observed, the relative variability is similar.  We conclude that the
degree of variability is the same for both lines, or larger for
\ion{Mg}{2}.    

We note that the spectra that we took are not accurately fluxed, so we
cannot compare the absorption variability with the continuum
variability directly. To compare the continuum variability with the
opacity, we downloaded the Catalina Sky Survey
(CSS)\footnote{http://nesssi.cacr.caltech.edu/DataRelease/} data for this 
object.  The data, shown in Fig.~\ref{fig15}, consists of the
error-weighted average per night. \citet{vivek14} also analyze these
data. It is noteworthy that the magnitudes plotted in their Appendix B
are about 0.5 magnitude fainter than the data we downloaded.  The origin of this
difference is not understood.   We speculate that the difference
originates in an updated calibration of the CSS data to the Johnson
{\it V}
band.  We also plot the V-band magnitude from the SDSS observation,
obtained using color correction 
terms from
\citet{jester05}\footnote{http://www.sdss3.org/dr8/algorithms/sdssUBVRITransform.php}. The
uncertainty was taken to be the color transformation RMS 
residual.  

The photometry shows that SDSS~J0850$+$4451 is very modestly variable;
the standard deviation on the photometry points is 0.11 mag,
corresponding to a factor of 10\%.  However, quasars are known to be
bluer when brighter, so a larger degree of variability may be present
in the photoionizing continuum.  

Comparing the photometry with the opacity, we see that in general the
line opacities are inversely related to the flux.  More specifically,
the V-band flux was relatively low when the \ion{Mg}{2} opacity was
highest, in 2010, and since then, the V-band magnitude has tended to
increase, and the line opacities to decrease.  This suggests that at
the time of the {\it HST} COS observation, the \ion{Mg}{2} and
\ion{He}{1}*$\lambda 3889$ opacities were no higher than observed
during the 2002 SDSS observations.  This implies that, if anything,
the \ion{Mg}{2} and \ion{He}{1}* lines were shallower the ones in the
spectra that we analyze in this paper.  Therefore the discrepancy
between the extrapolated spectral synthesis models and the observed
troughs shown in Fig.~\ref{fig13} would be potentially larger. 

In summary, the opacity and flux trends suggest that the differences
in partial covering that we measure are not an artifact of
variability.  However, because we do not have contemporaneous optical
and infrared spectra at the time of the {\it HST} spectrum, we cannot
rule this explanation out absolutely.  

\subsection{Predicted Variability Patterns}\label{predicted}

{ There has been an explosion in variability studies of broad
  absorption line quasars in recent years  
\citep[e.g.,][]{capellupo11,capellupo12,capellupo13,filizak13,vivek14,mcgraw15,mcgraw18}. 
Such studies can be used to constrain the distance of the absorber
from the central engine, and thereby constrain feedback metrics; for 
an example of such an analysis and additional references, see
\citet{mcgraw18}.  The distance can be constrained if there are changes
in covering fraction by assuming that Keplerian motion carries 
the absorber across the continuum source; \citet{mcgraw18} found
typical distances $r\lesssim 1$--10 pc in their sample of \ion{P}{5}
quasars.  If there are changes in ionization parameter, the distance
can be constrained by comparing the observed variability time scale
with the recombination time scale; \citet{mcgraw18} found a typical
range of $r \lesssim 100$--1000 pc.   

These variability studies have taken a general, qualitative, and
order-of-magnitude  approach to the estimation of the origin of BAL
variability.  {\it SimBAL}, in contrast, can place qualitative
constraints on the origin of variability.}
We use the {\it SimBAL} simulation grids to predict the kinds of
variability that might be observed in \ion{Mg}{2},
\ion{He}{1}*$\lambda 3889$, and \ion{He}{1}*$\lambda 10830$ absorption
lines as a function of a change in ionization parameter $\log U$
(equivalent to a change in ionizing flux for a fixed density), a
change in the column density $\log N_H$, or a change in the covering
fraction, parameterized by $\log a$, that might be equivalent to
transverse motion across the source.   

\begin{figure*}[!t]
\epsscale{1.0}
\begin{center}
\includegraphics[width=7.0 truein]{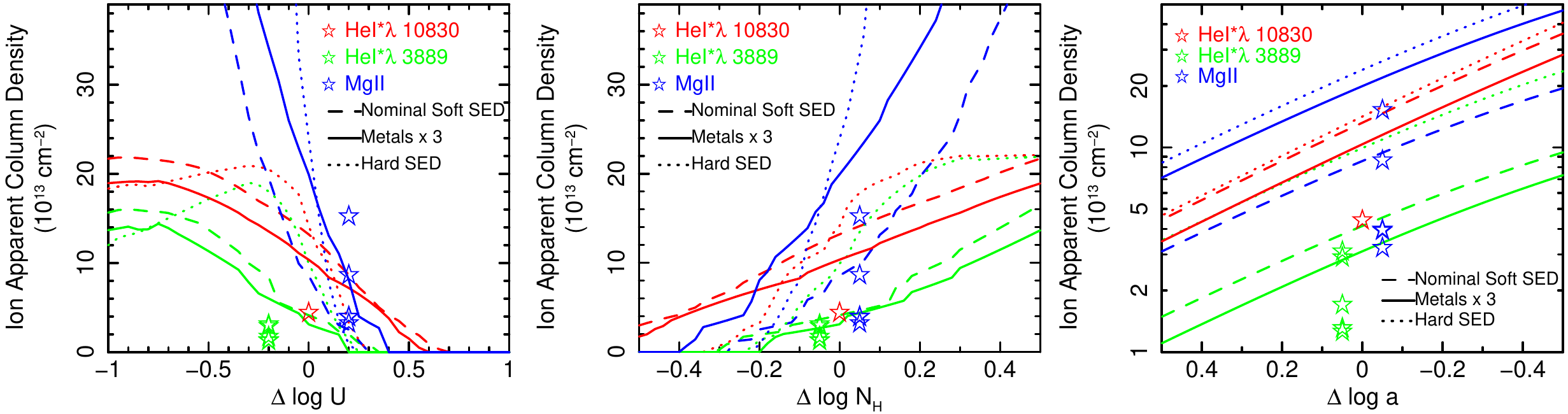}
\caption{The predicted variation of the apparent column densities of
  \ion{Mg}{2}, \ion{He}{1}*$\lambda 3889$, and \ion{He}{1}*$\lambda
  10830$ as a function of one varied parameter away from the
  best-fitting solution obtained from the UV spectrum.  Also shown as
  open stars are the apparent column densities measured directly from
  the spectrum for \ion{He}{1}*$\lambda 10830$, and from the model
  fits, for \ion{Mg}{2} and \ion{He}{1}*$\lambda 3889$.  These are 
  offset along the x-axis for clarity.   \label{fig16}}   
\end{center}
\end{figure*}

We chose the 11-bin models from Paper I for the second continuum for
the nominal soft SED, the hard SED, and the enhanced metallicity
cases. We varied one parameter at a time by the same amount in each of 
the eleven velocity bins.   At each deviation interval
away from the best fit, we created a synthetic spectrum, and then
estimated the apparent column density of \ion{Mg}{2},
\ion{He}{1}*$\lambda 3889$, and \ion{He}{1}*$\lambda 10830$,
by integrating over the line profile, using Eq.\ 9 in \citet{ss91}.
The results are shown in Fig.~\ref{fig16}. 

An increase in $\log U$ results in a decrease in the opacity for all
lines, but especially for \ion{Mg}{2}, because the regions in the slab
where Mg$^+$ and excited-state neutral helium lie are matter bounded.
So an increase in the ionization effectively pushes the regions where
these ions would be present beyond the back (i.e., opposite of the
illuminated face) of the cloud.
Increasing the thickness of the clouds causes the opacity for all the
lines to increase, but, again, especially for \ion{Mg}{2}.  The
interpretation is the same as above.  The column densities for the
best fitting model is matter-bounded for these ions, so increasing the
total column density increases the column density of these ions.  
The powerlaw covering fraction model has the property that all
opacities increase geometrically.  There are slight differences in
slope but those are not significant.
Finally, it is important to note that in all cases, the
\ion{He}{1}*$\lambda 10830$ line varies in concert with the
\ion{He}{1}*$\lambda 3889$ line. 

The data are insufficient to determine which of these
scenarios (variable ionization parameter, variable column density, or
variable covering fraction) is true.  But the pattern of variability
observed seems to weakly support variation in ionization parameter,
since, as discussed in  \S\ref{obs_var}, there is a possible 
anti-correlation observed between the opacity and continuum
flux, and a greater variability in \ion{Mg}{2} compared with
\ion{He}{1}*$\lambda 3889$. \citet{vivek14} suggested that ionization
variability might have   occurred in SDSS~J0850+4451.  

\section{The Host Galaxy Contribution to the Near-Infrared Continuum\label{host}}

As noted above, the observed \ion{He}{1}*$\lambda 10830$ absorption
line is much shallower than predicted by the model of the far-UV
spectrum.   This could mean that the covering fraction to the near-infrared
continuum is lower than it is to the ultraviolet continuum.  But
another possibility is that the host galaxy is contributing a
significant amount of the observed near-infrared continuum, making the
\ion{He}{1}*$\lambda 10830$ line appear shallower than it really is.
In this section, we estimate the plausible contribution of the host
galaxy to the near-infrared continuum to address this second
possibility. We found that the contribution of the host
galaxy is negligible.  

\subsection{SED Fitting}\label{sed_fitting}

\citet{lucy14} addressed the question of the host galaxy
contribution to the near-infrared continuum of FBQS~J1151$+$3822 via SED
modeling of broadband photometry.  They fit data from SDSS, 2MASS, and
WISE with a power law, elliptical galaxy template, and two blackbodies
to model both the warm ($T \approx 1200\ {\rm K}$) torus and cooler
($T \approx 300\ {\rm K}$) dust.

\begin{figure*}[!t]
\epsscale{1.0}
\begin{center}
\includegraphics[width=4.5truein]{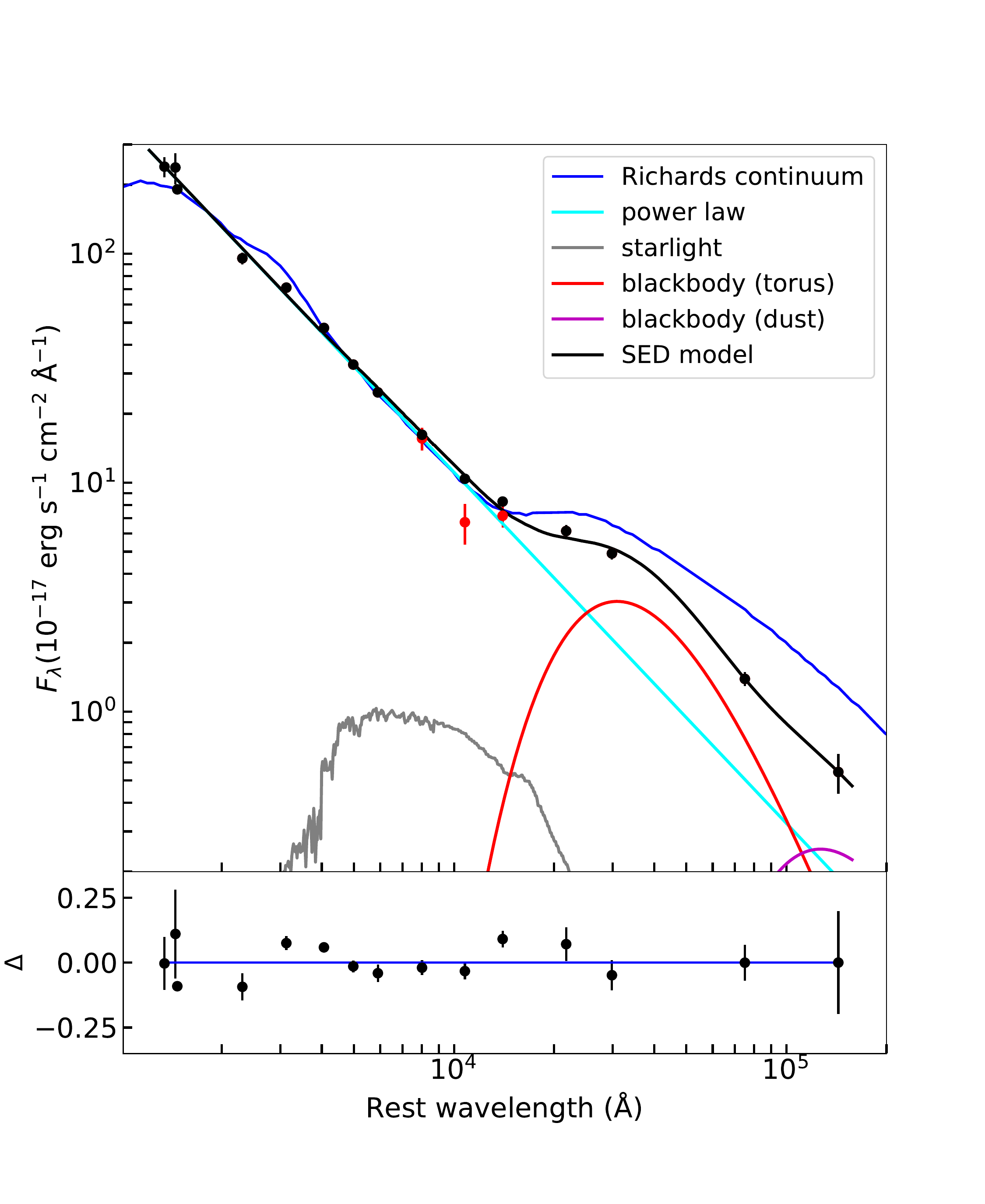}
\caption{Spectral energy distribution modeling to {\it SDSS}, {\it WISE},
    and observed JHK photometry. The {\it SDSS} photometry has been
    corrected for the contribution of emission lines \citep{lucy14}.
    The blue line shows the \citet{richards06} composite quasar
    spectrum fitted shortward of several microns, and extrapolated to
    long wavelengths, illustrating the lack of strong near-infrared
    emission from the torus compared with average quasars.  The
    contribution of the host galaxy to the continuum under
    negligible, and is consistent with the imaging analysis presented
    in Fig.~\ref{fig18}.    \label{sed_fitting_fig}} 
\end{center}
\end{figure*}

We follow that procedure here, but we use our new $JHK_s$ photometry,
add NUV photometry from {\it GALEX}, and include two ultraviolet
continuum points derived from the {\it HST} COS spectra (Paper I).
The SED is shown as black filled points in the upper panel of
Fig.~\ref{sed_fitting_fig}, along with the original 2MASS values in
red; these were not used in the SED fitting.  We first obtained an
estimate of the reddening in SDSS~J0850$+$4451 by building a 
simple model consisting only of the \citet{richards06} continuum.
This model has only two parameters: a multiplicative scaling of the
Richards continuum and the reddening intrinsic to the quasar.  This
model was fit to all points excluding those from {\it WISE}, and
yielded $E(B - V) = 0.03 \pm 0.03$.  The WISE points were excluded
because, compared to the Richards continuum, the torus in SDSS
J0850+4451 is considerably less bright, suggesting that this object is
deficient in hot dust \citep{hao10,hao11,lyu17}.

Also shown in the top panel of Fig.~\ref{sed_fitting_fig} is the
decomposition of the SED with components indicated in the figure
legend.  This solution has reddening $E(B - V) = 0.0$, meaning that
the solution hit the lower bound of allowed values.  The galaxy
contribution at $\lambda = 10830\ {\rm \AA}$ is $0.09 \pm 0.03$\%. The 
power-law index is $-1.53$, which is not unusual for quasars
\citep{krawczyk15}.  The residuals to the fit are shown in the lower
panel of the figure, where these are defined as $\Delta = $ (data $-$
model) / model.  The model fit is not particularly good, with reduced
$\chi^2 = 50$, but { much of this is likely to be due to the
  simplicity of the model; variability may play a role as well (see
  Fig.~\ref{fig15}).}  These 
results show that the contribution of the host galaxy emission to the
continuum under the \ion{He}{1}*$\lambda 10830$ absorption line is
negligible.   

\subsection{Image Analysis of the SDSS~J0850+4451 Host
  Galaxy}\label{image}

Coincidentally, an {\it HST} program\footnote{PI: Canalizo, ``The
  Nature of Low-Ionization BAL QSOs'', program
  number 11557} made WFC3 observations in the near-infrared of
SDSS~J0850+4451; the information about this observation is found in
Table~\ref{observations}.   We used these data to  provide a
complementary estimate of the contribution of the host galaxy to the
infrared continuum emission and the photometry.

{\it HST} images of quasars are dominated by the quasar, so it is
necessary to use the image point spread function (PSF) to isolate the 
emission of the host galaxy.  A PSF made
from observations of a star works better than a simulated one 
\citep[e.g.,][]{canalizo07}, and fortunately, a PSF star 
observation was made in association with this program.  The star was
the white dwarf GRW+70D5824, an object chosen to have $B-V$ color
similar to the average color of the sample.  It was observed with
three different exposure times in each filter in order to obtain a
broad dynamic range.  The star was placed on the same part of the
detector as the science observations. {\it HST} has variable focus 
(so-called ``breathing'').  Fortunately,  the star and the quasar were
observed during times when the focus was nearly the same.  The PSF is
undersampled by the detector, but that is accounted for by dithering
the telescope during the observation, and combining the individual
images  using the
MultiDrizzle\footnote{http://www.stsci.edu/hst/HST\_overview/documents/multidrizzle/multidrizzle\_cover.html}  
software, which also corrects for distortion.

Cosmic-ray rejection can remove photons from the core of the image
\citep{riess11}.  To determine whether this problem is present in the
multidrizzled images, we performed photometric analysis on the
flat-fielded images of the star.   We obtained a mean and standard
deviation of the net count rates for the star of $1.500 \pm 0.009
\times 10^{-15} \rm \, erg\, s^{-1}\, cm^{-2} \,$\AA\/$^{-1}$. 
The aperture correction for the F125W filter from 2 arc seconds to
infinity is 0.029 \citep{kalirai09}, resulting in a final estimated
flux density of $1.539 \pm 0.009 \times 10^{-15} \rm \, erg\, s^{-1}\,
cm^{-2} \,$\AA\/$^{-1}$.  The 2MASS $m_J$ is 13.248, corresponding to a
flux density of $1.57 \times 10^{-15} \rm erg\, s^{-1}\,
cm^{-2} \,$\AA\/$^{-1}$, less than 2\% from the measured value.  In
contrast, a similar extraction of the multidrizzled images (i.e.,
after cosmic-ray correction) yields an flux estimate of $1.392 \pm
0.006  \times 10^{-15} \rm \, erg\, s^{-1}\, cm^{-2}\,$\AA\/$^{-1}$,
11\% below the 2MASS value, indicating that indeed, cosmic-ray
rejection had removed source photons from the core of the image.  

We performed roughly the same analysis on the image of
SDSS~J0850+4451, but since we wished to compare with the MDM
photometry, we first blurred the image by convolution with a Gaussian 
to correspond to the 1.25 arc second seeing.  The resulting
measurement of the flux density was $8.75 \pm 0.18 \times 10^{-17}\,
\rm erg\, s^{-1}\, cm^{-2}\, $\AA\/$^{-1}$.  In contrast, the
multidrizzled image yields a measurement of the flux density of $6.90
\pm 0.12 \times 10^{-17}\, \rm erg\, s^{-1}\, cm^{-2}\, $\AA\/$^{-1}$,
about 20\% lower.   Nevertheless, this difference should only
influence the core of the quasar PSF, and not the extended host galaxy
emission.   

The image analysis was done using {\tt
  Sherpa}\footnote{http://cxc.harvard.edu/sherpa4.9/}
\citep{freeman01}.  This software requires three input images: the
target image, an error image, and a point spread function image. The
point spread image was constructed using the five observations of the
PSF star: three with exposures of 5.865 seconds, one with exposure of
11.729, and one with exposure 23.458 seconds.   We examined the images
and the data quality arrays and found no evidence 
for saturation.   Therefore, an exposure-weighted average of these
images was used for the PSF.  We smoothed the PSF image using the
method outlined by \citet{canalizo07}. To prepare the error images, we
followed the procedure outlined in {\it The DrizzlePac
  Handbook}\footnote{Page 89,
  http://documents.stsci.edu/hst/HST\_overview/documents/DrizzlePac/drizzlepac.pdf}. 

We modeled a circular region within 71 pixels (9.23 arc seconds),
excluding three faint sources near the edges of the region.
Initially, we chose a Gaussian model and a constant background.  This
model did not provide a good fit to the image, with a reduced
$\chi^2_\nu= 4.05$.  We next tried a model consisting of a Gaussian
profile, a Sersic model, and a constant.  The resulting reduced
$\chi^2_{\nu}$ was 1.34, a dramatic improvement in fit over the
Gaussian plus constant model that indicates clear evidence for the
detection of the host galaxy.   

\begin{figure*}[!t]
\epsscale{1.0}
\begin{center}
\includegraphics[width=6.5truein]{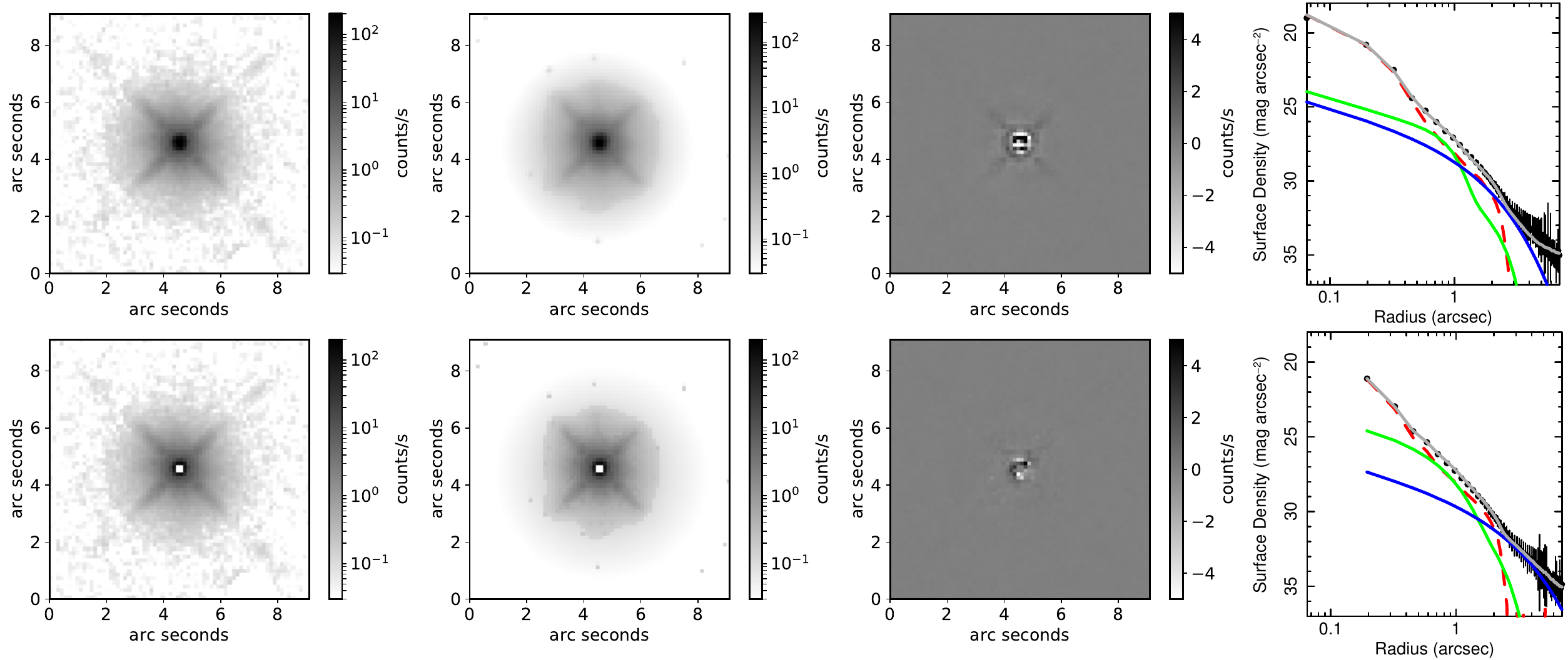}
\caption{The results of analysis of the {\it HST} IR F125W image of
  SDSS~J0850+4451. The model consisted of a two-dimensional Gaussian
  and two Sersic models.   From left to right: the data, the fitted
  model   convolved   with the PSF, the difference between the data
  image and the model,   and the radial profile from the data and the
  model.  The top panel   shows the 
  results from analysis of the whole image, and the bottom panel
  shows the results with the central four pixels removed from the
  fitting.  The host galaxy in SDSS~J0850+4451 is clearly
  detected.  \label{fig17}} 
\end{center}
\end{figure*}

Statistically, the Gaussian plus Sersic model provided a good
fit. However, examination of the radial profile showed positive
residuals between 2 and 5 arc seconds from the center, suggesting that
there is an additional larger-scale but fainter component.  We model this
component with an additional Sersic profile, but because the error
bars are large, we fixed the index to one (appropriate for a disk
galaxy) and the ellipticity to zero.  The $\chi^2$ decreased a small
amount, to $\chi^2_\nu = 1.27$, indicating that this component is not
statistically necessary.  The fit parameters are given
in Table \ref{image_fit_results}, and the image, best fitting model,
residuals, and profile are shown in Fig.~\ref{fig17}.

Noting the problems with cosmic-ray rejection in the core of the PSF
discussed above, we performed the same analysis but ignored the four
central pixels.  The reduced $\chi^2_\nu$ for this model was 0.58.
The same residuals were observed in the radial profile, so we added
another Sersic model as above, producing a reduced $\chi^2_\nu$ of
0.51. The model parameters are given in Table \ref{image_fit_results},
and fitting results are shown in Fig.~\ref{fig17}.

\begin{deluxetable}{lCC}
\tablewidth{0pt}
\tabletypesize{\small}
\tablecaption{Image Fitting Results}
\tablehead{
 \colhead{Parameter} & \colhead{Full Image} & \colhead{Central 4
   Pixels Removed}}
\startdata
Gaussian FWHM (pixels)\tablenotemark{a} & 1.36^{+0.006}_{-0.0.05} &
0.73^{+0.12}_{-0.002} \\
Gaussian Amplitude (counts s$^{-1}$)\tablenotemark{b} & 1332\pm 9 & 4250^{+17}_{-1100} \\
1 Sersic $R_0$ (pixels)\tablenotemark{a} & 5.0^{+0.06}_{-0.04} & 4.4^{+0.31}_{-0.04} \\
1 Sersic $R_0$ (kpc)\tablenotemark{c} & 4.1^{+0.05}_{-0.03} &
3.6^{+0.25}_{-0.03} \\
1 Sersic Eccentricity & 0.16^{+0.009}_{-0.008} & 0.12^{+0.02}_{-0.009} \\
1 Sersic Theta (radians) & 0.91\pm 0.02 & 1.0^{+0.04}_{-0.12} \\
1 Sersic Amplitude (counts s$^{-1}$)\tablenotemark{b} & 2.0^{+0.05}_{-0.07} &
2.5^{+0.04}_{-0.17} \\
1 Sersic $n$ & 0.21 \pm 0.02 & 0.53^{+0.01}_{-0.13} \\
2 Sersic $R_0$ (pixel)\tablenotemark{a} & 9.8^{+0.4}_{-0.5} & 17^{+2}_{-3} \\
2 Sersic $R_0$ (kpc)\tablenotemark{c} & 7.9^{+0.34}_{-0.40} &
14^{+1.6}_{-2.6} \\
2 Sersic Amplitude (counts s$^{-1}$)\tablenotemark{b} & 0.3^{+0.04}_{-0.03} &
0.07^{+0.06}_{-0.012} \\
Constant (counts s)$^{-1}$\tablenotemark{b} & 0.010^{+0.0007}_{-0.0006} & 0.006 \pm
0.001 \\
\enddata
\tablenotetext{a}{{\it HST} WFC3 IR channel has a plate scale of 0.13
  arc seconds per pixel.}
\tablenotetext{b}{{\it HST} WFC3 IR filter F125W has an inverse
  sensitivity of $2.2483 \times 10^{-20}\rm \, ergs\,
  cm^{-2}$\AA\/$\rm^{-1}\, electron^{-1}$, a pivot wavelength of
  $1.2486\rm \, \mu m$, and a RMS bandwidth of $0.0866 \rm \, \mu m$.}
\tablenotetext{c}{For $H_0=72,\rm km/s/Mpc$, $\Omega_M=0.27$,
    $\Omega_\Lambda=0.73$, and $z=0.5414$, the scale is $6.255\rm \,
    kpc/arc\, second$. }
\label{image_fit_results}
\end{deluxetable}

{\it Sherpa} allows us to save the unconvolved model component images.
The count rates from each component were obtained by summing 
over these images, and converted to flux densities using the inverse
sensitivity.  These values are given in Table \ref{image_fit_results}.
We note that the value obtained from the fit with the four central
pixels excluded ($8.5 \times 10^{-17}\rm \, erg\, s^{-1}\, cm^{-2}\,
$\AA\/$^{-1}$) is comparable to the value obtained using aperture
photometry on the flat-fielded images, as described above ($8.75
\times 10^{-17}\rm \, erg\, s^{-1}\, cm^{-2}\, $\AA\/$^{-1}$),
indicating that our model accounts for the  emission in the
image. Also, the flux densities from the galaxy component are
essentially the same whether we ignore or keep the four central
pixels, indicating that the issues stemming from the imperfect
cosmic-ray rejection did not affect the estimate of the host galaxy
flux.   

We convolved the model component images with the seeing and
applied an aperture to estimate the flux from each component
contributing to our photometry and spectroscopy.  We did this for two
cases.  The first case pertains to the MDM JHK photometry described in
\S\ref{mdmobs}. The seeing during those observations was about 1.25
arc seconds (about 
4 pixels) and the extraction aperture was 3.3 arc seconds in radius.
The second case pertained to the {\it Gemini} GNIRS spectroscopy.  The slit
width was 0.45 arc seconds, and the flux was extracted in an aperture
6 pixels in radius, corresponding to a total length of 1.8 arc seconds.
The orientation of the slit varied among the different spectroscopic
observations, so we chose a range of 
orientation angles to bracket the minimum and maximum fluxes.  The
seeing during the several {\it Gemini} observations was not available.  Our
program observing conditions requirement was  ``85\% to poor'',
corresponding to 0.85--1.55 arc seconds.  We assumed a seeing value of
0.91 arc seconds, corresponding to 3 pixels. The results are given in
Table \ref{image_model_flux_densities}.

\begin{deluxetable}{lccc}
\tablewidth{0pt}
\tabletypesize{\scriptsize}
\tablecaption{Image Deconvolution Results\label{image_decon}}
\tablehead{
 \colhead{Component} & \colhead{Intrinsic} & \colhead{MDM Aperture Photometry} &
 \colhead{{\it Gemini} GNIRS Spectroscopy} \\
& \colhead{($10^{-17}\rm \, erg\, s^{-1}\, cm^{-2}\,$\AA\/$^{-1}$)}
& \colhead{($10^{-17}\rm \, erg\, s^{-1}\, cm^{-2}\,$\AA\/$^{-1}$)}
& \colhead{($10^{-17}\rm \, erg\, s^{-1}\, cm^{-2}\,$\AA\/$^{-1}$)}}
\startdata
\multicolumn{4}{c}{Full Image} \\
Gaussian & 5.6 & 5.6 & 4.1--4.5 \\
1 Sersic & 0.66 & 0.66 & 0.20--0.22 \\
2 Sersic & 0.77 & 0.72 & 0.09 \\
Sersic+Sersic & 1.4 & 1.4 & 0.30--0.31 \\
Total & 7.1 & 7.0 & 4.4--4.8 \\
\hline 
\multicolumn{4}{c}{Central 4 Pixels Ignored} \\
Gaussian & 7.0 & 7.0 & 5.6--5.9 \\
1 Sersic & 0.90 & 0.90 & 0.29--0.30 \\
2 Sersic & 0.57 & 0.41 & 0.03 \\
Sersic+Sersic & 1.5 & 1.3 & 0.32--0.33 \\
Total & 8.5 & 8.3 & 5.9--6.2 \\
\enddata
\tablecomments{The flux densities were estimated from the unconvolved
  model images using the {\it HST} WFC3 IR filter F125W inverse
  sensitivity of $2.2483 \times 10^{-20}\rm \,  ergs\,
  cm^{-2}$\AA\/$\rm ^{-1}\, electron^{-1}$.  The pivot wavelength
  (observed frame) is $1.2486\rm \, \mu m$.  }
\label{image_model_flux_densities}
\end{deluxetable}

\begin{figure*}[!t]
\epsscale{1.0}
\begin{center}
\includegraphics[width=5.5truein]{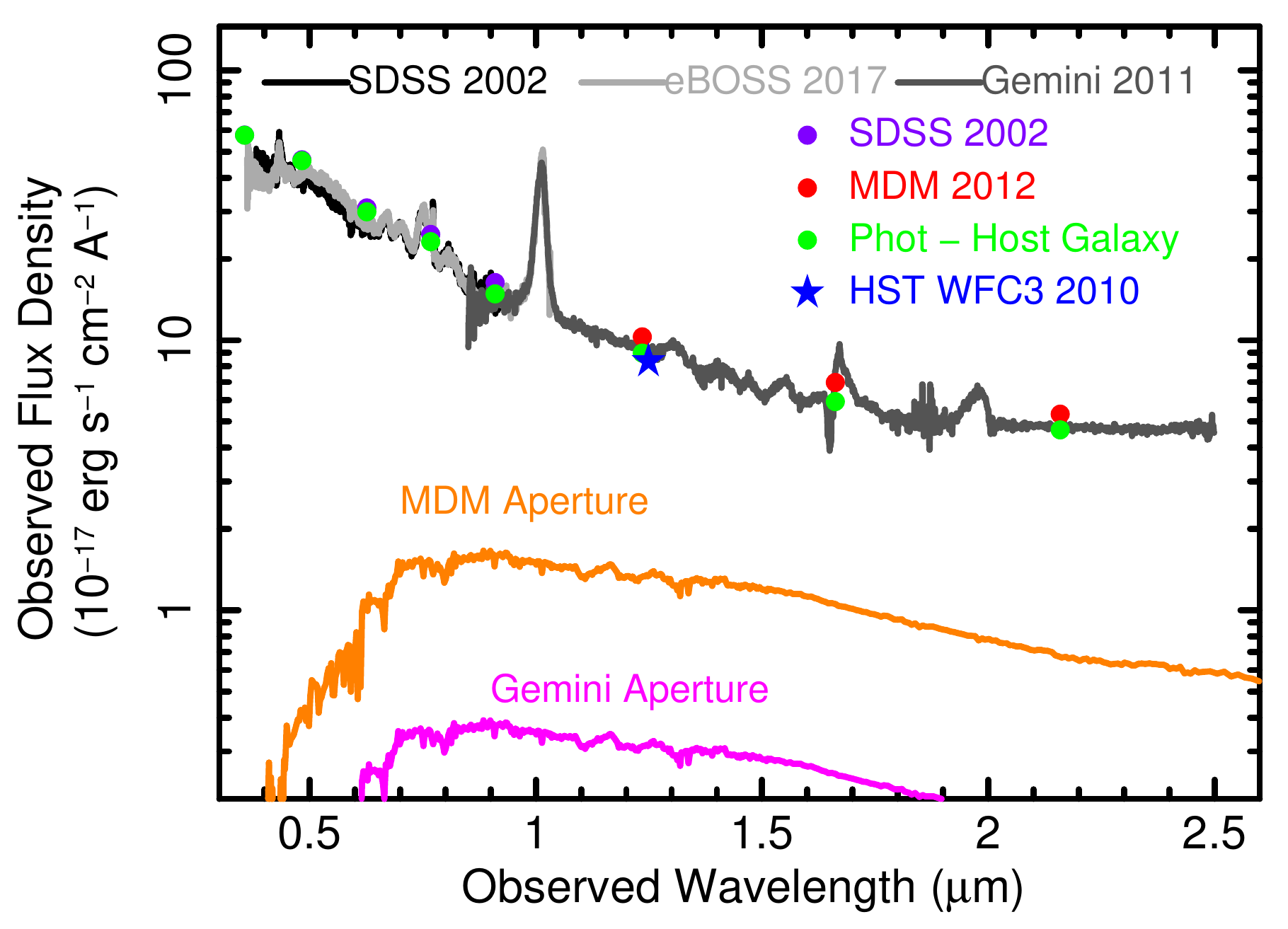}
\caption{The inferred contribution of the host galaxy to the MDM
  photometry and the {\it Gemini} spectroscopy.  The eBOSS spectrum was
  normalized to the SDSS spectrum, and then the {\it Gemini} spectrum was
  normalized to the result, taking advantage of the bright H$\alpha$
  line which is present in both the eBOSS and {\it Gemini} spectra.  The
  host galaxy is illustrated with a 5 Gyr-old template 
  \citep{polletta07}, redshifted to the observed frame and reddened
  for Milky Way attenuation,
  and then scaled with the aperture fluxes listed in Table
  \ref{image_model_flux_densities}.  We   conclude that the host
  galaxy contributes negligibly to the   continuum in {\it Gemini}  spectrum
  in the region of the   \ion{He}{1}*$\lambda 10830$ absorption
  line.  \label{fig18}} 
\end{center}
\end{figure*}

The results are displayed in Fig.~\ref{fig18}, which shows our
spectroscopy and photometry in the observed frame.  The 2017 eBOSS
spectrum was scaled to the SDSS 2002 spectrum by multiplying by a
factor of 0.91, and the {\it Gemini} spectrum was scaled to the result via
the overlapping H$\alpha$ line.  The SDSS photometry and near-IR
photometry from MDM were overlaid.  A 5 Gyr-old elliptical galaxy
template \citep{polletta07} was shifted to the observed frame and
reddened corresponding to Milky Way extinction, and then scaled to the
``Sersic+Sersic'' flux values listed in Table
\ref{image_model_flux_densities}.  This graph clearly shows that the
contribution of the host galaxy to the continuum under the
\ion{He}{1}*$\lambda 10830$ line is negligible.    

\subsection{How Does the Host Galaxy in SDSS~J0850+4451 Compare with
  Other Quasar Host Galaxies?}\label{host_comparison}

As a final check of our analysis, we briefly compare the host galaxy
properties with those from other low-redshift quasars.  For a fair
comparison, we recall the estimated black hole mass of $1.6 \times
10^9\rm \, M_\odot$ from Paper I.

\citet{bentz2009b,bentz2009a} analyze {\it HST} images from 
reverberation-mapped AGN and quasars.  All of these objects have
smaller black hole masses than estimated for SDSS~J0850+4451 ($\log
M_{bh}=9.2$), but several objects are close, including
PG~0804$+$761, PG 1226$+$023, PG~1426$+$015, and PG~1700$+$518, which
have log black hole masses of 8.84, 8.95, 9.11, and 8.89,
respectively.  We compared the derived properties of
SDSS~J0850$+$4451 with those obtained for these four galaxies.  The
Sersic radial scale factor, between
$\sim 4$ and $\sim 11 \rm \, kpc$, is similar to the four
comparison objects, which have scale factors between 3.3 to 12 kpc.  The
best fitting Sersic index is very low for SDSS~J0850$+$4511, between
0.2 and 0.5 for the smaller component, and fixed to 1 for the larger
component.  This may not be physical, noting that the inner region is
not well constrained due to the PSF.  The indices for the comparison
sample range from 1.0 to 5.6.  Integrating over the template scaled to the
total galaxy model flux and shifted into the rest frame yields a total
log luminosity of $10.9\rm \, [L_\odot]$, which is again similar to the
values of the comparison sample, which range from 10.6 to $11.2 \rm \,
[L_\odot]$ \citep{bentz2009a}.  

\citet[][their Figure 1]{landt11} present a graph showing
the enclosed luminosity density at 5100\AA\/ rest frame from a sample
of galaxies as a function of the extraction aperture.  At the redshift
of SDSS~J0850+4451, the {\it Gemini} GNIRS aperture encloses $31.7\rm \,
kpc^2$. The log luminosity density at 5100\AA\/ obtained from the scaled
5-Gyr-old elliptical template was $39.8\rm \, [erg\,
  s^{-1}$\AA\/$^{-1}]$.  This lies approximately 0.35 dex lower than
the regression line in \citet{landt11} Figure 1, corresponding to a
factor of $\sim 2$, but within the scatter around 
the regression line.

We therefore conclude that the host galaxy in SDSS~J0850$+$4511 is in
no way anomalous but is instead typical of a galaxy in a low-redshift 
quasar, and that our conclusion that the host galaxy contribution to
the continuum under the \ion{He}{1}*$\lambda 10830$ line is negligible
is robust.

\end{document}